\documentclass[twocolumn, 10pt, aps, superscriptaddress, showpacs, prl, citeautoscript, longbibliography]{revtex4-2}

\usepackage[ngerman, english]{babel}   
\usepackage[utf8]{inputenc}
\usepackage{lmodern}
\usepackage{amssymb}
\usepackage{amsmath}
\usepackage{graphicx}
\usepackage{diagbox} 
\usepackage{hhline}  

\usepackage{xr} 
\usepackage{xcite} 
  
\usepackage[usenames,dvipsnames]{xcolor} 
\usepackage[colorlinks, citecolor={blue}, urlcolor={blue}, linkcolor={blue},hypertexnames=false]{hyperref}

\graphicspath{{figures/}} 
\usepackage{tikz}             

\usepackage{tabularx}                    
\usepackage{multirow, bigdelim}  
\usepackage{longtable}
\usepackage{makecell}                   
\usepackage{cellspace}

\usepackage{mdwlist}      

\usepackage{verbatim}       
\usepackage{bm}
\usepackage{bbm}              
\usepackage{dsfont}			
\usepackage{nicefrac}        
\usepackage[vcentermath]{youngtab} 
\usepackage{cancel}

\usepackage{etoolbox} 

\newcolumntype{L}[1]{>{\raggedright\arraybackslash}p{#1}} 
\newcolumntype{C}[1]{>{\centering\arraybackslash}p{#1}} 
\newcolumntype{R}[1]{>{\raggedleft\arraybackslash}p{#1}} 
\newcommand{\doubleI}{\mathds{1}}

\newcommand{\Sec}[1]{Sec.~\ref{#1}}

\newcommand{\Eq}[1]{Eq.~\eqref{#1}}
\newcommand{\Eqs}[1]{Eqs.~\eqref{#1}}
\newcommand{\Fig}[1]{Fig.~\ref{#1}}
\newcommand{\Figs}[1]{Figs.~\ref{#1}}

\definecolor{darkgreen}{rgb}{0,0.5,0}
\definecolor{orange}{rgb}{1,0.5,.3}
\definecolor{darkred}{rgb}{.7,0,0}
\definecolor{purple}{rgb}{0.6,0,0.5}
\definecolor{darkpetrol}{RGB}{0,73,76}

\usepackage[normalem]{ulem} 

\newcommand{\setsmalltitle}[1]{{\textit{#1}.}---}

\newcommand{\dd}{\mathrm{d}}
\newcommand{\ee}{\mathrm{e}}
\newcommand{\ii}{\mathrm{i}}

\makeatletter
\def\maketitle{
	\@author@finish
	\title@column\titleblock@produce
	\suppressfloats[t]}
\makeatother

\begin{document}

\title{Probing Polaron Clouds by Rydberg Atom Spectroscopy}

\author{Marcel Gievers}
\affiliation{Arnold Sommerfeld Center for Theoretical Physics, 
	Center for NanoScience,\looseness=-1\,  and Munich 
	Center for Quantum Science and Technology,\looseness=-2\, Ludwig-Maximilians-Universität München, 80333 Munich, Germany}
\affiliation{Max Planck Institute  of  Quantum  Optics,  
            85748  Garching,  Germany}

\author{Marcel Wagner}
\affiliation{Institut für Theoretische Physik, Universität Heidelberg, 69120 Heidelberg, Germany}

\author{Richard Schmidt}
\affiliation{Institut für Theoretische Physik, Universität Heidelberg, 69120 Heidelberg, Germany}

\date{\today}

\begin{abstract}
In recent years, Rydberg excitations in atomic quantum gases have become a successful platform to explore quantum impurity problems.
A single impurity immersed in a Fermi gas leads to the formation of a {polaron}, a quasiparticle consisting of the impurity being dressed by the surrounding medium. With a radius of about the Fermi wavelength, the density profile of a polaron cannot be explored using \emph{in situ} optical imaging techniques. In this work, we propose a new experimental measurement technique that enables the \emph{in situ} imaging of the polaron cloud in ultracold quantum gases. {The impurity atom induces the formation of a polaron cloud and is then excited} to a Rydberg state. Because of the mesoscopic interaction range of Rydberg excitations, which can be tuned by the principal numbers of the Rydberg state,
atoms extracted from the polaron cloud  form dimers with the impurity. By performing first principle calculations of the absorption spectrum based on a functional determinant approach, we show how the occupation of the dimer state can be directly observed in spectroscopy experiments and can be mapped onto the density profile of the gas particles, hence providing a direct, real-time, and \emph{in situ} measure of the polaron cloud.
\end{abstract}

\maketitle

Mixtures of quantum particles are ubiquitous in physics, ranging from neutron matter \cite{Kutschera1993} and the BCS-BEC crossover in atomic gases \cite{zwerger2011bcs} to superconducting phases in solid-state physics \cite{highTc}. Quantum mixtures have been investigated for many decades, but more recently, drastic progress in the controllability of experiments with ultracold atomic gases allows for new insights into a plethora of physical phenomena. In the limit of an extremely imbalanced quantum mixture, a single impurity is immersed in a Fermi gas. This leads to the formation of Fermi polarons, quasiparticles formed of gas particles which dress the impurity \cite{Prokofev1998,prokof2008fermi,schirotzek2009observation,Massignan2014}. Even though many polaronic properties are understood to a great extent \cite{Prokofev1998,Chevy2006,Comescot2007,prokof2008fermi,Mora2009,Punk2009,schirotzek2009observation,nascimbene_collective_2009,Zoellner2011,koschorreck2012attractive,kohstall_metastability_2012,wenz2013,Massignan2014,Vlietinck2013,Ong2015,Christensen2015b,Scazza2016,sidler2017fermi,schmidt2018universal,Yan2019Fermi,Oppong2019,Ness2020,Adlong2020,fritsche2021stability,Dolgirev2021,scazza2022repulsive}, a direct observation of the polaron dressing cloud in continuum systems has been out of reach.

Fermi polarons are generated by the short-range interaction between a Fermi gas and impurity particles, which in cold atoms can be tuned by Feshbach resonances.  The size of the resulting polaron dressing cloud is of the order of the Fermi wavelength, i.e., $r_c\sim k_\mathrm{F}^{-1}$. For typical densities of ultracold atoms, i.e., $\rho_0 = 10^{11}{-}10^{13}\,\mathrm{cm}^{-3}$, the relevant length scales lie in the suboptical regime, $r_c \sim 100{-}500\,\mathrm{nm}$, which hinders gaining insight into the real-space structure of these fundamentally important quasiparticles.

In this Letter, we demonstrate how to overcome this challenge enabling an \emph{in situ} measurement of the polaron cloud in cold atom experiments.  To this end, we propose a new measurement technique to explore the density profile around the impurity by use of atomic Rydberg states. Key to the idea is the use of the long-range interaction between the Rydberg atom and the bath particles. This interaction is generated by the outermost electron on the Rydberg orbit \cite{du1987interaction} and induces the formation of ultralong-range Rydberg molecules (ULRMs), i.e., deeply bound states of atoms inside the interaction potential \cite{greene2000creation,bendkowsky2009observation,butscher2010atom,DeSalvo2015Ultralong,camargo2016lifetimes,camargo2018creation,kleinbach2018ionic,Whalen2019}. Intriguingly, the extent of the Rydberg atoms of $r_\mathrm{Ryd}=50{-}500\,\mathrm{nm}$ matches precisely the typical size of the polaron dressing cloud. Hence, by tuning the principal quantum numbers $n_\mathrm{Ryd}$ of the Rydberg excitation, the binding length of ULRMs is tuned through the polaron cloud (cf.~\Fig{fig:Experimental-procedure}). ULRMs can thus serve as a precision sensor inserted into the polaron cloud. The occupation of ULRMs is detected via a straightforward measurement of the \textit{optical} linear response absorption and can be mapped onto the \textit{suboptical} size of the polaron cloud. Due to its fermionic nature, our method differs from recent probing of a BEC with Rydberg impurities, as for the fermionic systems studied here, the number of particles in the Rydberg radius remains always small \cite{schlagmuller2016probing,kleinbach2018ionic,veit2021pulsed}.

We calculate Rydberg absorption spectra in the presence of a polaron cloud around the impurity using a functional determinant approach\cite{knap2012time,schmidt2018theory,sous2020rydberg}. The Rydberg blockade mechanism \cite{low2012experimental}
ensures that our single-impurity calculations are applicable. While our approach becomes exact in the limit of heavy impurities immersed in a gas of lighter atoms, the idea of Rydberg sensing of polaron clouds can be extended to arbitrary mass ratios \cite{cetina2016ultrafast,liu2019variational}.
We show that, when the Rydberg excitation is immersed in a polaron cloud, the weight of the peak in the Rydberg absorption spectrum corresponding to the ULRM ground state gives
direct access to the density evaluated at a distance $r_{\mathrm{Ryd}}$ from the impurity. This way, the complete density profile of polaron clouds, which so far eluded experimental observations, can be mapped out by use of a simple Rydberg spectroscopy experiment.

\setsmalltitle{Model}
We consider a Fermi gas combined with a single charge-neutral and immobile impurity atom. The impurity can be brought into three states $\sigma \in \{ 0, 1, \mathrm{R}\}$. For $\sigma=0$, the impurity is not interacting with the bath particles. For $\sigma=1$, the impurity interacts with the bath particles via a short-range interaction
that induces the formation of a Fermi polaron \footnote{We use here the terminology of polarons also in the infinite impurity mass limit where  Anderson's orthogonality leads formally to a vanishing  polaron quasiparticle weight \cite{schmidt2018universal}. For the physics of the polaron cloud formation this is, however, not relevant here.}. For $\sigma = \mathrm{R}$, the impurity is in the Rydberg state, which evokes the long-range interaction with the bath particles.
The Hamiltonian reads:
\begin{subequations}
	\begin{align}
		\hat H &=
		\hat{\mathds{1}}\otimes\hat H_0 + \sum_{\sigma = 1, \text{R}}|\sigma\rangle\langle\sigma|\otimes\hat V_\sigma ,\\
		\hat H_0 &= \sum_{\bm{k}}\varepsilon_{\bm{k}}\hat{c}^\dagger_{\bm{k}}\hat c_{\bm{k}},\quad\hat V_\sigma = \int_{\bm{r}}V_\sigma(\bm{r})\hat c^\dagger_{\bm{r}}\hat c_{\bm{r}}.
	\end{align}
\end{subequations}
Here, $\hat c^\dagger_{\bm{k}},\hat c_{\bm{k}}$ are fermionic operators of the gas and $|\sigma\rangle$ is the state of the impurity. When the impurity is in the state $|\sigma\rangle$, the time evolution of the fermionic gas is given by the Hamiltonian
$\hat{H}_\sigma = \hat H_0 + \hat V_\sigma$,
acting only on the fermionic subspace.

\begin{figure}
 \includegraphics[width=\linewidth]{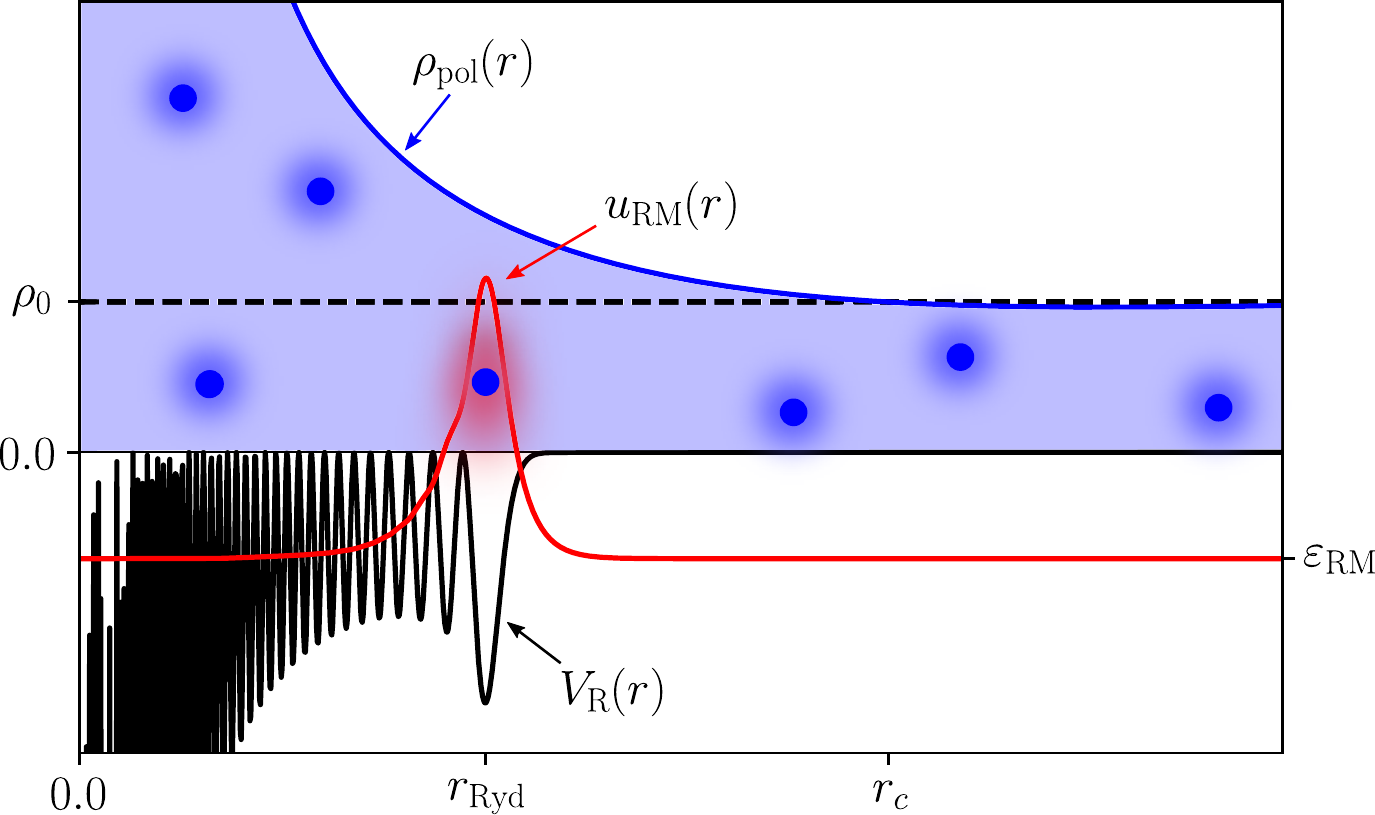}
	\caption{
    A Rydberg atom in a polaron cloud. The bath density $\rho_\mathrm{pol}(r)$ (blue line) is increased at the center compared to the background density $\rho_0$ (dashed line). The Rydberg potential $V_\mathrm{R}(r)$ (black line) is tuned such that the outermost bound state $u_\mathrm{RM}(r)$ (red line) at $r_\mathrm{Ryd}$ is situated near the polaron cloud radius $r_c$.}
	\label{fig:Experimental-procedure}
\end{figure}

We propose the following procedure.
At the beginning, the impurity is in the noninteracting state $|0\rangle$ and the bath particles form a Fermi sea $|\mathrm{FS}\rangle$, yielding the many-body state $|\psi(t_0)\rangle = {|0\rangle\otimes|\mathrm{FS}\rangle}$.
By a radio-frequency (rf) pulse, the impurity is then
switched to the short-range interacting state, i.e.,  ${|\psi(t_0+0^+)\rangle} = |1\rangle\otimes |\mathrm{FS}\rangle$. Time evolution leads to the formation of the polaron cloud around the impurity atom. After sufficiently long dephasing time, the system is well described by $|\psi(t_1)\rangle = |1\rangle \otimes |\mathrm{pol}\rangle$. Finally, by driving an optical transition, the impurity atom is transferred to the Rydberg state, i.e., ${|\psi(t_1+0^+)\rangle} = |\mathrm{R}\rangle\otimes|\mathrm{pol}\rangle$. This way, the Rydberg atom is, by construction, exactly placed in the center of the polaron cloud as illustrated in \Fig{fig:Experimental-procedure}.

We simulate the impurity in the $|1\rangle$ state by a delta potential with an $s$-wave scattering length $a$ \cite{knap2012time,supplement}.\nocite{knap2012time,schmidt2018theory,sous2020rydberg,sakurai1995modern,bloch2008many,du1987interaction,Burkhardt2006,eiles2018formation,klich2003elementary} The potential in the $|\mathrm{R}\rangle$ state, which is generated by scattering of the Rydberg electron with the bath particles, is given by \cite{du1987interaction}
\begin{align}
	V_{\mathrm{R}}(\bm{r}) = \frac{2\pi\hbar^2 a_e}{m_e}|\psi_{n_\mathrm{Ryd}}(\bm{r})|^2.
    \label{eq:Rydberg-potential-general}
\end{align}
Here, $\psi_{n_\mathrm{Ryd}}(\bm{r})$ is the wave function of an $s$-wave Rydberg electron with principal number $n_\mathrm{Ryd}$ and scattering length $a_e$ between an electron and neutral atoms of the background gas \cite{supplement}. Because of the nonlocal potential $V_\mathrm{R}(\bm{r})$, there is a finite overlap between the polaron cloud and bound states inside $V_\mathrm{R}(\bm{r})$ (cf.~\Fig{fig:Experimental-procedure}).

For the calculation of physical quantities, we use the functional determinant approach, which is a standard method for determining spectra \cite{cetina2016ultrafast}.
Specifically, the density $\rho_\sigma(\bm{r},t)$ around the impurity in state $|\sigma\rangle$
is obtained by a Klich formula \cite{supplement,klich2003elementary,schonhammer2007full}:
\begin{align}
	\rho_\sigma(\bm{r},t) &= \mathrm{tr}[\hat\rho(t)\hat c^\dagger_{\bm{r}}\hat c_{\bm{r}}] = \langle\bm{r}|\ee^{-\ii\hat h_\sigma t}n_\mathrm{F}(\hat h_0)\ee^{\ii\hat h_\sigma t}|\bm{r}\rangle,
    \label{eq:time-dependent-density}
\end{align}
where $n_\mathrm{F}(\varepsilon)=(\ee^{\beta(\varepsilon-\mu)}+1)^{-1}$ is the Fermi-Dirac distribution with inverse temperature $\beta$ and chemical potential $\mu$ and we set $\hbar = 1 = k_\mathrm{B}$. The single-particle operator $\hat h_\sigma$ corresponds to the Hamiltonian of the gas particles $\hat H_\sigma$. 

The absorption spectrum of the Rydberg atom inside the polaron is obtained from Fermi's golden rule,
\begin{align}
    \nonumber A_\mathrm{pol}(\omega) &= 2\pi \sum_f|\langle f|\mathrm{pol}\rangle|^2\delta[\omega-(E_f-E_i)]\\
    &= 2\mathrm{Re}\int_0^\infty\dd t\,\ee^{\ii\omega t}\langle\mathrm{pol}|\ee^{\ii\hat H_1t}\ee^{-\ii\hat H_\mathrm{R}t}|\mathrm{pol}\rangle,
    \label{eq:Absorption-spectrum}
\end{align}
where  the gas is initially in the polaron state $|\mathrm{pol}\rangle$ with total energy $E_i$ and $|f\rangle$ represent the complete set of final states of the gas in presence of the Rydberg impurity $|\mathrm{R}\rangle$ with total energies $E_f$.

We obtain the absorption spectrum as the Fourier transform of the Ramsey signal \cite{supplement,schmidt2018universal}, which is calculated as a time-dependent Slater determinant \cite{supplement,knap2012time,sous2020rydberg}. Through $\hat H_1$ and $\hat H_\mathrm{R}$, $A_\mathrm{pol}(\omega)$ depends on both the Rydberg principal quantum number $n_\mathrm{Ryd}$ and the scattering length $a$ of the polaron. In the following, we consider the system at zero temperature \footnote{To be precise, we have run our calculations at $T=0.001\,\varepsilon_\mathrm{F}$.}
and express
physical quantities in terms of the Fermi momentum $k_\mathrm{F}$ and Fermi energy $\varepsilon_\mathrm{F}$, respectively \cite{supplement}. Still, our method is robust against finite temperature as it only depends on the weight of spectral peaks and not their widths \cite{supplement}.

\setsmalltitle{Fermi polaron cloud}
Before turning to its observation, we first describe the formation of the polaron cloud in the initial state of the system.
In particular, we are interested in the stationary density profile, which is established after a hold time $t_1\gg 1/\varepsilon_\mathrm{F}$ \cite{supplement}. 
In that limit, the density profile \Eq{eq:time-dependent-density} 
 close to the impurity is well described by
the ground state of the Hamiltonian $\hat{H}_1$ and 
given by
\begin{align}
	\rho_\mathrm{pol}(\bm{r}) &= \langle \bm{r} |n_\mathrm{F}(\hat h_1) |\bm{r}\rangle.
	\label{eq:stationary-density}
\end{align}
Because of spherical symmetry, the density depends only on the distance to the impurity, i.e., $\rho_\mathrm{pol}=\rho_\mathrm{pol}(r)$.

\begin{figure}
    \raggedright
    (a)\\
    \vspace{-0.4cm}
    \centering 
    \includegraphics[width=\linewidth]{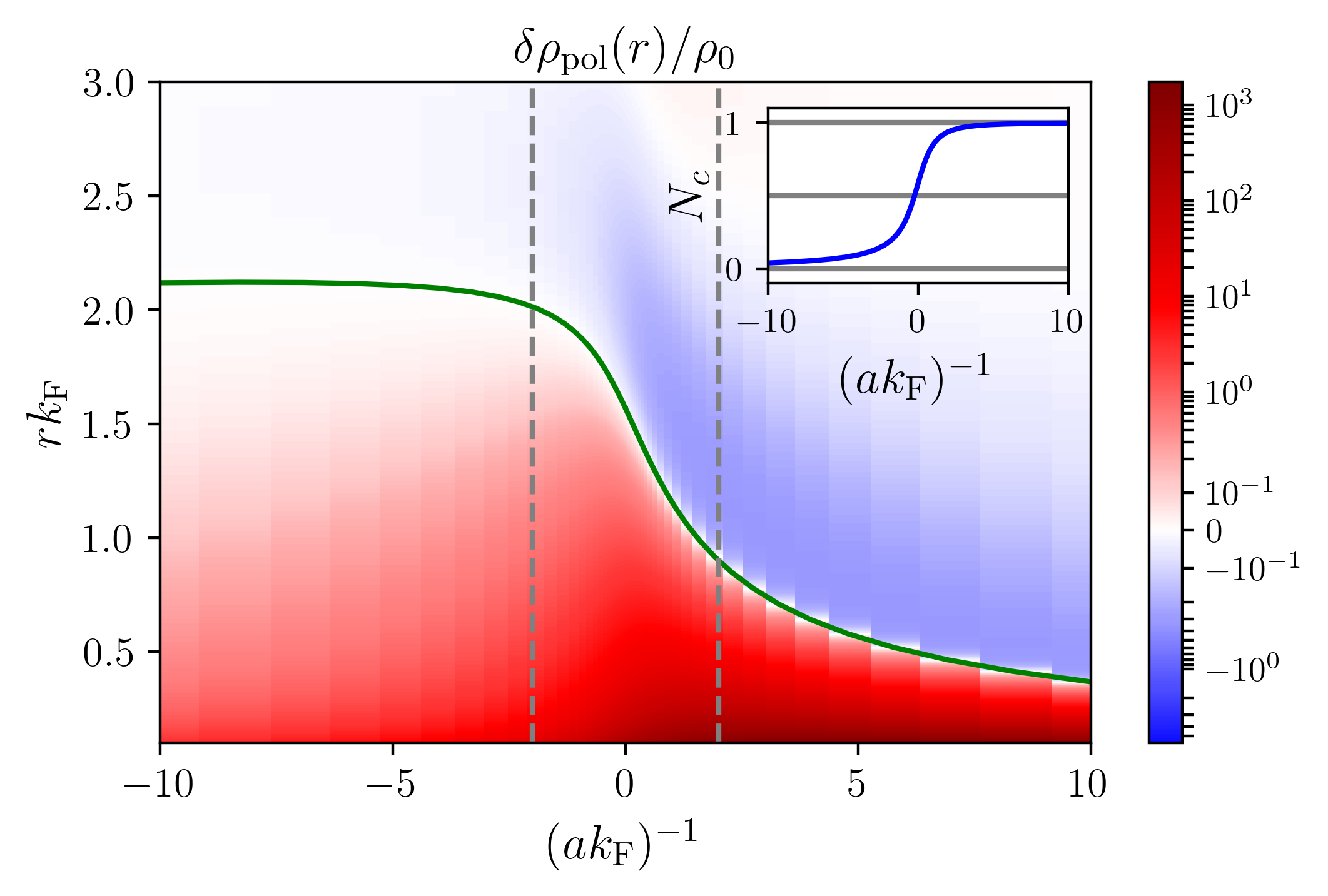}
    \raggedright
    (b)\\
    \vspace{-0.4cm}
    \centering
    \includegraphics[width=\linewidth]{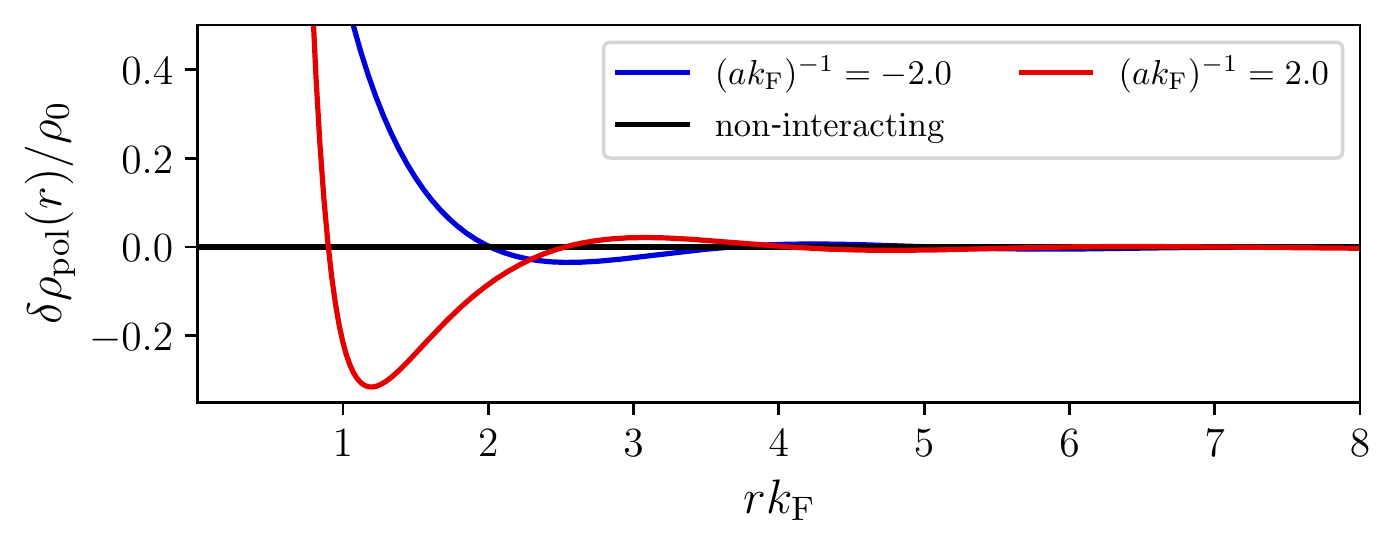}
	\caption{
 (a) Polaron density profiles $\delta\rho_\mathrm{pol}(r) = \rho_\mathrm{pol}(r) - \rho_0$ in dependence on the inverse scattering lengths $(a k_\mathrm{F})^{-1}$ measured in terms of $\rho_0$. 
 The size of the polaron cloud $r_c$ is marked in green. The integrated number of excess particles in the cloud $N_c$
 is shown in the inset. Note that the colorplot is semi-logarithmic.
 (b) Polaron cloud density for  two exemplary scattering lengths marked as gray dashed lines in the upper plot.
    }  
	\label{fig:polaron_clouds}
\end{figure}

Figure \ref{fig:polaron_clouds} shows the density of the polaron cloud
as a function of the inverse  dimensionless scattering length $(ak_\mathrm{F})^{-1}$ measured with respect to the background density, $\rho_0=k_\mathrm{F}^3/(6\pi^2)$.
For $a<0$, the
single-particle wave functions are drawn toward the impurity, resulting in a density enhancement near $r=0$. This enhancement is accompanied  by Friedel-like oscillations farther away from the impurity [see also \Fig{fig:polaron_clouds}(b)]. On the contrary, for $a>0$, the
single-particle scattering wave functions are pushed away from the impurity. However, the bound state emerging at positive scattering length still leads to an overall enhancement of the density near the impurity \cite{schmidt2018theory}. Note that
the particle density is formally divergent at $r=0$, which is an artifact of the contact interaction and not present for physical finite-range potentials \cite{supplement}.
However, for all our considerations, the delta impurity is a valid approximation as the usual van der Waals length of the atoms is much shorter than the size of the Rydberg state.
Importantly, the integrated number of particles in the polaron cloud converges to a well-defined, finite number, also in the limit of contact interaction.

We define the region with an increased particle density around the impurity as the \emph{polaron cloud}. The size $r_c$ of the polaron cloud, visualized by a green line in \Fig{fig:polaron_clouds}(a),  is determined by the first crossing of $\rho_\mathrm{pol}(r_c)=\rho_0$ and it is of the order of the Fermi wavelength.

The number of particles contributing to the polaron cloud $N_c$ is given by the integrated number of excess atoms within the volume defined by $r_c$, i.e., $N_c = N_\mathrm{pol}(r_c) - N_0(r_c)$ \cite{supplement}.
Note that the number of particles contributing to the polaron cloud is at most one particle despite the infinitely many particle-hole excitations required to obtain the exact many-body solution \cite{anderson1967infrared}. This is confirmed by a thermodynamic consideration using Fumi's theorem \cite{schmidt2018universal,supplement}.
However, although the  number of contributing particles is small, it is  the enormous density increase at the center that results in a significant effect in absorption spectroscopy in the presence of the Rydberg excitation. 

We note that the real-time evolution of the polaron cloud formation can be obtained in a similar fashion by directly applying \Eq{eq:time-dependent-density}. While the corresponding absorption spectra, which then track the real-time formation of the polaron cloud, can also be calculated  using linear response theory,
in this work we focus on the quasistationary limit [cf.~\Eq{eq:stationary-density}].

\setsmalltitle{Rydberg atom spectroscopy}
Let us now describe the Rydberg absorption spectra in the presence of a polaron cloud.
The potential $V_\mathrm{R}(r)$ in \Eq{eq:Rydberg-potential-general}
features a pronounced minimum at the Rydberg radius $r_\mathrm{Ryd}$ (see~\Fig{fig:Experimental-procedure}). This minimum 
supports a spatially confined bound state leading to a prominent dimer peak in the absorption spectrum \cite{bendkowsky2009observation,butscher2010atom,low2012experimental,Whalen2019,sous2020rydberg,supplement}.
We refer to that
state as the \emph{Rydberg molecule (RM)} to differentiate it from higher excited bound states. The Rydberg radius $r_\mathrm{Ryd}$ can be tuned by the principal quantum number $n_\mathrm{Ryd}$ 
and is characteristic for the atomic species of the impurity (in our case $^{87}$Rb).

\begin{figure}
    \includegraphics[width=\linewidth]{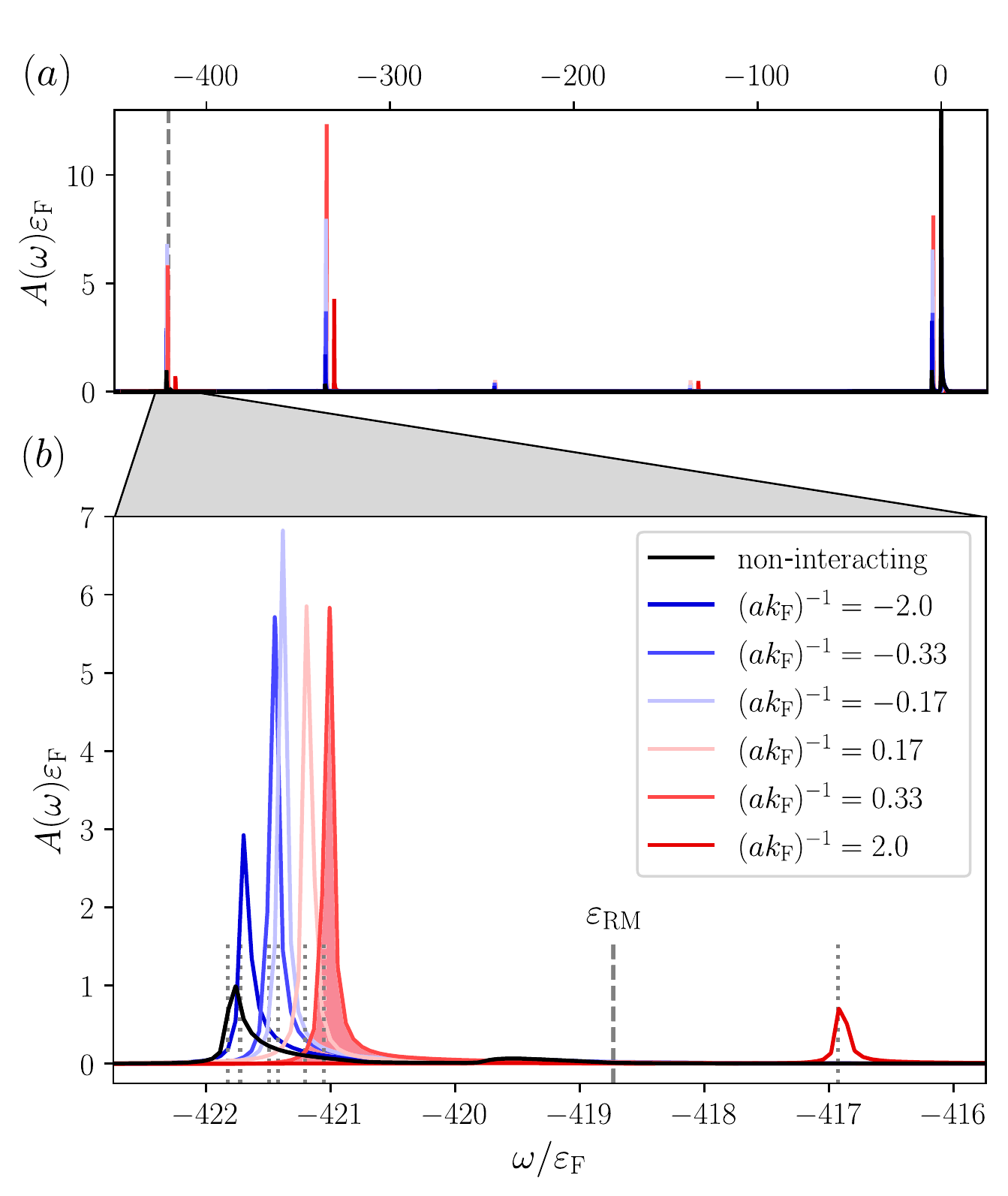}
	\caption{(a) Absorption spectrum of a Rydberg impurity with $n_\mathrm{Ryd}=60$ generated in polarons of different inverse scattering lengths $(ak_\mathrm{F})^{-1}$ at background density $\rho_0 = 5\times 10^{11}\,\mathrm{cm}^{-3}$. The binding energy $\varepsilon_\mathrm{RM}$ is marked as a dashed gray line. (b) Magnification of the RM peak: Peak positions $\omega_\mathrm{peak}$ \Eq{eq:peak-position} are marked by gray dotted lines. The red shaded region below the curve for $(ak_\mathrm{F})^{-1}=0.33$ indicates the value of the corresponding peak weight $I_\mathrm{pol}$. The calculations are performed for a $^{87}$Rb impurity in $^{40}$K particles.
    }  
	\label{fig:Rydberg-polarons}
\end{figure}

A typical absorption spectrum  calculated in the presence of the polaron cloud is shown in \Fig{fig:Rydberg-polarons}(a).
The visible peaks correspond to the various bound states between the Rydberg atom and bath particles. For  $a=0$, the density $\rho_0$ is spatially constant in the initial state and one recovers the results for a single Rydberg impurity in a Fermi gas \cite{Whalen2019,sous2020rydberg}. Because of its good overlap with the scattering states of the Fermi gas, the RM with binding energy $\varepsilon_\mathrm{RM}$ (indicated by the dashed vertical line) has the largest oscillator strength.

For $a\neq 0$, a Fermi polaron is formed in the initial state. In this case the RM peak
does not necessarily remain the most prominent peak of the spectrum. This can be understood from  the increased density close to the impurity that causes bound states localized more closely to the impurity to have a larger overlap with the polaron's scattering states.

\begin{figure}
	\includegraphics[width=\linewidth]{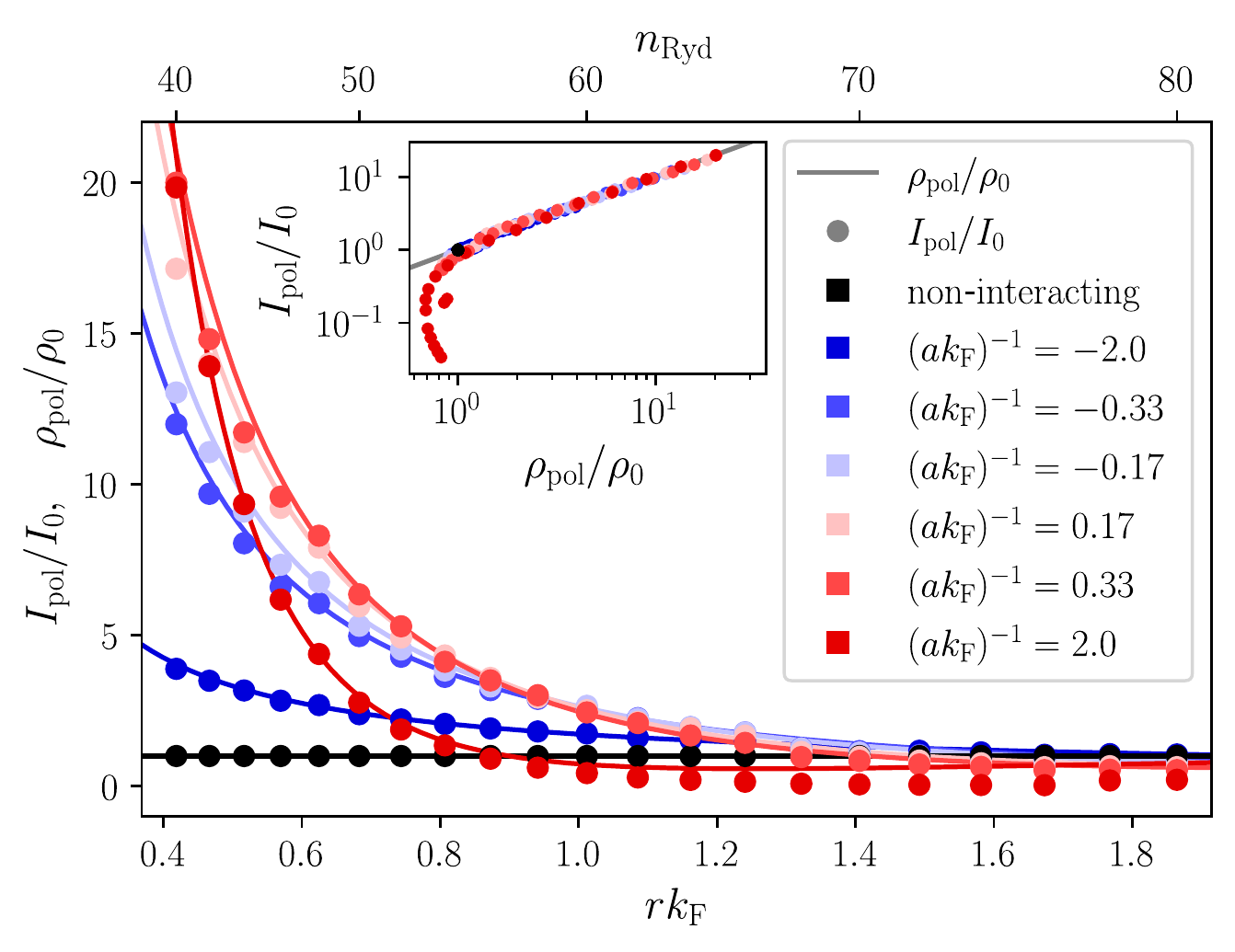}
	\caption{Normalized density profiles $\rho_\mathrm{pol}(r)/\rho_0$ \Eq{eq:stationary-density} for polaron clouds formed at inverse scattering lengths $(ak_\mathrm{F})^{-1}$ (solid lines) compared to the integrated dimer response $I_\mathrm{pol}(r_\mathrm{Ryd})/I_0(r_\mathrm{Ryd})$ \Eq{eq:I_pol-integral} (dots). The latter corresponds to the Rydberg radius through the respective principal numbers, i.e., $r_\mathrm{Ryd}(n_\mathrm{Ryd})$. We use a fixed $\rho_0={5\times 10^{11}\,\mathrm{cm}^{-3}}$. The inset shows the dependence between $I_\mathrm{pol}/I_0$ and the densities $\rho_\mathrm{pol}/\rho_0$. Apart from deviations for small densities, the data points show a tight relation, which is underlined by the gray line marking identity. The calculations are performed for a $^{87}$Rb impurity in $^{40}$K particles.}
	\label{fig:Rydberg-densities-delta-as}
\end{figure}

Looking closer at the RM response [cf.~\Fig{fig:Rydberg-polarons}(b)], we observe two key changes in the spectrum that enable the spectroscopy of  polarons: a) the weight of the peaks is modified and b) their position is shifted.
In order to associate the weight of the RM peak with the polaron cloud density at  distance $r_\mathrm{Ryd}$,
we perform an integral over a frequency window that encompasses the dimer peak [see exemplary shaded region in  \Fig{fig:Rydberg-polarons}(b)]:
\begin{align}
    I_\mathrm{pol}(n_\mathrm{Ryd}) = \int_{\mathrm{peak}}\dd\omega\, A_\mathrm{pol}(n_\mathrm{Ryd},\omega).
    \label{eq:I_pol-integral}
\end{align}
For absorption spectra at different principal numbers and scattering lengths, we calculate the integrated weight as a function of the principal number, i.e., $I_\mathrm{pol}(n_\mathrm{Ryd})$. Crucially, since $n_\mathrm{Ryd}$ is directly related to $r_\mathrm{Ryd}$, these values are associated with the density $\rho_\mathrm{pol}$ at the respective distance. The comparison of $I_\mathrm{pol}(n_\mathrm{Ryd})=I_\mathrm{pol}(r_\mathrm{Ryd})$ with $\rho_\mathrm{pol}(r)$ is shown in \Fig{fig:Rydberg-densities-delta-as}, where we normalized the signal strength and density  
by the noninteracting values $I_0(r_\mathrm{Ryd})$ and $\rho_0$. We find striking agreement between the integrated weights of the dimer peaks $I_\mathrm{pol}(r_\mathrm{Ryd})/I_0(r_\mathrm{Ryd})$ and the densities $\rho_\mathrm{pol}(r)/\rho_0$.
The strong correlation between both quantities is further analyzed in the inset of \Fig{fig:Rydberg-densities-delta-as}. Especially for larger density values, the mapping from absorption response to polaron cloud densities works exceptionally well. As our procedure only depends on the integrated spectral weight, it is robust against broadening effects on the spectrum, i.e., finite temperatures \cite{cetina2016ultrafast,supplement}, mobile impurities \cite{liu2019variational} and the finite lifetime of the Rydberg excitation \cite{camargo2016lifetimes}.

Our predictions are made in units of the Fermi momentum, i.e., $r_c = r_c(k_\mathrm{F})$. Hence, another way of tuning the location of the Rydberg radius $r_\mathrm{Ryd}$ relative to the polaron cloud is by changing the overall density of the medium $\rho_0=k_\mathrm{F}^3/(6\pi^2)$. The discussion of density profiles reconstructed using this alternative method is provided in the Supplemental Material \cite{supplement}.

The shifted positions of the RM peaks (cf.~\Fig{fig:Rydberg-polarons}) allow us to directly measure the energy of attractive Fermi polarons. The peak position is given by
\begin{align}
    \omega_\mathrm{peak} &= \varepsilon_\mathrm{RM} - E_{\mathrm{pol}}(a) + E_{\mathrm{pol},\mathrm{R}}(n_\mathrm{Ryd}),
    \label{eq:peak-position}
\end{align}
which makes the argument evident. First, the RM peak energy is reduced by the energy of the attractive polaron $E_{\mathrm{pol}}(a)$ in the initial state \cite{schmidt2018universal}.
Second, the presence of the Rydberg impurity shifts the single-particle energies of the Fermi gas itself once more, resulting in an additional polaron shift given by
$E_{\mathrm{pol},\mathrm{R}}(n_\mathrm{Ryd})$. It is quite noticeable that our absorption spectra thus simultaneously provide two complementary properties of the polaron: it resolves its energy as well as its real-space density profile.

\setsmalltitle{Conclusion}
We have proposed a new technique for probing the dressing cloud of polarons  using Rydberg spectroscopy.
ULRMs are the key feature enabling the approach. They are formed at a specific distance from the impurity and lead to dimer peaks that can be uniquely identified in absorption spectra. When tuning the principal number $n_\mathrm{Ryd}$ of the Rydberg state, the location of the ULRM is changed and the integrated weight of its dimer peaks directly corresponds to the density of gas particles. 
While we focused on the case of Fermi polarons, our procedure is general and can be extended to mobile impurities featuring molaron states \cite{diessel2022probing}, and other  correlated many-body states such as Bose polarons \cite{jorgensen2016,hu2016,Yan2019,Ardila2019} or polarons created in BCS superfluids \cite{wang2022exact}. Thus, the proposed technique paves the way to completely new observations in experiments with ultracold atoms that can  probe length scales beyond the optical regime in an \emph{in situ} fashion.

Finally, by employing  deeply bound states as a probe, the proposed spectroscopy allows for observing dynamics on  timescales that are ultrafast compared to the typical scales of the underlying many-body system. This allows us, for example, to investigate the formation of the polaron cloud in real time. In theory, this can be simulated by considering the linear response to the switch-on of the impurity similar to pump-probe spectroscopy. 

\noindent
\\
\setsmalltitle{Acknowledgments}
We thank Hossein Sadeghpour and Márton Kanász-Nagy for fruitful discussions.  We acknowledge funding for M.G. from the International Max Planck Research School for Quantum Science and Technology (IMPRS-QST). M.W. and R.S. acknowledge support by the DFG Priority Programme of Giant Interactions in Rydberg Systems (GiRyd) and by the Deutsche Forschungsgemeinschaft under Germany's Excellence Strategy EXC 2181/1 - 390900948 (the Heidelberg STRUCTURES Excellence Cluster). The numerical simulations were performed on the Linux clusters at the Leibniz Supercomputing Center.

\bibliographystyle{apsrev4-2}
\bibliography{bibliography}

\begin{thebibliography}{63}%
\makeatletter
\providecommand \@ifxundefined [1]{%
 \@ifx{#1\undefined}
}%
\providecommand \@ifnum [1]{%
 \ifnum #1\expandafter \@firstoftwo
 \else \expandafter \@secondoftwo
 \fi
}%
\providecommand \@ifx [1]{%
 \ifx #1\expandafter \@firstoftwo
 \else \expandafter \@secondoftwo
 \fi
}%
\providecommand \natexlab [1]{#1}%
\providecommand \enquote  [1]{``#1''}%
\providecommand \bibnamefont  [1]{#1}%
\providecommand \bibfnamefont [1]{#1}%
\providecommand \citenamefont [1]{#1}%
\providecommand \href@noop [0]{\@secondoftwo}%
\providecommand \href [0]{\begingroup \@sanitize@url \@href}%
\providecommand \@href[1]{\@@startlink{#1}\@@href}%
\providecommand \@@href[1]{\endgroup#1\@@endlink}%
\providecommand \@sanitize@url [0]{\catcode `\\12\catcode `\$12\catcode
  `\&12\catcode `\#12\catcode `\^12\catcode `\_12\catcode `\%12\relax}%
\providecommand \@@startlink[1]{}%
\providecommand \@@endlink[0]{}%
\providecommand \url  [0]{\begingroup\@sanitize@url \@url }%
\providecommand \@url [1]{\endgroup\@href {#1}{\urlprefix }}%
\providecommand \urlprefix  [0]{URL }%
\providecommand \Eprint [0]{\href }%
\providecommand \doibase [0]{https://doi.org/}%
\providecommand \selectlanguage [0]{\@gobble}%
\providecommand \bibinfo  [0]{\@secondoftwo}%
\providecommand \bibfield  [0]{\@secondoftwo}%
\providecommand \translation [1]{[#1]}%
\providecommand \BibitemOpen [0]{}%
\providecommand \bibitemStop [0]{}%
\providecommand \bibitemNoStop [0]{.\EOS\space}%
\providecommand \EOS [0]{\spacefactor3000\relax}%
\providecommand \BibitemShut  [1]{\csname bibitem#1\endcsname}%
\let\auto@bib@innerbib\@empty
\bibitem [{\citenamefont {Kutschera}\ and\ \citenamefont
  {W\'ojcik}(1993)}]{Kutschera1993}%
  \BibitemOpen
  \bibfield  {author} {\bibinfo {author} {\bibfnamefont {M.}~\bibnamefont
  {Kutschera}}\ and\ \bibinfo {author} {\bibfnamefont {W.}~\bibnamefont
  {W\'ojcik}},\ }\href {https://doi.org/10.1103/PhysRevC.47.1077} {\bibfield
  {journal} {\bibinfo  {journal} {Phys. Rev. C}\ }\textbf {\bibinfo {volume}
  {47}},\ \bibinfo {pages} {1077} (\bibinfo {year} {1993})}\BibitemShut
  {NoStop}%
\bibitem [{\citenamefont {Zwerger}(2011)}]{zwerger2011bcs}%
  \BibitemOpen
  \bibfield  {author} {\bibinfo {author} {\bibfnamefont {W.}~\bibnamefont
  {Zwerger}},\ }\href@noop {} {\emph {\bibinfo {title} {The BCS-BEC crossover
  and the unitary Fermi gas}}},\ Vol.\ \bibinfo {volume} {836}\ (\bibinfo
  {publisher} {Springer Science \& Business Media},\ \bibinfo {year}
  {2011})\BibitemShut {NoStop}%
\bibitem [{\citenamefont {Schrieffer}\ and\ \citenamefont
  {Brooks}(2007)}]{highTc}%
  \BibitemOpen
  \bibfield  {author} {\bibinfo {author} {\bibfnamefont {J.~R.}\ \bibnamefont
  {Schrieffer}}\ and\ \bibinfo {author} {\bibfnamefont {J.~S.}\ \bibnamefont
  {Brooks}},\ }\href@noop {} {\emph {\bibinfo {title} {Handbook of
  high-temperature superconductivity}}}\ (\bibinfo  {publisher} {Springer},\
  \bibinfo {year} {2007})\BibitemShut {NoStop}%
\bibitem [{\citenamefont {Prokof'ev}\ and\ \citenamefont
  {Svistunov}(1998)}]{Prokofev1998}%
  \BibitemOpen
  \bibfield  {author} {\bibinfo {author} {\bibfnamefont {N.~V.}\ \bibnamefont
  {Prokof'ev}}\ and\ \bibinfo {author} {\bibfnamefont {B.~V.}\ \bibnamefont
  {Svistunov}},\ }\href {https://doi.org/10.1103/PhysRevLett.81.2514}
  {\bibfield  {journal} {\bibinfo  {journal} {Phys. Rev. Lett.}\ }\textbf
  {\bibinfo {volume} {81}},\ \bibinfo {pages} {2514} (\bibinfo {year}
  {1998})}\BibitemShut {NoStop}%
\bibitem [{\citenamefont {Prokof’ev}\ and\ \citenamefont
  {Svistunov}(2008)}]{prokof2008fermi}%
  \BibitemOpen
  \bibfield  {author} {\bibinfo {author} {\bibfnamefont {N.}~\bibnamefont
  {Prokof’ev}}\ and\ \bibinfo {author} {\bibfnamefont {B.}~\bibnamefont
  {Svistunov}},\ }\href
  {https://journals.aps.org/prb/abstract/10.1103/PhysRevB.77.020408} {\bibfield
   {journal} {\bibinfo  {journal} {Phys. Rev. B}\ }\textbf {\bibinfo {volume}
  {77}},\ \bibinfo {pages} {020408} (\bibinfo {year} {2008})}\BibitemShut
  {NoStop}%
\bibitem [{\citenamefont {Schirotzek}\ \emph {et~al.}(2009)\citenamefont
  {Schirotzek}, \citenamefont {Wu}, \citenamefont {Sommer},\ and\ \citenamefont
  {Zwierlein}}]{schirotzek2009observation}%
  \BibitemOpen
  \bibfield  {author} {\bibinfo {author} {\bibfnamefont {A.}~\bibnamefont
  {Schirotzek}}, \bibinfo {author} {\bibfnamefont {C.-H.}\ \bibnamefont {Wu}},
  \bibinfo {author} {\bibfnamefont {A.}~\bibnamefont {Sommer}},\ and\ \bibinfo
  {author} {\bibfnamefont {M.~W.}\ \bibnamefont {Zwierlein}},\ }\href
  {https://journals.aps.org/prl/abstract/10.1103/PhysRevLett.102.230402}
  {\bibfield  {journal} {\bibinfo  {journal} {Phys. Rev. Lett.}\ }\textbf
  {\bibinfo {volume} {102}},\ \bibinfo {pages} {230402} (\bibinfo {year}
  {2009})}\BibitemShut {NoStop}%
\bibitem [{\citenamefont {Massignan}\ \emph {et~al.}(2014)\citenamefont
  {Massignan}, \citenamefont {Zaccanti},\ and\ \citenamefont
  {Bruun}}]{Massignan2014}%
  \BibitemOpen
  \bibfield  {author} {\bibinfo {author} {\bibfnamefont {P.}~\bibnamefont
  {Massignan}}, \bibinfo {author} {\bibfnamefont {M.}~\bibnamefont
  {Zaccanti}},\ and\ \bibinfo {author} {\bibfnamefont {G.~M.}\ \bibnamefont
  {Bruun}},\ }\href {http://stacks.iop.org/0034-4885/77/i=3/a=034401}
  {\bibfield  {journal} {\bibinfo  {journal} {Reports on Progress in Physics}\
  }\textbf {\bibinfo {volume} {77}},\ \bibinfo {pages} {034401} (\bibinfo
  {year} {2014})}\BibitemShut {NoStop}%
\bibitem [{\citenamefont {Chevy}(2006)}]{Chevy2006}%
  \BibitemOpen
  \bibfield  {author} {\bibinfo {author} {\bibfnamefont {F.}~\bibnamefont
  {Chevy}},\ }\href {https://doi.org/10.1103/PhysRevA.74.063628} {\bibfield
  {journal} {\bibinfo  {journal} {Phys. Rev. A}\ }\textbf {\bibinfo {volume}
  {74}},\ \bibinfo {pages} {063628} (\bibinfo {year} {2006})}\BibitemShut
  {NoStop}%
\bibitem [{\citenamefont {Combescot}\ \emph {et~al.}(2007)\citenamefont
  {Combescot}, \citenamefont {Recati}, \citenamefont {Lobo},\ and\
  \citenamefont {Chevy}}]{Comescot2007}%
  \BibitemOpen
  \bibfield  {author} {\bibinfo {author} {\bibfnamefont {R.}~\bibnamefont
  {Combescot}}, \bibinfo {author} {\bibfnamefont {A.}~\bibnamefont {Recati}},
  \bibinfo {author} {\bibfnamefont {C.}~\bibnamefont {Lobo}},\ and\ \bibinfo
  {author} {\bibfnamefont {F.}~\bibnamefont {Chevy}},\ }\href
  {https://doi.org/10.1103/PhysRevLett.98.180402} {\bibfield  {journal}
  {\bibinfo  {journal} {Phys. Rev. Lett.}\ }\textbf {\bibinfo {volume} {98}},\
  \bibinfo {pages} {180402} (\bibinfo {year} {2007})}\BibitemShut {NoStop}%
\bibitem [{\citenamefont {Mora}\ and\ \citenamefont {Chevy}(2009)}]{Mora2009}%
  \BibitemOpen
  \bibfield  {author} {\bibinfo {author} {\bibfnamefont {C.}~\bibnamefont
  {Mora}}\ and\ \bibinfo {author} {\bibfnamefont {F.}~\bibnamefont {Chevy}},\
  }\href {https://doi.org/10.1103/PhysRevA.80.033607} {\bibfield  {journal}
  {\bibinfo  {journal} {Phys. Rev. A}\ }\textbf {\bibinfo {volume} {80}},\
  \bibinfo {pages} {033607} (\bibinfo {year} {2009})}\BibitemShut {NoStop}%
\bibitem [{\citenamefont {Punk}\ \emph {et~al.}(2009)\citenamefont {Punk},
  \citenamefont {Dumitrescu},\ and\ \citenamefont {Zwerger}}]{Punk2009}%
  \BibitemOpen
  \bibfield  {author} {\bibinfo {author} {\bibfnamefont {M.}~\bibnamefont
  {Punk}}, \bibinfo {author} {\bibfnamefont {P.~T.}\ \bibnamefont
  {Dumitrescu}},\ and\ \bibinfo {author} {\bibfnamefont {W.}~\bibnamefont
  {Zwerger}},\ }\href {https://doi.org/10.1103/PhysRevA.80.053605} {\bibfield
  {journal} {\bibinfo  {journal} {Phys. Rev. A}\ }\textbf {\bibinfo {volume}
  {80}},\ \bibinfo {pages} {053605} (\bibinfo {year} {2009})}\BibitemShut
  {NoStop}%
\bibitem [{\citenamefont {Nascimb{\`e}ne}\ \emph {et~al.}(2009)\citenamefont
  {Nascimb{\`e}ne}, \citenamefont {Navon}, \citenamefont {Jiang}, \citenamefont
  {Tarruell}, \citenamefont {Teichmann}, \citenamefont {{McKeever}},
  \citenamefont {Chevy},\ and\ \citenamefont
  {Salomon}}]{nascimbene_collective_2009}%
  \BibitemOpen
  \bibfield  {author} {\bibinfo {author} {\bibfnamefont {S.}~\bibnamefont
  {Nascimb{\`e}ne}}, \bibinfo {author} {\bibfnamefont {N.}~\bibnamefont
  {Navon}}, \bibinfo {author} {\bibfnamefont {K.~J.}\ \bibnamefont {Jiang}},
  \bibinfo {author} {\bibfnamefont {L.}~\bibnamefont {Tarruell}}, \bibinfo
  {author} {\bibfnamefont {M.}~\bibnamefont {Teichmann}}, \bibinfo {author}
  {\bibfnamefont {J.}~\bibnamefont {{McKeever}}}, \bibinfo {author}
  {\bibfnamefont {F.}~\bibnamefont {Chevy}},\ and\ \bibinfo {author}
  {\bibfnamefont {C.}~\bibnamefont {Salomon}},\ }\href
  {https://doi.org/10.1103/PhysRevLett.103.170402} {\bibfield  {journal}
  {\bibinfo  {journal} {Phys. Rev. Lett.}\ }\textbf {\bibinfo {volume} {103}},\
  \bibinfo {pages} {170402} (\bibinfo {year} {2009})}\BibitemShut {NoStop}%
\bibitem [{\citenamefont {Z\"ollner}\ \emph {et~al.}(2011)\citenamefont
  {Z\"ollner}, \citenamefont {Bruun},\ and\ \citenamefont
  {Pethick}}]{Zoellner2011}%
  \BibitemOpen
  \bibfield  {author} {\bibinfo {author} {\bibfnamefont {S.}~\bibnamefont
  {Z\"ollner}}, \bibinfo {author} {\bibfnamefont {G.~M.}\ \bibnamefont
  {Bruun}},\ and\ \bibinfo {author} {\bibfnamefont {C.~J.}\ \bibnamefont
  {Pethick}},\ }\href {https://doi.org/10.1103/PhysRevA.83.021603} {\bibfield
  {journal} {\bibinfo  {journal} {Phys. Rev. A}\ }\textbf {\bibinfo {volume}
  {83}},\ \bibinfo {pages} {021603} (\bibinfo {year} {2011})}\BibitemShut
  {NoStop}%
\bibitem [{\citenamefont {Koschorreck}\ \emph {et~al.}(2012)\citenamefont
  {Koschorreck}, \citenamefont {Pertot}, \citenamefont {Vogt}, \citenamefont
  {Fr{\"o}hlich}, \citenamefont {Feld},\ and\ \citenamefont
  {K{\"o}hl}}]{koschorreck2012attractive}%
  \BibitemOpen
  \bibfield  {author} {\bibinfo {author} {\bibfnamefont {M.}~\bibnamefont
  {Koschorreck}}, \bibinfo {author} {\bibfnamefont {D.}~\bibnamefont {Pertot}},
  \bibinfo {author} {\bibfnamefont {E.}~\bibnamefont {Vogt}}, \bibinfo {author}
  {\bibfnamefont {B.}~\bibnamefont {Fr{\"o}hlich}}, \bibinfo {author}
  {\bibfnamefont {M.}~\bibnamefont {Feld}},\ and\ \bibinfo {author}
  {\bibfnamefont {M.}~\bibnamefont {K{\"o}hl}},\ }\href
  {https://www.nature.com/articles/nature11151} {\bibfield  {journal} {\bibinfo
   {journal} {Nature}\ }\textbf {\bibinfo {volume} {485}},\ \bibinfo {pages}
  {619} (\bibinfo {year} {2012})}\BibitemShut {NoStop}%
\bibitem [{\citenamefont {Kohstall}\ \emph {et~al.}(2012)\citenamefont
  {Kohstall}, \citenamefont {Zaccanti}, \citenamefont {Jag}, \citenamefont
  {Trenkwalder}, \citenamefont {Massignan}, \citenamefont {Bruun},
  \citenamefont {Schreck},\ and\ \citenamefont
  {Grimm}}]{kohstall_metastability_2012}%
  \BibitemOpen
  \bibfield  {author} {\bibinfo {author} {\bibfnamefont {C.}~\bibnamefont
  {Kohstall}}, \bibinfo {author} {\bibfnamefont {M.}~\bibnamefont {Zaccanti}},
  \bibinfo {author} {\bibfnamefont {M.}~\bibnamefont {Jag}}, \bibinfo {author}
  {\bibfnamefont {A.}~\bibnamefont {Trenkwalder}}, \bibinfo {author}
  {\bibfnamefont {P.}~\bibnamefont {Massignan}}, \bibinfo {author}
  {\bibfnamefont {G.~M.}\ \bibnamefont {Bruun}}, \bibinfo {author}
  {\bibfnamefont {F.}~\bibnamefont {Schreck}},\ and\ \bibinfo {author}
  {\bibfnamefont {R.}~\bibnamefont {Grimm}},\ }\href
  {https://doi.org/10.1038/nature11065} {\bibfield  {journal} {\bibinfo
  {journal} {Nature}\ }\textbf {\bibinfo {volume} {485}},\ \bibinfo {pages}
  {615} (\bibinfo {year} {2012})}\BibitemShut {NoStop}%
\bibitem [{\citenamefont {Wenz}\ \emph {et~al.}(2013)\citenamefont {Wenz},
  \citenamefont {Z{\"u}rn}, \citenamefont {Murmann}, \citenamefont {Brouzos},
  \citenamefont {Lompe},\ and\ \citenamefont {Jochim}}]{wenz2013}%
  \BibitemOpen
  \bibfield  {author} {\bibinfo {author} {\bibfnamefont {A.}~\bibnamefont
  {Wenz}}, \bibinfo {author} {\bibfnamefont {G.}~\bibnamefont {Z{\"u}rn}},
  \bibinfo {author} {\bibfnamefont {S.}~\bibnamefont {Murmann}}, \bibinfo
  {author} {\bibfnamefont {I.}~\bibnamefont {Brouzos}}, \bibinfo {author}
  {\bibfnamefont {T.}~\bibnamefont {Lompe}},\ and\ \bibinfo {author}
  {\bibfnamefont {S.}~\bibnamefont {Jochim}},\ }\href
  {https://www.science.org/doi/10.1126/science.1240516} {\bibfield  {journal}
  {\bibinfo  {journal} {Science}\ }\textbf {\bibinfo {volume} {342}},\ \bibinfo
  {pages} {457} (\bibinfo {year} {2013})}\BibitemShut {NoStop}%
\bibitem [{\citenamefont {Vlietinck}\ \emph {et~al.}(2013)\citenamefont
  {Vlietinck}, \citenamefont {Ryckebusch},\ and\ \citenamefont
  {Van~Houcke}}]{Vlietinck2013}%
  \BibitemOpen
  \bibfield  {author} {\bibinfo {author} {\bibfnamefont {J.}~\bibnamefont
  {Vlietinck}}, \bibinfo {author} {\bibfnamefont {J.}~\bibnamefont
  {Ryckebusch}},\ and\ \bibinfo {author} {\bibfnamefont {K.}~\bibnamefont
  {Van~Houcke}},\ }\href {https://doi.org/10.1103/PhysRevB.87.115133}
  {\bibfield  {journal} {\bibinfo  {journal} {Phys. Rev. B}\ }\textbf {\bibinfo
  {volume} {87}},\ \bibinfo {pages} {115133} (\bibinfo {year}
  {2013})}\BibitemShut {NoStop}%
\bibitem [{\citenamefont {Ong}\ \emph {et~al.}(2015)\citenamefont {Ong},
  \citenamefont {Cheng}, \citenamefont {Arakelyan},\ and\ \citenamefont
  {Thomas}}]{Ong2015}%
  \BibitemOpen
  \bibfield  {author} {\bibinfo {author} {\bibfnamefont {W.}~\bibnamefont
  {Ong}}, \bibinfo {author} {\bibfnamefont {C.}~\bibnamefont {Cheng}}, \bibinfo
  {author} {\bibfnamefont {I.}~\bibnamefont {Arakelyan}},\ and\ \bibinfo
  {author} {\bibfnamefont {J.~E.}\ \bibnamefont {Thomas}},\ }\href
  {https://doi.org/10.1103/PhysRevLett.114.110403} {\bibfield  {journal}
  {\bibinfo  {journal} {Phys. Rev. Lett.}\ }\textbf {\bibinfo {volume} {114}},\
  \bibinfo {pages} {110403} (\bibinfo {year} {2015})}\BibitemShut {NoStop}%
\bibitem [{\citenamefont {Christensen}\ and\ \citenamefont
  {Bruun}(2015)}]{Christensen2015b}%
  \BibitemOpen
  \bibfield  {author} {\bibinfo {author} {\bibfnamefont {R.~S.}\ \bibnamefont
  {Christensen}}\ and\ \bibinfo {author} {\bibfnamefont {G.~M.}\ \bibnamefont
  {Bruun}},\ }\href {https://doi.org/10.1103/PhysRevA.91.042702} {\bibfield
  {journal} {\bibinfo  {journal} {Phys. Rev. A}\ }\textbf {\bibinfo {volume}
  {91}},\ \bibinfo {pages} {042702} (\bibinfo {year} {2015})}\BibitemShut
  {NoStop}%
\bibitem [{\citenamefont {Scazza}\ \emph {et~al.}(2017)\citenamefont {Scazza},
  \citenamefont {Valtolina}, \citenamefont {Massignan}, \citenamefont {Recati},
  \citenamefont {Amico}, \citenamefont {Burchianti}, \citenamefont {Fort},
  \citenamefont {Inguscio}, \citenamefont {Zaccanti},\ and\ \citenamefont
  {Roati}}]{Scazza2016}%
  \BibitemOpen
  \bibfield  {author} {\bibinfo {author} {\bibfnamefont {F.}~\bibnamefont
  {Scazza}}, \bibinfo {author} {\bibfnamefont {G.}~\bibnamefont {Valtolina}},
  \bibinfo {author} {\bibfnamefont {P.}~\bibnamefont {Massignan}}, \bibinfo
  {author} {\bibfnamefont {A.}~\bibnamefont {Recati}}, \bibinfo {author}
  {\bibfnamefont {A.}~\bibnamefont {Amico}}, \bibinfo {author} {\bibfnamefont
  {A.}~\bibnamefont {Burchianti}}, \bibinfo {author} {\bibfnamefont
  {C.}~\bibnamefont {Fort}}, \bibinfo {author} {\bibfnamefont {M.}~\bibnamefont
  {Inguscio}}, \bibinfo {author} {\bibfnamefont {M.}~\bibnamefont {Zaccanti}},\
  and\ \bibinfo {author} {\bibfnamefont {G.}~\bibnamefont {Roati}},\ }\href
  {https://doi.org/10.1103/PhysRevLett.118.083602} {\bibfield  {journal}
  {\bibinfo  {journal} {Phys. Rev. Lett.}\ }\textbf {\bibinfo {volume} {118}},\
  \bibinfo {pages} {083602} (\bibinfo {year} {2017})}\BibitemShut {NoStop}%
\bibitem [{\citenamefont {Sidler}\ \emph {et~al.}(2017)\citenamefont {Sidler},
  \citenamefont {Back}, \citenamefont {Cotlet}, \citenamefont {Srivastava},
  \citenamefont {Fink}, \citenamefont {Kroner}, \citenamefont {Demler},\ and\
  \citenamefont {Imamoglu}}]{sidler2017fermi}%
  \BibitemOpen
  \bibfield  {author} {\bibinfo {author} {\bibfnamefont {M.}~\bibnamefont
  {Sidler}}, \bibinfo {author} {\bibfnamefont {P.}~\bibnamefont {Back}},
  \bibinfo {author} {\bibfnamefont {O.}~\bibnamefont {Cotlet}}, \bibinfo
  {author} {\bibfnamefont {A.}~\bibnamefont {Srivastava}}, \bibinfo {author}
  {\bibfnamefont {T.}~\bibnamefont {Fink}}, \bibinfo {author} {\bibfnamefont
  {M.}~\bibnamefont {Kroner}}, \bibinfo {author} {\bibfnamefont
  {E.}~\bibnamefont {Demler}},\ and\ \bibinfo {author} {\bibfnamefont
  {A.}~\bibnamefont {Imamoglu}},\ }\href
  {https://www.nature.com/articles/nphys3949} {\bibfield  {journal} {\bibinfo
  {journal} {Nature Physics}\ }\textbf {\bibinfo {volume} {13}},\ \bibinfo
  {pages} {255} (\bibinfo {year} {2017})}\BibitemShut {NoStop}%
\bibitem [{\citenamefont {Schmidt}\ \emph
  {et~al.}(2018{\natexlab{a}})\citenamefont {Schmidt}, \citenamefont {Knap},
  \citenamefont {Ivanov}, \citenamefont {You}, \citenamefont {Cetina},\ and\
  \citenamefont {Demler}}]{schmidt2018universal}%
  \BibitemOpen
  \bibfield  {author} {\bibinfo {author} {\bibfnamefont {R.}~\bibnamefont
  {Schmidt}}, \bibinfo {author} {\bibfnamefont {M.}~\bibnamefont {Knap}},
  \bibinfo {author} {\bibfnamefont {D.~A.}\ \bibnamefont {Ivanov}}, \bibinfo
  {author} {\bibfnamefont {J.-S.}\ \bibnamefont {You}}, \bibinfo {author}
  {\bibfnamefont {M.}~\bibnamefont {Cetina}},\ and\ \bibinfo {author}
  {\bibfnamefont {E.}~\bibnamefont {Demler}},\ }\href
  {https://iopscience.iop.org/article/10.1088/1361-6633/aa9593/meta} {\bibfield
   {journal} {\bibinfo  {journal} {Reports on Progress in Physics}\ }\textbf
  {\bibinfo {volume} {81}},\ \bibinfo {pages} {024401} (\bibinfo {year}
  {2018}{\natexlab{a}})}\BibitemShut {NoStop}%
\bibitem [{\citenamefont {Yan}\ \emph {et~al.}(2019)\citenamefont {Yan},
  \citenamefont {Patel}, \citenamefont {Mukherjee}, \citenamefont {Fletcher},
  \citenamefont {Struck},\ and\ \citenamefont {Zwierlein}}]{Yan2019Fermi}%
  \BibitemOpen
  \bibfield  {author} {\bibinfo {author} {\bibfnamefont {Z.}~\bibnamefont
  {Yan}}, \bibinfo {author} {\bibfnamefont {P.~B.}\ \bibnamefont {Patel}},
  \bibinfo {author} {\bibfnamefont {B.}~\bibnamefont {Mukherjee}}, \bibinfo
  {author} {\bibfnamefont {R.~J.}\ \bibnamefont {Fletcher}}, \bibinfo {author}
  {\bibfnamefont {J.}~\bibnamefont {Struck}},\ and\ \bibinfo {author}
  {\bibfnamefont {M.~W.}\ \bibnamefont {Zwierlein}},\ }\href
  {https://doi.org/10.1103/PhysRevLett.122.093401} {\bibfield  {journal}
  {\bibinfo  {journal} {Phys. Rev. Lett.}\ }\textbf {\bibinfo {volume} {122}},\
  \bibinfo {pages} {093401} (\bibinfo {year} {2019})}\BibitemShut {NoStop}%
\bibitem [{\citenamefont {Darkwah~Oppong}\ \emph {et~al.}(2019)\citenamefont
  {Darkwah~Oppong}, \citenamefont {Riegger}, \citenamefont {Bettermann},
  \citenamefont {H\"ofer}, \citenamefont {Levinsen}, \citenamefont {Parish},
  \citenamefont {Bloch},\ and\ \citenamefont {F\"olling}}]{Oppong2019}%
  \BibitemOpen
  \bibfield  {author} {\bibinfo {author} {\bibfnamefont {N.}~\bibnamefont
  {Darkwah~Oppong}}, \bibinfo {author} {\bibfnamefont {L.}~\bibnamefont
  {Riegger}}, \bibinfo {author} {\bibfnamefont {O.}~\bibnamefont {Bettermann}},
  \bibinfo {author} {\bibfnamefont {M.}~\bibnamefont {H\"ofer}}, \bibinfo
  {author} {\bibfnamefont {J.}~\bibnamefont {Levinsen}}, \bibinfo {author}
  {\bibfnamefont {M.~M.}\ \bibnamefont {Parish}}, \bibinfo {author}
  {\bibfnamefont {I.}~\bibnamefont {Bloch}},\ and\ \bibinfo {author}
  {\bibfnamefont {S.}~\bibnamefont {F\"olling}},\ }\href
  {https://doi.org/10.1103/PhysRevLett.122.193604} {\bibfield  {journal}
  {\bibinfo  {journal} {Phys. Rev. Lett.}\ }\textbf {\bibinfo {volume} {122}},\
  \bibinfo {pages} {193604} (\bibinfo {year} {2019})}\BibitemShut {NoStop}%
\bibitem [{\citenamefont {Ness}\ \emph {et~al.}(2020)\citenamefont {Ness},
  \citenamefont {Shkedrov}, \citenamefont {Florshaim}, \citenamefont {Diessel},
  \citenamefont {von Milczewski}, \citenamefont {Schmidt},\ and\ \citenamefont
  {Sagi}}]{Ness2020}%
  \BibitemOpen
  \bibfield  {author} {\bibinfo {author} {\bibfnamefont {G.}~\bibnamefont
  {Ness}}, \bibinfo {author} {\bibfnamefont {C.}~\bibnamefont {Shkedrov}},
  \bibinfo {author} {\bibfnamefont {Y.}~\bibnamefont {Florshaim}}, \bibinfo
  {author} {\bibfnamefont {O.~K.}\ \bibnamefont {Diessel}}, \bibinfo {author}
  {\bibfnamefont {J.}~\bibnamefont {von Milczewski}}, \bibinfo {author}
  {\bibfnamefont {R.}~\bibnamefont {Schmidt}},\ and\ \bibinfo {author}
  {\bibfnamefont {Y.}~\bibnamefont {Sagi}},\ }\href
  {https://doi.org/10.1103/PhysRevX.10.041019} {\bibfield  {journal} {\bibinfo
  {journal} {Phys. Rev. X}\ }\textbf {\bibinfo {volume} {10}},\ \bibinfo
  {pages} {041019} (\bibinfo {year} {2020})}\BibitemShut {NoStop}%
\bibitem [{\citenamefont {Adlong}\ \emph {et~al.}(2020)\citenamefont {Adlong},
  \citenamefont {Liu}, \citenamefont {Scazza}, \citenamefont {Zaccanti},
  \citenamefont {Oppong}, \citenamefont {F\"olling}, \citenamefont {Parish},\
  and\ \citenamefont {Levinsen}}]{Adlong2020}%
  \BibitemOpen
  \bibfield  {author} {\bibinfo {author} {\bibfnamefont {H.~S.}\ \bibnamefont
  {Adlong}}, \bibinfo {author} {\bibfnamefont {W.~E.}\ \bibnamefont {Liu}},
  \bibinfo {author} {\bibfnamefont {F.}~\bibnamefont {Scazza}}, \bibinfo
  {author} {\bibfnamefont {M.}~\bibnamefont {Zaccanti}}, \bibinfo {author}
  {\bibfnamefont {N.~D.}\ \bibnamefont {Oppong}}, \bibinfo {author}
  {\bibfnamefont {S.}~\bibnamefont {F\"olling}}, \bibinfo {author}
  {\bibfnamefont {M.~M.}\ \bibnamefont {Parish}},\ and\ \bibinfo {author}
  {\bibfnamefont {J.}~\bibnamefont {Levinsen}},\ }\href
  {https://doi.org/10.1103/PhysRevLett.125.133401} {\bibfield  {journal}
  {\bibinfo  {journal} {Phys. Rev. Lett.}\ }\textbf {\bibinfo {volume} {125}},\
  \bibinfo {pages} {133401} (\bibinfo {year} {2020})}\BibitemShut {NoStop}%
\bibitem [{\citenamefont {Fritsche}\ \emph {et~al.}(2021)\citenamefont
  {Fritsche}, \citenamefont {Baroni}, \citenamefont {Dobler}, \citenamefont
  {Kirilov}, \citenamefont {Huang}, \citenamefont {Grimm}, \citenamefont
  {Bruun}, \citenamefont {Massignan} \emph {et~al.}}]{fritsche2021stability}%
  \BibitemOpen
  \bibfield  {author} {\bibinfo {author} {\bibfnamefont {I.}~\bibnamefont
  {Fritsche}}, \bibinfo {author} {\bibfnamefont {C.}~\bibnamefont {Baroni}},
  \bibinfo {author} {\bibfnamefont {E.}~\bibnamefont {Dobler}}, \bibinfo
  {author} {\bibfnamefont {E.}~\bibnamefont {Kirilov}}, \bibinfo {author}
  {\bibfnamefont {B.}~\bibnamefont {Huang}}, \bibinfo {author} {\bibfnamefont
  {R.}~\bibnamefont {Grimm}}, \bibinfo {author} {\bibfnamefont {G.~M.}\
  \bibnamefont {Bruun}}, \bibinfo {author} {\bibfnamefont {P.}~\bibnamefont
  {Massignan}}, \emph {et~al.},\ }\href
  {https://journals.aps.org/pra/abstract/10.1103/PhysRevA.103.053314#fulltext}
  {\bibfield  {journal} {\bibinfo  {journal} {Phys. Rev. A}\ }\textbf {\bibinfo
  {volume} {103}},\ \bibinfo {pages} {053314} (\bibinfo {year}
  {2021})}\BibitemShut {NoStop}%
\bibitem [{\citenamefont {Dolgirev}\ \emph {et~al.}(2021)\citenamefont
  {Dolgirev}, \citenamefont {Qu}, \citenamefont {Zvonarev}, \citenamefont
  {Shi},\ and\ \citenamefont {Demler}}]{Dolgirev2021}%
  \BibitemOpen
  \bibfield  {author} {\bibinfo {author} {\bibfnamefont {P.~E.}\ \bibnamefont
  {Dolgirev}}, \bibinfo {author} {\bibfnamefont {Y.-F.}\ \bibnamefont {Qu}},
  \bibinfo {author} {\bibfnamefont {M.~B.}\ \bibnamefont {Zvonarev}}, \bibinfo
  {author} {\bibfnamefont {T.}~\bibnamefont {Shi}},\ and\ \bibinfo {author}
  {\bibfnamefont {E.}~\bibnamefont {Demler}},\ }\href
  {https://doi.org/10.1103/PhysRevX.11.041015} {\bibfield  {journal} {\bibinfo
  {journal} {Phys. Rev. X}\ }\textbf {\bibinfo {volume} {11}},\ \bibinfo
  {pages} {041015} (\bibinfo {year} {2021})}\BibitemShut {NoStop}%
\bibitem [{\citenamefont {Scazza}\ \emph {et~al.}(2022)\citenamefont {Scazza},
  \citenamefont {Zaccanti}, \citenamefont {Massignan}, \citenamefont {Parish},\
  and\ \citenamefont {Levinsen}}]{scazza2022repulsive}%
  \BibitemOpen
  \bibfield  {author} {\bibinfo {author} {\bibfnamefont {F.}~\bibnamefont
  {Scazza}}, \bibinfo {author} {\bibfnamefont {M.}~\bibnamefont {Zaccanti}},
  \bibinfo {author} {\bibfnamefont {P.}~\bibnamefont {Massignan}}, \bibinfo
  {author} {\bibfnamefont {M.~M.}\ \bibnamefont {Parish}},\ and\ \bibinfo
  {author} {\bibfnamefont {J.}~\bibnamefont {Levinsen}},\ }\href
  {https://www.mdpi.com/2218-2004/10/2/55} {\bibfield  {journal} {\bibinfo
  {journal} {Atoms}\ }\textbf {\bibinfo {volume} {10}},\ \bibinfo {pages} {55}
  (\bibinfo {year} {2022})}\BibitemShut {NoStop}%
\bibitem [{\citenamefont {Du}\ and\ \citenamefont
  {Greene}(1987)}]{du1987interaction}%
  \BibitemOpen
  \bibfield  {author} {\bibinfo {author} {\bibfnamefont {N.~Y.}\ \bibnamefont
  {Du}}\ and\ \bibinfo {author} {\bibfnamefont {C.~H.}\ \bibnamefont
  {Greene}},\ }\href
  {https://journals.aps.org/pra/abstract/10.1103/PhysRevA.36.971} {\bibfield
  {journal} {\bibinfo  {journal} {Phys. Rev. A}\ }\textbf {\bibinfo {volume}
  {36}},\ \bibinfo {pages} {971} (\bibinfo {year} {1987})}\BibitemShut
  {NoStop}%
\bibitem [{\citenamefont {Greene}\ \emph {et~al.}(2000)\citenamefont {Greene},
  \citenamefont {Dickinson},\ and\ \citenamefont
  {Sadeghpour}}]{greene2000creation}%
  \BibitemOpen
  \bibfield  {author} {\bibinfo {author} {\bibfnamefont {C.~H.}\ \bibnamefont
  {Greene}}, \bibinfo {author} {\bibfnamefont {A.}~\bibnamefont {Dickinson}},\
  and\ \bibinfo {author} {\bibfnamefont {H.}~\bibnamefont {Sadeghpour}},\
  }\href {https://journals.aps.org/prl/abstract/10.1103/PhysRevLett.85.2458}
  {\bibfield  {journal} {\bibinfo  {journal} {Phys. Rev. Lett.}\ }\textbf
  {\bibinfo {volume} {85}},\ \bibinfo {pages} {2458} (\bibinfo {year}
  {2000})}\BibitemShut {NoStop}%
\bibitem [{\citenamefont {Bendkowsky}\ \emph {et~al.}(2009)\citenamefont
  {Bendkowsky}, \citenamefont {Butscher}, \citenamefont {Nipper}, \citenamefont
  {Shaffer}, \citenamefont {L{\"o}w},\ and\ \citenamefont
  {Pfau}}]{bendkowsky2009observation}%
  \BibitemOpen
  \bibfield  {author} {\bibinfo {author} {\bibfnamefont {V.}~\bibnamefont
  {Bendkowsky}}, \bibinfo {author} {\bibfnamefont {B.}~\bibnamefont
  {Butscher}}, \bibinfo {author} {\bibfnamefont {J.}~\bibnamefont {Nipper}},
  \bibinfo {author} {\bibfnamefont {J.~P.}\ \bibnamefont {Shaffer}}, \bibinfo
  {author} {\bibfnamefont {R.}~\bibnamefont {L{\"o}w}},\ and\ \bibinfo {author}
  {\bibfnamefont {T.}~\bibnamefont {Pfau}},\ }\href
  {https://www.nature.com/articles/nature07945} {\bibfield  {journal} {\bibinfo
   {journal} {Nature}\ }\textbf {\bibinfo {volume} {458}},\ \bibinfo {pages}
  {1005} (\bibinfo {year} {2009})}\BibitemShut {NoStop}%
\bibitem [{\citenamefont {Butscher}\ \emph {et~al.}(2010)\citenamefont
  {Butscher}, \citenamefont {Nipper}, \citenamefont {Balewski}, \citenamefont
  {Kukota}, \citenamefont {Bendkowsky}, \citenamefont {L{\"o}w},\ and\
  \citenamefont {Pfau}}]{butscher2010atom}%
  \BibitemOpen
  \bibfield  {author} {\bibinfo {author} {\bibfnamefont {B.}~\bibnamefont
  {Butscher}}, \bibinfo {author} {\bibfnamefont {J.}~\bibnamefont {Nipper}},
  \bibinfo {author} {\bibfnamefont {J.~B.}\ \bibnamefont {Balewski}}, \bibinfo
  {author} {\bibfnamefont {L.}~\bibnamefont {Kukota}}, \bibinfo {author}
  {\bibfnamefont {V.}~\bibnamefont {Bendkowsky}}, \bibinfo {author}
  {\bibfnamefont {R.}~\bibnamefont {L{\"o}w}},\ and\ \bibinfo {author}
  {\bibfnamefont {T.}~\bibnamefont {Pfau}},\ }\href
  {https://www.nature.com/articles/nphys1828} {\bibfield  {journal} {\bibinfo
  {journal} {Nature Physics}\ }\textbf {\bibinfo {volume} {6}},\ \bibinfo
  {pages} {970} (\bibinfo {year} {2010})}\BibitemShut {NoStop}%
\bibitem [{\citenamefont {DeSalvo}\ \emph {et~al.}(2015)\citenamefont
  {DeSalvo}, \citenamefont {Aman}, \citenamefont {Dunning}, \citenamefont
  {Killian}, \citenamefont {Sadeghpour}, \citenamefont {Yoshida},\ and\
  \citenamefont {Burgd\"orfer}}]{DeSalvo2015Ultralong}%
  \BibitemOpen
  \bibfield  {author} {\bibinfo {author} {\bibfnamefont {B.~J.}\ \bibnamefont
  {DeSalvo}}, \bibinfo {author} {\bibfnamefont {J.~A.}\ \bibnamefont {Aman}},
  \bibinfo {author} {\bibfnamefont {F.~B.}\ \bibnamefont {Dunning}}, \bibinfo
  {author} {\bibfnamefont {T.~C.}\ \bibnamefont {Killian}}, \bibinfo {author}
  {\bibfnamefont {H.~R.}\ \bibnamefont {Sadeghpour}}, \bibinfo {author}
  {\bibfnamefont {S.}~\bibnamefont {Yoshida}},\ and\ \bibinfo {author}
  {\bibfnamefont {J.}~\bibnamefont {Burgd\"orfer}},\ }\href
  {https://doi.org/10.1103/PhysRevA.92.031403} {\bibfield  {journal} {\bibinfo
  {journal} {Phys. Rev. A}\ }\textbf {\bibinfo {volume} {92}},\ \bibinfo
  {pages} {031403} (\bibinfo {year} {2015})}\BibitemShut {NoStop}%
\bibitem [{\citenamefont {Camargo}\ \emph {et~al.}(2016)\citenamefont
  {Camargo}, \citenamefont {Whalen}, \citenamefont {Ding}, \citenamefont
  {Sadeghpour}, \citenamefont {Yoshida}, \citenamefont {Burgd\"orfer},
  \citenamefont {Dunning},\ and\ \citenamefont
  {Killian}}]{camargo2016lifetimes}%
  \BibitemOpen
  \bibfield  {author} {\bibinfo {author} {\bibfnamefont {F.}~\bibnamefont
  {Camargo}}, \bibinfo {author} {\bibfnamefont {J.~D.}\ \bibnamefont {Whalen}},
  \bibinfo {author} {\bibfnamefont {R.}~\bibnamefont {Ding}}, \bibinfo {author}
  {\bibfnamefont {H.~R.}\ \bibnamefont {Sadeghpour}}, \bibinfo {author}
  {\bibfnamefont {S.}~\bibnamefont {Yoshida}}, \bibinfo {author} {\bibfnamefont
  {J.}~\bibnamefont {Burgd\"orfer}}, \bibinfo {author} {\bibfnamefont {F.~B.}\
  \bibnamefont {Dunning}},\ and\ \bibinfo {author} {\bibfnamefont {T.~C.}\
  \bibnamefont {Killian}},\ }\href {https://doi.org/10.1103/PhysRevA.93.022702}
  {\bibfield  {journal} {\bibinfo  {journal} {Phys. Rev. A}\ }\textbf {\bibinfo
  {volume} {93}},\ \bibinfo {pages} {022702} (\bibinfo {year}
  {2016})}\BibitemShut {NoStop}%
\bibitem [{\citenamefont {Camargo}\ \emph {et~al.}(2018)\citenamefont
  {Camargo}, \citenamefont {Schmidt}, \citenamefont {Whalen}, \citenamefont
  {Ding}, \citenamefont {Woehl~Jr}, \citenamefont {Yoshida}, \citenamefont
  {Burgd{\"o}rfer}, \citenamefont {Dunning}, \citenamefont {Sadeghpour},
  \citenamefont {Demler} \emph {et~al.}}]{camargo2018creation}%
  \BibitemOpen
  \bibfield  {author} {\bibinfo {author} {\bibfnamefont {F.}~\bibnamefont
  {Camargo}}, \bibinfo {author} {\bibfnamefont {R.}~\bibnamefont {Schmidt}},
  \bibinfo {author} {\bibfnamefont {J.}~\bibnamefont {Whalen}}, \bibinfo
  {author} {\bibfnamefont {R.}~\bibnamefont {Ding}}, \bibinfo {author}
  {\bibfnamefont {G.}~\bibnamefont {Woehl~Jr}}, \bibinfo {author}
  {\bibfnamefont {S.}~\bibnamefont {Yoshida}}, \bibinfo {author} {\bibfnamefont
  {J.}~\bibnamefont {Burgd{\"o}rfer}}, \bibinfo {author} {\bibfnamefont
  {F.}~\bibnamefont {Dunning}}, \bibinfo {author} {\bibfnamefont
  {H.}~\bibnamefont {Sadeghpour}}, \bibinfo {author} {\bibfnamefont
  {E.}~\bibnamefont {Demler}}, \emph {et~al.},\ }\href
  {https://journals.aps.org/prl/abstract/10.1103/PhysRevLett.120.083401}
  {\bibfield  {journal} {\bibinfo  {journal} {Phys. Rev. Lett.}\ }\textbf
  {\bibinfo {volume} {120}},\ \bibinfo {pages} {083401} (\bibinfo {year}
  {2018})}\BibitemShut {NoStop}%
\bibitem [{\citenamefont {Kleinbach}\ \emph {et~al.}(2018)\citenamefont
  {Kleinbach}, \citenamefont {Engel}, \citenamefont {Dieterle}, \citenamefont
  {L{\"o}w}, \citenamefont {Pfau},\ and\ \citenamefont
  {Meinert}}]{kleinbach2018ionic}%
  \BibitemOpen
  \bibfield  {author} {\bibinfo {author} {\bibfnamefont {K.~S.}\ \bibnamefont
  {Kleinbach}}, \bibinfo {author} {\bibfnamefont {F.}~\bibnamefont {Engel}},
  \bibinfo {author} {\bibfnamefont {T.}~\bibnamefont {Dieterle}}, \bibinfo
  {author} {\bibfnamefont {R.}~\bibnamefont {L{\"o}w}}, \bibinfo {author}
  {\bibfnamefont {T.}~\bibnamefont {Pfau}},\ and\ \bibinfo {author}
  {\bibfnamefont {F.}~\bibnamefont {Meinert}},\ }\href
  {https://journals.aps.org/prl/abstract/10.1103/PhysRevLett.120.193401}
  {\bibfield  {journal} {\bibinfo  {journal} {Phys. Rev. Lett.}\ }\textbf
  {\bibinfo {volume} {120}},\ \bibinfo {pages} {193401} (\bibinfo {year}
  {2018})}\BibitemShut {NoStop}%
\bibitem [{\citenamefont {Whalen}\ \emph {et~al.}(2019)\citenamefont {Whalen},
  \citenamefont {Kanungo}, \citenamefont {Ding}, \citenamefont {Wagner},
  \citenamefont {Schmidt}, \citenamefont {Sadeghpour}, \citenamefont {Yoshida},
  \citenamefont {Burgd\"orfer}, \citenamefont {Dunning},\ and\ \citenamefont
  {Killian}}]{Whalen2019}%
  \BibitemOpen
  \bibfield  {author} {\bibinfo {author} {\bibfnamefont {J.~D.}\ \bibnamefont
  {Whalen}}, \bibinfo {author} {\bibfnamefont {S.~K.}\ \bibnamefont {Kanungo}},
  \bibinfo {author} {\bibfnamefont {R.}~\bibnamefont {Ding}}, \bibinfo {author}
  {\bibfnamefont {M.}~\bibnamefont {Wagner}}, \bibinfo {author} {\bibfnamefont
  {R.}~\bibnamefont {Schmidt}}, \bibinfo {author} {\bibfnamefont {H.~R.}\
  \bibnamefont {Sadeghpour}}, \bibinfo {author} {\bibfnamefont
  {S.}~\bibnamefont {Yoshida}}, \bibinfo {author} {\bibfnamefont
  {J.}~\bibnamefont {Burgd\"orfer}}, \bibinfo {author} {\bibfnamefont {F.~B.}\
  \bibnamefont {Dunning}},\ and\ \bibinfo {author} {\bibfnamefont {T.~C.}\
  \bibnamefont {Killian}},\ }\href
  {https://doi.org/10.1103/PhysRevA.100.011402} {\bibfield  {journal} {\bibinfo
   {journal} {Phys. Rev. A}\ }\textbf {\bibinfo {volume} {100}},\ \bibinfo
  {pages} {011402} (\bibinfo {year} {2019})}\BibitemShut {NoStop}%
\bibitem [{\citenamefont {Schlagm{\"u}ller}\ \emph {et~al.}(2016)\citenamefont
  {Schlagm{\"u}ller}, \citenamefont {Liebisch}, \citenamefont {Nguyen},
  \citenamefont {Lochead}, \citenamefont {Engel}, \citenamefont {B{\"o}ttcher},
  \citenamefont {Westphal}, \citenamefont {Kleinbach}, \citenamefont {L{\"o}w},
  \citenamefont {Hofferberth} \emph {et~al.}}]{schlagmuller2016probing}%
  \BibitemOpen
  \bibfield  {author} {\bibinfo {author} {\bibfnamefont {M.}~\bibnamefont
  {Schlagm{\"u}ller}}, \bibinfo {author} {\bibfnamefont {T.~C.}\ \bibnamefont
  {Liebisch}}, \bibinfo {author} {\bibfnamefont {H.}~\bibnamefont {Nguyen}},
  \bibinfo {author} {\bibfnamefont {G.}~\bibnamefont {Lochead}}, \bibinfo
  {author} {\bibfnamefont {F.}~\bibnamefont {Engel}}, \bibinfo {author}
  {\bibfnamefont {F.}~\bibnamefont {B{\"o}ttcher}}, \bibinfo {author}
  {\bibfnamefont {K.~M.}\ \bibnamefont {Westphal}}, \bibinfo {author}
  {\bibfnamefont {K.~S.}\ \bibnamefont {Kleinbach}}, \bibinfo {author}
  {\bibfnamefont {R.}~\bibnamefont {L{\"o}w}}, \bibinfo {author} {\bibfnamefont
  {S.}~\bibnamefont {Hofferberth}}, \emph {et~al.},\ }\href
  {https://journals.aps.org/prl/abstract/10.1103/PhysRevLett.116.053001}
  {\bibfield  {journal} {\bibinfo  {journal} {Phys. Rev. Lett.}\ }\textbf
  {\bibinfo {volume} {116}},\ \bibinfo {pages} {053001} (\bibinfo {year}
  {2016})}\BibitemShut {NoStop}%
\bibitem [{\citenamefont {Veit}\ \emph {et~al.}(2021)\citenamefont {Veit},
  \citenamefont {Zuber}, \citenamefont {Herrera-Sancho}, \citenamefont
  {Anasuri}, \citenamefont {Schmid}, \citenamefont {Meinert}, \citenamefont
  {L{\"o}w},\ and\ \citenamefont {Pfau}}]{veit2021pulsed}%
  \BibitemOpen
  \bibfield  {author} {\bibinfo {author} {\bibfnamefont {C.}~\bibnamefont
  {Veit}}, \bibinfo {author} {\bibfnamefont {N.}~\bibnamefont {Zuber}},
  \bibinfo {author} {\bibfnamefont {O.}~\bibnamefont {Herrera-Sancho}},
  \bibinfo {author} {\bibfnamefont {V.}~\bibnamefont {Anasuri}}, \bibinfo
  {author} {\bibfnamefont {T.}~\bibnamefont {Schmid}}, \bibinfo {author}
  {\bibfnamefont {F.}~\bibnamefont {Meinert}}, \bibinfo {author} {\bibfnamefont
  {R.}~\bibnamefont {L{\"o}w}},\ and\ \bibinfo {author} {\bibfnamefont
  {T.}~\bibnamefont {Pfau}},\ }\href
  {https://journals.aps.org/prx/abstract/10.1103/PhysRevX.11.011036} {\bibfield
   {journal} {\bibinfo  {journal} {Physical Review X}\ }\textbf {\bibinfo
  {volume} {11}},\ \bibinfo {pages} {011036} (\bibinfo {year}
  {2021})}\BibitemShut {NoStop}%
\bibitem [{\citenamefont {Knap}\ \emph {et~al.}(2012)\citenamefont {Knap},
  \citenamefont {Shashi}, \citenamefont {Nishida}, \citenamefont {Imambekov},
  \citenamefont {Abanin},\ and\ \citenamefont {Demler}}]{knap2012time}%
  \BibitemOpen
  \bibfield  {author} {\bibinfo {author} {\bibfnamefont {M.}~\bibnamefont
  {Knap}}, \bibinfo {author} {\bibfnamefont {A.}~\bibnamefont {Shashi}},
  \bibinfo {author} {\bibfnamefont {Y.}~\bibnamefont {Nishida}}, \bibinfo
  {author} {\bibfnamefont {A.}~\bibnamefont {Imambekov}}, \bibinfo {author}
  {\bibfnamefont {D.~A.}\ \bibnamefont {Abanin}},\ and\ \bibinfo {author}
  {\bibfnamefont {E.}~\bibnamefont {Demler}},\ }\href
  {https://journals.aps.org/prx/abstract/10.1103/PhysRevX.2.041020} {\bibfield
  {journal} {\bibinfo  {journal} {Phys. Rev. X}\ }\textbf {\bibinfo {volume}
  {2}},\ \bibinfo {pages} {041020} (\bibinfo {year} {2012})}\BibitemShut
  {NoStop}%
\bibitem [{\citenamefont {Schmidt}\ \emph
  {et~al.}(2018{\natexlab{b}})\citenamefont {Schmidt}, \citenamefont {Whalen},
  \citenamefont {Ding}, \citenamefont {Camargo}, \citenamefont {Woehl{, Jr.}},
  \citenamefont {Yoshida}, \citenamefont {Burgd{\"o}rfer}, \citenamefont
  {Dunning}, \citenamefont {Demler}, \citenamefont {Sadeghpour},\ and\
  \citenamefont {Killian}}]{schmidt2018theory}%
  \BibitemOpen
  \bibfield  {author} {\bibinfo {author} {\bibfnamefont {R.}~\bibnamefont
  {Schmidt}}, \bibinfo {author} {\bibfnamefont {J.~D.}\ \bibnamefont {Whalen}},
  \bibinfo {author} {\bibfnamefont {R.}~\bibnamefont {Ding}}, \bibinfo {author}
  {\bibfnamefont {F.}~\bibnamefont {Camargo}}, \bibinfo {author} {\bibfnamefont
  {G.}~\bibnamefont {Woehl{, Jr.}}}, \bibinfo {author} {\bibfnamefont
  {S.}~\bibnamefont {Yoshida}}, \bibinfo {author} {\bibfnamefont
  {J.}~\bibnamefont {Burgd{\"o}rfer}}, \bibinfo {author} {\bibfnamefont
  {F.~B.}\ \bibnamefont {Dunning}}, \bibinfo {author} {\bibfnamefont
  {E.}~\bibnamefont {Demler}}, \bibinfo {author} {\bibfnamefont {H.~R.}\
  \bibnamefont {Sadeghpour}},\ and\ \bibinfo {author} {\bibfnamefont {T.~C.}\
  \bibnamefont {Killian}},\ }\href
  {https://journals.aps.org/pra/abstract/10.1103/PhysRevA.97.022707} {\bibfield
   {journal} {\bibinfo  {journal} {Phys. Rev. A}\ }\textbf {\bibinfo {volume}
  {97}},\ \bibinfo {pages} {022707} (\bibinfo {year}
  {2018}{\natexlab{b}})}\BibitemShut {NoStop}%
\bibitem [{\citenamefont {Sous}\ \emph {et~al.}(2020)\citenamefont {Sous},
  \citenamefont {Sadeghpour}, \citenamefont {Killian}, \citenamefont {Demler},\
  and\ \citenamefont {Schmidt}}]{sous2020rydberg}%
  \BibitemOpen
  \bibfield  {author} {\bibinfo {author} {\bibfnamefont {J.}~\bibnamefont
  {Sous}}, \bibinfo {author} {\bibfnamefont {H.~R.}\ \bibnamefont
  {Sadeghpour}}, \bibinfo {author} {\bibfnamefont {T.~C.}\ \bibnamefont
  {Killian}}, \bibinfo {author} {\bibfnamefont {E.}~\bibnamefont {Demler}},\
  and\ \bibinfo {author} {\bibfnamefont {R.}~\bibnamefont {Schmidt}},\ }\href
  {https://journals.aps.org/prresearch/abstract/10.1103/PhysRevResearch.2.023021}
  {\bibfield  {journal} {\bibinfo  {journal} {Phys. Rev. Res.}\ }\textbf
  {\bibinfo {volume} {2}},\ \bibinfo {pages} {023021} (\bibinfo {year}
  {2020})}\BibitemShut {NoStop}%
\bibitem [{\citenamefont {L{\"o}w}\ \emph {et~al.}(2012)\citenamefont
  {L{\"o}w}, \citenamefont {Weimer}, \citenamefont {Nipper}, \citenamefont
  {Balewski}, \citenamefont {Butscher}, \citenamefont {B{\"u}chler},\ and\
  \citenamefont {Pfau}}]{low2012experimental}%
  \BibitemOpen
  \bibfield  {author} {\bibinfo {author} {\bibfnamefont {R.}~\bibnamefont
  {L{\"o}w}}, \bibinfo {author} {\bibfnamefont {H.}~\bibnamefont {Weimer}},
  \bibinfo {author} {\bibfnamefont {J.}~\bibnamefont {Nipper}}, \bibinfo
  {author} {\bibfnamefont {J.~B.}\ \bibnamefont {Balewski}}, \bibinfo {author}
  {\bibfnamefont {B.}~\bibnamefont {Butscher}}, \bibinfo {author}
  {\bibfnamefont {H.~P.}\ \bibnamefont {B{\"u}chler}},\ and\ \bibinfo {author}
  {\bibfnamefont {T.}~\bibnamefont {Pfau}},\ }\href
  {https://iopscience.iop.org/article/10.1088/0953-4075/45/11/113001/meta}
  {\bibfield  {journal} {\bibinfo  {journal} {Journal of Physics B: Atomic,
  Molecular and Optical Physics}\ }\textbf {\bibinfo {volume} {45}},\ \bibinfo
  {pages} {113001} (\bibinfo {year} {2012})}\BibitemShut {NoStop}%
\bibitem [{\citenamefont {Cetina}\ \emph {et~al.}(2016)\citenamefont {Cetina},
  \citenamefont {Jag}, \citenamefont {Lous}, \citenamefont {Fritsche},
  \citenamefont {Walraven}, \citenamefont {Grimm}, \citenamefont {Levinsen},
  \citenamefont {Parish}, \citenamefont {Schmidt}, \citenamefont {Knap} \emph
  {et~al.}}]{cetina2016ultrafast}%
  \BibitemOpen
  \bibfield  {author} {\bibinfo {author} {\bibfnamefont {M.}~\bibnamefont
  {Cetina}}, \bibinfo {author} {\bibfnamefont {M.}~\bibnamefont {Jag}},
  \bibinfo {author} {\bibfnamefont {R.~S.}\ \bibnamefont {Lous}}, \bibinfo
  {author} {\bibfnamefont {I.}~\bibnamefont {Fritsche}}, \bibinfo {author}
  {\bibfnamefont {J.~T.}\ \bibnamefont {Walraven}}, \bibinfo {author}
  {\bibfnamefont {R.}~\bibnamefont {Grimm}}, \bibinfo {author} {\bibfnamefont
  {J.}~\bibnamefont {Levinsen}}, \bibinfo {author} {\bibfnamefont {M.~M.}\
  \bibnamefont {Parish}}, \bibinfo {author} {\bibfnamefont {R.}~\bibnamefont
  {Schmidt}}, \bibinfo {author} {\bibfnamefont {M.}~\bibnamefont {Knap}}, \emph
  {et~al.},\ }\href {https://www.science.org/doi/abs/10.1126/science.aaf5134}
  {\bibfield  {journal} {\bibinfo  {journal} {Science}\ }\textbf {\bibinfo
  {volume} {354}},\ \bibinfo {pages} {96} (\bibinfo {year} {2016})}\BibitemShut
  {NoStop}%
\bibitem [{\citenamefont {Liu}\ \emph {et~al.}(2019)\citenamefont {Liu},
  \citenamefont {Levinsen},\ and\ \citenamefont {Parish}}]{liu2019variational}%
  \BibitemOpen
  \bibfield  {author} {\bibinfo {author} {\bibfnamefont {W.~E.}\ \bibnamefont
  {Liu}}, \bibinfo {author} {\bibfnamefont {J.}~\bibnamefont {Levinsen}},\ and\
  \bibinfo {author} {\bibfnamefont {M.~M.}\ \bibnamefont {Parish}},\ }\href
  {https://journals.aps.org/prl/abstract/10.1103/PhysRevLett.122.205301}
  {\bibfield  {journal} {\bibinfo  {journal} {Physical Review Letters}\
  }\textbf {\bibinfo {volume} {122}},\ \bibinfo {pages} {205301} (\bibinfo
  {year} {2019})}\BibitemShut {NoStop}%
\bibitem [{Note1()}]{Note1}%
  \BibitemOpen
  \bibinfo {note} {We use here the terminology of polarons also in the infinite
  impurity mass limit where Anderson's orthogonality leads formally to a
  vanishing polaron quasiparticle weight \cite {schmidt2018universal}. For the
  physics of the polaron cloud formation this is, however, not relevant
  here.}\BibitemShut {Stop}%
\bibitem [{sup()}]{supplement}%
  \BibitemOpen
  \href@noop {} {}\bibinfo {note} {See supplemental material for additional
  details; the supplemental material includes
  Refs.~\cite{knap2012time,schmidt2018theory,sous2020rydberg,sakurai1995modern,bloch2008many,du1987interaction,Burkhardt2006,eiles2018formation,klich2003elementary,schmidt2018universal,lous2017thermometry}}\BibitemShut
  {NoStop}%
\bibitem [{\citenamefont {Sakurai}\ and\ \citenamefont
  {Napolitano}(2011)}]{sakurai1995modern}%
  \BibitemOpen
  \bibfield  {author} {\bibinfo {author} {\bibfnamefont {J.~J.}\ \bibnamefont
  {Sakurai}}\ and\ \bibinfo {author} {\bibfnamefont {J.}~\bibnamefont
  {Napolitano}},\ }\href@noop {} {\emph {\bibinfo {title} {Modern quantum
  mechanics, revised edition}}}\ (\bibinfo  {publisher} {Addison-Wesley,
  Boston},\ \bibinfo {year} {2011})\BibitemShut {NoStop}%
\bibitem [{\citenamefont {Bloch}\ \emph {et~al.}(2008)\citenamefont {Bloch},
  \citenamefont {Dalibard},\ and\ \citenamefont {Zwerger}}]{bloch2008many}%
  \BibitemOpen
  \bibfield  {author} {\bibinfo {author} {\bibfnamefont {I.}~\bibnamefont
  {Bloch}}, \bibinfo {author} {\bibfnamefont {J.}~\bibnamefont {Dalibard}},\
  and\ \bibinfo {author} {\bibfnamefont {W.}~\bibnamefont {Zwerger}},\ }\href
  {https://journals.aps.org/rmp/abstract/10.1103/RevModPhys.80.885} {\bibfield
  {journal} {\bibinfo  {journal} {Rev. Mod. Phys.}\ }\textbf {\bibinfo {volume}
  {80}},\ \bibinfo {pages} {885} (\bibinfo {year} {2008})}\BibitemShut
  {NoStop}%
\bibitem [{\citenamefont {Burkhardt}\ and\ \citenamefont
  {Leventhal}(2006)}]{Burkhardt2006}%
  \BibitemOpen
  \bibfield  {author} {\bibinfo {author} {\bibfnamefont {C.~E.}\ \bibnamefont
  {Burkhardt}}\ and\ \bibinfo {author} {\bibfnamefont {J.~J.}\ \bibnamefont
  {Leventhal}},\ }\bibinfo {title} {The quantum defect},\ in\ \href
  {https://doi.org/10.1007/0-387-31074-6_11} {\emph {\bibinfo {booktitle}
  {Topics in Atomic Physics}}}\ (\bibinfo  {publisher} {Springer},\ \bibinfo
  {address} {New York},\ \bibinfo {year} {2006})\ pp.\ \bibinfo {pages}
  {214--229}\BibitemShut {NoStop}%
\bibitem [{\citenamefont {Eiles}(2018)}]{eiles2018formation}%
  \BibitemOpen
  \bibfield  {author} {\bibinfo {author} {\bibfnamefont {M.~T.}\ \bibnamefont
  {Eiles}},\ }\href
  {https://journals.aps.org/pra/abstract/10.1103/PhysRevA.98.042706} {\bibfield
   {journal} {\bibinfo  {journal} {Phys. Rev. A}\ }\textbf {\bibinfo {volume}
  {98}},\ \bibinfo {pages} {042706} (\bibinfo {year} {2018})}\BibitemShut
  {NoStop}%
\bibitem [{\citenamefont {Klich}(2003)}]{klich2003elementary}%
  \BibitemOpen
  \bibfield  {author} {\bibinfo {author} {\bibfnamefont {I.}~\bibnamefont
  {Klich}},\ }\bibinfo {title} {An elementary derivation of {L}evitov's
  formula},\ in\ \href {https://doi.org/10.1007/978-94-010-0089-5_19} {\emph
  {\bibinfo {booktitle} {Quantum Noise in Mesoscopic Physics}}},\ \bibinfo
  {editor} {edited by\ \bibinfo {editor} {\bibfnamefont {Y.~V.}\ \bibnamefont
  {Nazarov}}}\ (\bibinfo  {publisher} {Springer Netherlands},\ \bibinfo
  {address} {Dordrecht},\ \bibinfo {year} {2003})\ pp.\ \bibinfo {pages}
  {397--402}\BibitemShut {NoStop}%
\bibitem [{\citenamefont {Sch{\"o}nhammer}(2007)}]{schonhammer2007full}%
  \BibitemOpen
  \bibfield  {author} {\bibinfo {author} {\bibfnamefont {K.}~\bibnamefont
  {Sch{\"o}nhammer}},\ }\href
  {https://journals.aps.org/prb/abstract/10.1103/PhysRevB.75.205329} {\bibfield
   {journal} {\bibinfo  {journal} {Phys. Rev. B}\ }\textbf {\bibinfo {volume}
  {75}},\ \bibinfo {pages} {205329} (\bibinfo {year} {2007})}\BibitemShut
  {NoStop}%
\bibitem [{Note2()}]{Note2}%
  \BibitemOpen
  \bibinfo {note} {To be precise, we have run our calculations at
  $T=0.001\protect \,\varepsilon _\protect \mathrm {F}$.}\BibitemShut {Stop}%
\bibitem [{\citenamefont {Anderson}(1967)}]{anderson1967infrared}%
  \BibitemOpen
  \bibfield  {author} {\bibinfo {author} {\bibfnamefont {P.~W.}\ \bibnamefont
  {Anderson}},\ }\href
  {https://journals.aps.org/prl/abstract/10.1103/PhysRevLett.18.1049}
  {\bibfield  {journal} {\bibinfo  {journal} {Phys. Rev. Lett.}\ }\textbf
  {\bibinfo {volume} {18}},\ \bibinfo {pages} {1049} (\bibinfo {year}
  {1967})}\BibitemShut {NoStop}%
\bibitem [{\citenamefont {Diessel}\ \emph {et~al.}(2022)\citenamefont
  {Diessel}, \citenamefont {von Milczewski}, \citenamefont {Christianen},\ and\
  \citenamefont {Schmidt}}]{diessel2022probing}%
  \BibitemOpen
  \bibfield  {author} {\bibinfo {author} {\bibfnamefont {O.~K.}\ \bibnamefont
  {Diessel}}, \bibinfo {author} {\bibfnamefont {J.}~\bibnamefont {von
  Milczewski}}, \bibinfo {author} {\bibfnamefont {A.}~\bibnamefont
  {Christianen}},\ and\ \bibinfo {author} {\bibfnamefont {R.}~\bibnamefont
  {Schmidt}},\ }\href {https://arxiv.org/abs/2209.11758} {\bibfield  {journal}
  {\bibinfo  {journal} {arXiv:2209.11758 [cond-mat.quant-gas]}\ } (\bibinfo
  {year} {2022})}\BibitemShut {NoStop}%
\bibitem [{\citenamefont {J\o{}rgensen}\ \emph {et~al.}(2016)\citenamefont
  {J\o{}rgensen}, \citenamefont {Wacker}, \citenamefont {Skalmstang},
  \citenamefont {Parish}, \citenamefont {Levinsen}, \citenamefont
  {Christensen}, \citenamefont {Bruun},\ and\ \citenamefont
  {Arlt}}]{jorgensen2016}%
  \BibitemOpen
  \bibfield  {author} {\bibinfo {author} {\bibfnamefont {N.~B.}\ \bibnamefont
  {J\o{}rgensen}}, \bibinfo {author} {\bibfnamefont {L.}~\bibnamefont
  {Wacker}}, \bibinfo {author} {\bibfnamefont {K.~T.}\ \bibnamefont
  {Skalmstang}}, \bibinfo {author} {\bibfnamefont {M.~M.}\ \bibnamefont
  {Parish}}, \bibinfo {author} {\bibfnamefont {J.}~\bibnamefont {Levinsen}},
  \bibinfo {author} {\bibfnamefont {R.~S.}\ \bibnamefont {Christensen}},
  \bibinfo {author} {\bibfnamefont {G.~M.}\ \bibnamefont {Bruun}},\ and\
  \bibinfo {author} {\bibfnamefont {J.~J.}\ \bibnamefont {Arlt}},\ }\href
  {https://doi.org/10.1103/PhysRevLett.117.055302} {\bibfield  {journal}
  {\bibinfo  {journal} {Phys. Rev. Lett.}\ }\textbf {\bibinfo {volume} {117}},\
  \bibinfo {pages} {055302} (\bibinfo {year} {2016})}\BibitemShut {NoStop}%
\bibitem [{\citenamefont {Hu}\ \emph {et~al.}(2016)\citenamefont {Hu},
  \citenamefont {Van~de Graaff}, \citenamefont {Kedar}, \citenamefont {Corson},
  \citenamefont {Cornell},\ and\ \citenamefont {Jin}}]{hu2016}%
  \BibitemOpen
  \bibfield  {author} {\bibinfo {author} {\bibfnamefont {M.-G.}\ \bibnamefont
  {Hu}}, \bibinfo {author} {\bibfnamefont {M.~J.}\ \bibnamefont {Van~de
  Graaff}}, \bibinfo {author} {\bibfnamefont {D.}~\bibnamefont {Kedar}},
  \bibinfo {author} {\bibfnamefont {J.~P.}\ \bibnamefont {Corson}}, \bibinfo
  {author} {\bibfnamefont {E.~A.}\ \bibnamefont {Cornell}},\ and\ \bibinfo
  {author} {\bibfnamefont {D.~S.}\ \bibnamefont {Jin}},\ }\href
  {https://doi.org/10.1103/PhysRevLett.117.055301} {\bibfield  {journal}
  {\bibinfo  {journal} {Phys. Rev. Lett.}\ }\textbf {\bibinfo {volume} {117}},\
  \bibinfo {pages} {055301} (\bibinfo {year} {2016})}\BibitemShut {NoStop}%
\bibitem [{\citenamefont {Yan}\ \emph {et~al.}(2020)\citenamefont {Yan},
  \citenamefont {Ni}, \citenamefont {Robens},\ and\ \citenamefont
  {Zwierlein}}]{Yan2019}%
  \BibitemOpen
  \bibfield  {author} {\bibinfo {author} {\bibfnamefont {Z.~Z.}\ \bibnamefont
  {Yan}}, \bibinfo {author} {\bibfnamefont {Y.}~\bibnamefont {Ni}}, \bibinfo
  {author} {\bibfnamefont {C.}~\bibnamefont {Robens}},\ and\ \bibinfo {author}
  {\bibfnamefont {M.~W.}\ \bibnamefont {Zwierlein}},\ }\href
  {https://doi.org/10.1126/science.aax5850} {\bibfield  {journal} {\bibinfo
  {journal} {Science}\ }\textbf {\bibinfo {volume} {368}},\ \bibinfo {pages}
  {190} (\bibinfo {year} {2020})}\BibitemShut {NoStop}%
\bibitem [{\citenamefont {Pe\~na Ardila}\ \emph {et~al.}(2019)\citenamefont
  {Pe\~na Ardila}, \citenamefont {J\o{}rgensen}, \citenamefont {Pohl},
  \citenamefont {Giorgini}, \citenamefont {Bruun},\ and\ \citenamefont
  {Arlt}}]{Ardila2019}%
  \BibitemOpen
  \bibfield  {author} {\bibinfo {author} {\bibfnamefont {L.~A.}\ \bibnamefont
  {Pe\~na Ardila}}, \bibinfo {author} {\bibfnamefont {N.~B.}\ \bibnamefont
  {J\o{}rgensen}}, \bibinfo {author} {\bibfnamefont {T.}~\bibnamefont {Pohl}},
  \bibinfo {author} {\bibfnamefont {S.}~\bibnamefont {Giorgini}}, \bibinfo
  {author} {\bibfnamefont {G.~M.}\ \bibnamefont {Bruun}},\ and\ \bibinfo
  {author} {\bibfnamefont {J.~J.}\ \bibnamefont {Arlt}},\ }\href
  {https://doi.org/10.1103/PhysRevA.99.063607} {\bibfield  {journal} {\bibinfo
  {journal} {Phys. Rev. A}\ }\textbf {\bibinfo {volume} {99}},\ \bibinfo
  {pages} {063607} (\bibinfo {year} {2019})}\BibitemShut {NoStop}%
\bibitem [{\citenamefont {Wang}\ \emph {et~al.}(2022)\citenamefont {Wang},
  \citenamefont {Liu},\ and\ \citenamefont {Hu}}]{wang2022exact}%
  \BibitemOpen
  \bibfield  {author} {\bibinfo {author} {\bibfnamefont {J.}~\bibnamefont
  {Wang}}, \bibinfo {author} {\bibfnamefont {X.-J.}\ \bibnamefont {Liu}},\ and\
  \bibinfo {author} {\bibfnamefont {H.}~\bibnamefont {Hu}},\ }\href
  {https://journals.aps.org/prl/abstract/10.1103/PhysRevLett.128.175301}
  {\bibfield  {journal} {\bibinfo  {journal} {Phys. Rev. Lett.}\ }\textbf
  {\bibinfo {volume} {128}},\ \bibinfo {pages} {175301} (\bibinfo {year}
  {2022})}\BibitemShut {NoStop}%
\bibitem [{\citenamefont {Lous}\ \emph {et~al.}(2017)\citenamefont {Lous},
  \citenamefont {Fritsche}, \citenamefont {Jag}, \citenamefont {Huang},\ and\
  \citenamefont {Grimm}}]{lous2017thermometry}%
  \BibitemOpen
  \bibfield  {author} {\bibinfo {author} {\bibfnamefont {R.~S.}\ \bibnamefont
  {Lous}}, \bibinfo {author} {\bibfnamefont {I.}~\bibnamefont {Fritsche}},
  \bibinfo {author} {\bibfnamefont {M.}~\bibnamefont {Jag}}, \bibinfo {author}
  {\bibfnamefont {B.}~\bibnamefont {Huang}},\ and\ \bibinfo {author}
  {\bibfnamefont {R.}~\bibnamefont {Grimm}},\ }\href
  {https://doi.org/10.1103/PhysRevA.95.053627} {\bibfield  {journal} {\bibinfo
  {journal} {Phys. Rev. A}\ }\textbf {\bibinfo {volume} {95}},\ \bibinfo
  {pages} {053627} (\bibinfo {year} {2017})}\BibitemShut {NoStop}%
\end{thebibliography}%

\clearpage

\title{Supplemental material: Probing Polaron Clouds by Rydberg Atom Spectroscopy}
\date{\today}
\maketitle

\setcounter{equation}{0}
\setcounter{figure}{0}
\setcounter{table}{0}
\setcounter{page}{1}
\makeatletter
\renewcommand{\theequation}{S\arabic{equation}}
\renewcommand{\thesection}{S\arabic{section}}
\renewcommand{\thefigure}{S\arabic{figure}}

In this document, we provide  details on the methods and parameters we have used in order to produce the data shown in the main text. In \Sec{sec:Centro-symmetric-potentials} we give the single-particle eigenfunctions for the system in presence of different impurities.
\Sec{sec:Klich} motivates the functional determinant approach (FDA) expressions for the investigated quantities: density profiles discussed in \Sec{sec:Density-details} and Ramsey signals which are used to compute the absorption spectra discussed in \Sec{sec:Details-spectra}. We conclude this supplement in \Sec{sec:Numerical_accuracy} with a brief comment on our numerical accuracy.

\section{Spherically symmetric potentials}\label{sec:Centro-symmetric-potentials}

All the FDA calculations \cite{knap2012time,schmidt2018theory,sous2020rydberg} are performed in the bases of single-particle solutions for the Schrödinger equations with the respective impurities. The solution of single-particle impurity problems is a standard quantum-mechanical exercise \cite{sakurai1995modern}. Still, for the sake of completeness and to settle our notation, we provide here the respective eigenfunctions and overlaps used in our FDA computations.

We consider a single-particle Hamiltonian $\hat h_\sigma$ with a spherically symmetric potential $\hat v_\sigma$:
\begin{subequations}
	\begin{align}
		\hat h_\sigma &= \frac{\hat{\bm{p}}^2}{2m}+ \hat v_\sigma,\\
		\frac{\hat{\bm{p}}^2}{2m} &= - \frac{1}{2m}\bm{\nabla}^2 = -\frac{1}{2mr^2}\left[\partial_r(r^2\partial_r)-\hat{\bm{l}}^2\right].
	\end{align}
\end{subequations}
The eigenfunctions of the angular-momentum operator $\hat{\bm{l}}^2$ are the spherical harmonics $\hat{\bm{l}}^2 Y_{lm} = l(l+1)Y_{lm}$. With the wave function
\begin{align}
	\langle\bm{r}|klm\rangle &= Y_{lm}(\Omega_{\bm{r}})\frac{u_{kl}(r)}{r},
	\label{eq:wavefunction-general}
\end{align}
the problem is reduced to a one-dimensional Schrödinger equation:
\begin{subequations}
	\label{eq:Schrödinger-radial}
	\begin{align}
		&u_{kl}''(r) + {2m}\left[E-V_{\sigma,\mathrm{eff}}(r)\right]u_{kl}(r) = 0,\\
		&V_{\sigma,\mathrm{eff}}(r) = V_\sigma(r) + \frac{1}{2mr^2}l(l+1).
	\end{align}
\end{subequations}
Its radial solutions $u_{kl}(r)$ form an orthonormal basis and depend on the respective impurity potential $\hat v_\sigma$. We write the eigenfunctions of the free problem as $|\bm{n}\rangle$, i.e., $\hat h_0|\bm{n}\rangle=\varepsilon_n|\bm{n}\rangle$, and the solutions of the impurity problem as $|\bm{\alpha}\rangle$, i.e., $\hat h_\sigma|\bm{\alpha}\rangle = E_{\sigma,\bm{\alpha}}|\bm{\alpha}\rangle$. Here, $\bm{n},\bm{\alpha}$ include the quantum numbers referring to momentum and angular momentum.

\subsection{Noninteracting case}

For a noninteracting impurity $|0\rangle$ with $V_0 (r) = 0$, we use the basis of free states
\begin{align}
	u_{nl}(r) &= \mathcal{N}_{nl}\, r\, j_{l}(k_{nl}r),
	\label{eq:radial-wavefunction}
\end{align}
with the spherical Bessel functions $j_l(x)$. As our system is limited to a sphere of radius $R$, we enforce ${\psi_{klm}(|\bm{r}|= R)}=0$ such that the momenta are discretized,
\begin{align}
	k_{nl} &= z_{nl}/R.
\end{align}
Here, $z_{nl}$ is the $n$th zero of the $l$th spherical Bessel functions, i.e., $j_l(z_{nl})=0$. The normalization factor $\mathcal{N}_{nl}$ yields
\begin{align}
	\mathcal{N}_{nl} &= \frac{\sqrt{2}}{\sqrt{R^3[\,j_{l+1}(z_{nl})]^2}},
\end{align}
which comes from the definite integral over spherical Bessel functions
\begin{align}
	\int_0^R\dd r\,r^2j_l(k_{nl}r)j_l(k_{n'l}r) &= \delta_{nn'}\frac{R^3}{2}[\,j_{l+1}(z_{nl})]^2.
\end{align}

For zero angular momentum, we have $Y_{00} = 1/\sqrt{4\pi}$, $j_0(x) = \sin x/x$, $z_{n0} = n\pi$ and $\mathcal{N}_{n0}=\sqrt{2/R}\,k_{n0}$. So the total wave function reads:
\begin{subequations}
	\begin{align}
		u_{n0}(r) &=\sqrt{ \frac{2}{R}}\sin(k_n r),
		\label{eq:u_n0}\\
		\langle\bm{r}|n00\rangle &= \frac{1}{\sqrt{2\pi R}}\frac{\sin(k_n r)}{r}, \quad k_n = \frac{n\pi}{R}.
	\end{align}
\end{subequations}

The general integral over two radial wave functions is given by:
\begin{align}
	\nonumber&\int_0^r\dd r'\,|u_{nl}(r')|^2 = \mathcal{N}_{nl}^2\int_0^r\dd r'\,r'^2 [\,j_l(k_{nl}r')]^2\\
	&=\mathcal{N}_{nl}^2 \frac{r^3}{2}\left\{[\,j_l(k_{nl}r)]^2-j_{l-1}(k_{nl}r)j_{l+1}(k_{nl}r)\right\}.
	\label{eq:general-integral-for-number}
\end{align}
This is used to calculated the number of particles within a certain radius $N(r)$, see \Eq{eq:number_final} below.

For the $s$-wave case, \Eq{eq:general-integral-for-number} can be simplified to:
\begin{align}
	\int_0^r\dd r'\,|u_{n0}(r')|^2 = \frac{2}{R}\left[\frac{r}{2}-\frac{\sin 2k_n r}{4k_n}\right].
	\label{eq:integral_u_n0}
\end{align}

\subsection{Delta impurity} \label{sec:Delta-impurity}

For the interacting impurity $|1\rangle$ described by a delta potential $V_1(r) = \frac{a}{2mr^2}\delta(r)\partial_r(r...)$ \cite{bloch2008many},
the radial wave function is given by a combination of spherical Bessel functions $j_l(kr)$ and Neumann functions $y_l(kr)$
\begin{align}
	u_{k l}(r) &= \mathcal{N}_{k l} r \left[\cos\delta_l(k) j_l(kr) - \sin\delta_l(k) y_l(kr)\right].
	\label{eq:Bessel-and-Neumann}
\end{align}
Again, the confinement of the sphere only allows discrete momenta $k_{\alpha l}$.

As the delta impurity is local, it affects only the $s$-wave contribution. With $y_0(x) = -\cos(x)/x$, the corresponding radial wave function is given by:
\begin{subequations}
	\begin{align}
		u_{\alpha 0}(r) &= \sqrt{\frac{2}{R}} A_{\alpha} \sin(k_\alpha r + \delta_\alpha),
		\label{eq:radial-u-delta}\\
		A_\alpha &= \sqrt{\frac{R}{2}}\frac{\mathcal{N}_\alpha}{k_\alpha}=\left(1+\frac{\sin 2\delta_\alpha}{2k_\alpha R}\right)^{-1/2}.
	\end{align}
\end{subequations}
The phase shift $\delta_\alpha$ is connected to the $s$-wave scattering length $a$ and fulfills the following conditions together with the momentum $k_\alpha$:
\begin{align}
	k_\alpha R + \delta_\alpha = \alpha \pi, \quad \delta_\alpha = -\arctan(k_\alpha a).
	\label{eq:phase-shift}
\end{align}
The total wave function becomes:
\begin{align}
	\langle\bm{r}|\alpha 00\rangle &= A_\alpha \frac{1}{\sqrt{2\pi R}}\frac{\sin(k_\alpha r + \delta_\alpha)}{r}.
\end{align}

The overlap integrals of the radial functions \eqref{eq:radial-u-delta} with the noninteracting ones \eqref{eq:u_n0} are given by:
\begin{align}
	\langle n | \alpha \rangle &= \int_0^R\dd r\, u_{n0}(r)u_{\alpha 0}(r) =  \frac{2A_\alpha}{R} \frac{k_n}{k_n^2-k_\alpha^2}\sin\delta_\alpha.
	\label{eq:overlap-n-alpha-delta}
\end{align}
Here and in the following, we use the abbreviation $\langle n|\alpha\rangle\equiv\langle n00|\alpha 00\rangle$.

For positive scattering lengths $a>0$, we have an additional bound state with a negative binding energy. The corresponding radial wave function is derived by replacing $k_\alpha$ by $\ii\kappa = \ii/a$. With $\delta = -\ii \kappa R = -\ii\, \mathrm{artanh}(\kappa a)$, we conclude:
\begin{subequations}
	\begin{align}
		u_b(r) &= -\sqrt{\frac{2}{R}}A_b\sinh[\kappa(r-R)],
		\label{eq:u-b-delta}\\
		A_b    &= \left(\frac{\sinh(2\kappa R)}{2\kappa R} -1\right)^{-1/2}\\
		\Rightarrow u_b(r) &= \frac{\sqrt{2\kappa}(\ee^{-\kappa r}-\ee^{-\kappa(2R-r)})}{\sqrt{1-4\kappa R\ee^{-2\kappa R}-\ee^{-4\kappa R}}} .
	\end{align}
\end{subequations}
The second form is crucial for the numerics as large numbers $\sim\ee^{\kappa R}$ are canceled.

Similar to \Eq{eq:overlap-n-alpha-delta}, the overlap integrals of the radial bound state function \Eq{eq:u-b-delta} with the noninteracting ones are given by:
\begin{align}
	\nonumber\langle n | b\rangle &= \frac{2A_b}{R} \frac{k_n}{k_n^2+\kappa^2}\sinh\kappa R\\
	&= \frac{2}{R}\frac{k_n}{k_n^2+\kappa^2}\frac{\sqrt{\kappa R}(1-\ee^{-2\kappa R})}{(1-4\kappa R \ee^{-2\kappa R}-\ee^{-4\kappa R})^{1/2}}.
	\label{eq:overlap-n-b-delta}
\end{align}

In analogy to \Eq{eq:integral_u_n0}, the integrals over two radial wave functions yield:
\begin{subequations}
	\label{eq:integrated-wavefunctions-delta}
	\begin{align}
		\nonumber&\int_0^r\dd r'\,|u_{\alpha 0}(r')|^2\\
		&= \frac{2}{R}A_\alpha^2 \left(\frac{r}{2}-\frac{\sin[2(k_\alpha r+\delta_\alpha)]-\sin(2\delta_\alpha)}{4k_\alpha}\right),\\
		\nonumber&\int_0^r\dd r'\,|u_{b 0}(r')|^2\\
		&=  \frac{1-\ee^{-2\kappa r}-\ee^{-2\kappa R}(\ee^{-2\kappa R}-\ee^{-2\kappa(R-r)}+4\kappa r)}{1-\ee^{-4\kappa R}-4\kappa R\ee^{-2\kappa R}}.
	\end{align}
\end{subequations}

\subsection{Spherical-well potential}

As mentioned in the main text, the density in presence of a delta impurity diverges at the center. To make sure that this does not affect the spectra, we have performed analogous computations of spectra corresponding to a polaron formed in presence of a spherical-well potential. Here, the divergent density is cured. A spherical well of radius $d$ is described by the potential $V_1(r<d) = V_0 < 0$ and $V_1(r>d) = 0$. Inside the well $r<d$, the radial functions $u_{Kl}(r)$ are given by the spherical Bessel functions with momentum $K=\sqrt{2m(E-V_0)}$ [cf.~\Eq{eq:radial-wavefunction}]. Outside the well $r>d$, the radial functions $u_{kl}(r)$ are given by a linear combination of spherical Bessel and Neumann functions [cf.~\Eq{eq:Bessel-and-Neumann}]
\begin{subequations}
	\label{eq:SQWL-ukl}
	\begin{align}
		u_{Kl}(r<d) &= A_{Kl} r \, j_l (Kr),\\
		u_{kl}(r>d) &= B_{kl} r \, j_l (kr) + C_{kl} r \, y_l (k r).
	\end{align}
\end{subequations}
From the continuous differentiability at the well edge ${r=d}$ and the boundary condition at ${r=R}$, we receive the following general relations for the coefficients $A_{Kl}, B_{kl}$ and $C_{kl}$
\begin{subequations}
	\label{eq:SQWL-conditions}
	\begin{align}
		A_{Kl}\,j_l(Kd) &= B_{kl}j_l(kd)+C_{kl}y_l(d)\,\\
		A_{Kl} K \, j_{l+1}(Kd) &= k \left[B_{kl}j_{l+1}(kd)+C_{kl}y_{l+1}(kd)\right],\\
		0 &= B_{kl}\,j_l(kR) + C_{kl}\, y_l(kR),
	\end{align}
\end{subequations}
where we have used $\dd_x j_l(x) = l/x \cdot j_l(x) - j_{l+1}(x)$ as property of differentiated spherical Bessel functions. For $l=0$, which is in particular interesting for local impurities, the conditions \eqref{eq:SQWL-ukl}-\eqref{eq:SQWL-conditions} lead to the following relation
\begin{align}
	\nonumber &k_\alpha \cos[k_\alpha(R-d)]\sin(K_\alpha d)&\\ &+ K_\alpha \cos(K_\alpha d)\sin[k_\alpha(R-d)] = 0,
	\label{eq:SQWL-condition-l=0}
\end{align}
which is valid for discrete values of momenta $k_\alpha$ and $K_\alpha$ with the corresponding energies $E_\alpha = k_\alpha^2/(2m) = K_\alpha^2/(2m)+V_0$. \Eq{eq:SQWL-condition-l=0} is solved numerically and the momenta are inserted into the radial functions:
\begin{subequations}
	\label{eq:u_alpha_SQWL}
	\begin{align}
		u_{\alpha 0}(r) &= {A_\alpha}\left\{
		\begin{matrix}
			\sin(K_\alpha r)/K_\alpha & r<d\\
			\frac{\cos(K_\alpha d)}{\cos[k_\alpha(R-d)]}\sin[k_\alpha(r-R)]/k_\alpha & r> d
		\end{matrix}\right.\,,\\
		\nonumber A_\alpha &=  \sqrt{8(K_\alpha k_\alpha)^3}\left\{4k_\alpha^3K_\alpha d-2k_\alpha^3\sin(2K_\alpha d)\right.\\
		\nonumber&\phantom{=}+4K_\alpha k_\alpha (R-d)[K_\alpha^2\cos^2(K_\alpha d)\\
		\nonumber&\phantom{=}+k_\alpha^2\sin^2(K_\alpha d)]+2k_\alpha K_\alpha^2\sin(2K_\alpha d)\\
		\nonumber&\phantom{=}-2K_\alpha^2k_\alpha\cos[2k_\alpha(R-d)]\sin(2K_\alpha d)\\
		\nonumber&\phantom{=}-K_\alpha [K_\alpha^2-k_\alpha^2+(K_\alpha^2+k_\alpha^2)\cos(2K_\alpha d)]\\
		&\phantom{=-K_\alpha\quad\cdot}\left.\times\sin[2k_\alpha(R-d)]\right\}^{-1/2}.
	\end{align}
\end{subequations}
The value for the coefficient $A_\alpha$ is determined from normalization of the wave function.

The scattering length $a$ of the spherical well can be determined from the relations $\tan\delta_l(k) = -C_{kl}/B_{kl}$ and $\lim_{k\to 0} k\, \mathrm{cot}\,\delta_0(k)=-a^{-1}$, which gives:
\begin{align}
	a = \lim_{k\to 0}\frac{1}{k}\frac{C_{k0}}{B_{k0}} = d + \frac{2mV_0\tan(\sqrt{-2mV_0}d)}{\sqrt{-2mV_0}^3}.
\end{align}
We solve this equation numerically for $V_0$ at a fixed value of $d$ to obtain the scattering length $a$. Due to the periodicity of the tangent, negative scattering lengths $a$ are found for $2mV_0\in(-\pi^2/(4d^2),0)$ and positive ones for $2mV_0\in(-9\pi^2/(4d^2),-\pi^2/(4d^2))$.

There are bound states with $V_0<E_b<0$ where $k_\alpha$ is replaced by the imaginary number $\ii\kappa_b=\sqrt{-2mE_b}$ while $K_b=\sqrt{2m(E_b-V_0)}$ stays positive. The analogous expressions of \Eqs{eq:SQWL-condition-l=0}-\eqref{eq:u_alpha_SQWL} are
\begin{subequations}
	\label{eq:u_b_SQWL}
	\begin{align}
		\kappa_b & \sin(K_b d)+K_b \cos(K_b d)\tanh[\kappa_b(R-d)]=0,\\
		u_{b 0}(r) &= {A_b}\left\{\begin{matrix}
			\sin(K_b r)/K_b & r<d\\
			-\frac{\cos(K_b d)}{\kappa_b}\ee^{-\kappa_b(r-d)}\frac{1-\ee^{-2\kappa_b(R-r)}}{1+\ee^{-2\kappa_b(R-d)}} & r> d
		\end{matrix}\right.\,,\\
		\nonumber A_b &= 2\sqrt{\kappa_b^3K_b^3}(1+\ee^{-2\kappa_b(R-d)})\\
		\nonumber &\phantom{=}\!\times\!\left\{2K_b^3\cos^2(K_bd)\left[1-\ee^{-4\kappa_b(R-d)}\right.\right.\\
		\nonumber&\phantom{=\times\quad 2K_b^3\cos^2(K_bd)}\left.-4\kappa_b(R-d)\ee^{-2\kappa_b(R-d)}\right]\\
		&\phantom{=}\!\left.+{\kappa_b}^3\left(1\!+\!\ee^{-2\kappa_b(R-d)}\right)^2\!\left[2K_bd-\sin(2K_bd)\right]\right\}^{-1/2}\!\!\!.
	\end{align}
\end{subequations}
At most, there can be $\lfloor\sqrt{-2mV_0}d/\pi\rfloor$ bound states due to the periodicity of the oscillating function. We take potential depths $V_0$ such that there is only one bound state for positive scattering lengths. 

Analogously to \Eqs{eq:overlap-n-alpha-delta} and \eqref{eq:overlap-n-b-delta}, the overlaps for these wave functions have an analytical expression
\begin{widetext}
	\begin{align}
		\nonumber\langle n|\alpha\rangle &= \sqrt{\frac{2 }{R}}A_\alpha\left[\frac{k_n\cos k_n d \sin K_\alpha d - K_\alpha \cos K_\alpha d \sin k_n d}{K_\alpha(K_\alpha^2-k_n^2)}\right.\\
		&\phantom{= \sqrt{\frac{2}{R}}A_\alpha}\left.+\cos K_\alpha d \frac{-k_\alpha\cos[k_\alpha(R-d)]\sin k_n d - k_n\cos k_n d \sin [k_\alpha(R-d)]+k_\alpha \sin k_n R}{k_\alpha\cos[k_\alpha(R-d)](k_n^2-k^2_\alpha)}\right],\\
		\nonumber\langle n|b\rangle &= \sqrt{\frac{2 }{R}}A_b\left[\frac{K_b\cos K_b d \sin k_n d - k_n \cos k_n d \sin K_b d}{K_b(k_n^2-K_b^2)}\right.\\
		&\phantom{= \sqrt{\frac{2 }{R}}A_b}\left.-\cos K_b d \frac{(1+\ee^{-2\kappa_b(R-d)})\kappa_b\sin(k_nd)+(1-\ee^{-2\kappa_b(R-d)})k_n\cos k_n d-2\ee^{-\kappa_b(R-d)}\kappa_b\sin k_n R}{\kappa_b(k_n^2+\kappa_b^2)(1+\ee^{-2\kappa_b(R-d)})}\right].
	\end{align}
\end{widetext}

\subsection{Rydberg impurity}\label{sec:Rydberg-system}

The potential $V_\mathrm{R}(\bm{r})$, generated by the impurity in the Rydberg state, depends on the wave function of the Rydberg electron $\psi_{\mathrm{R},nlm}(\bm{r}) = Y_{lm}(\Omega_{\bm{r}})u_{\mathrm{R},nl}(r)/r$, where the radial wave functions can be approximated in terms of Whittaker functions \cite{du1987interaction}
\begin{align}
	u_{\mathrm{R},nl}(r) = \frac{W_{\nu,l+1/2}[2r/(\nu a_0)]}{\sqrt{a_0\nu^2\Gamma(\nu+l+1)\Gamma(\nu-l)}}, \text{ with } \nu = n - \delta_l.
\end{align}
The level shifts of the energies $E_{nl} \propto -1/[2(n-\delta_l)^2]$ are specific for the atom type. For an impurity of $^{87}\mathrm{Rb}$, we have $\delta_{l=0}=3.13$ \cite{Burkhardt2006}.

\begin{figure}
	\raggedright
	(a)\\
	\vspace{-0.4cm}
	\centering
	\includegraphics[width=\linewidth]{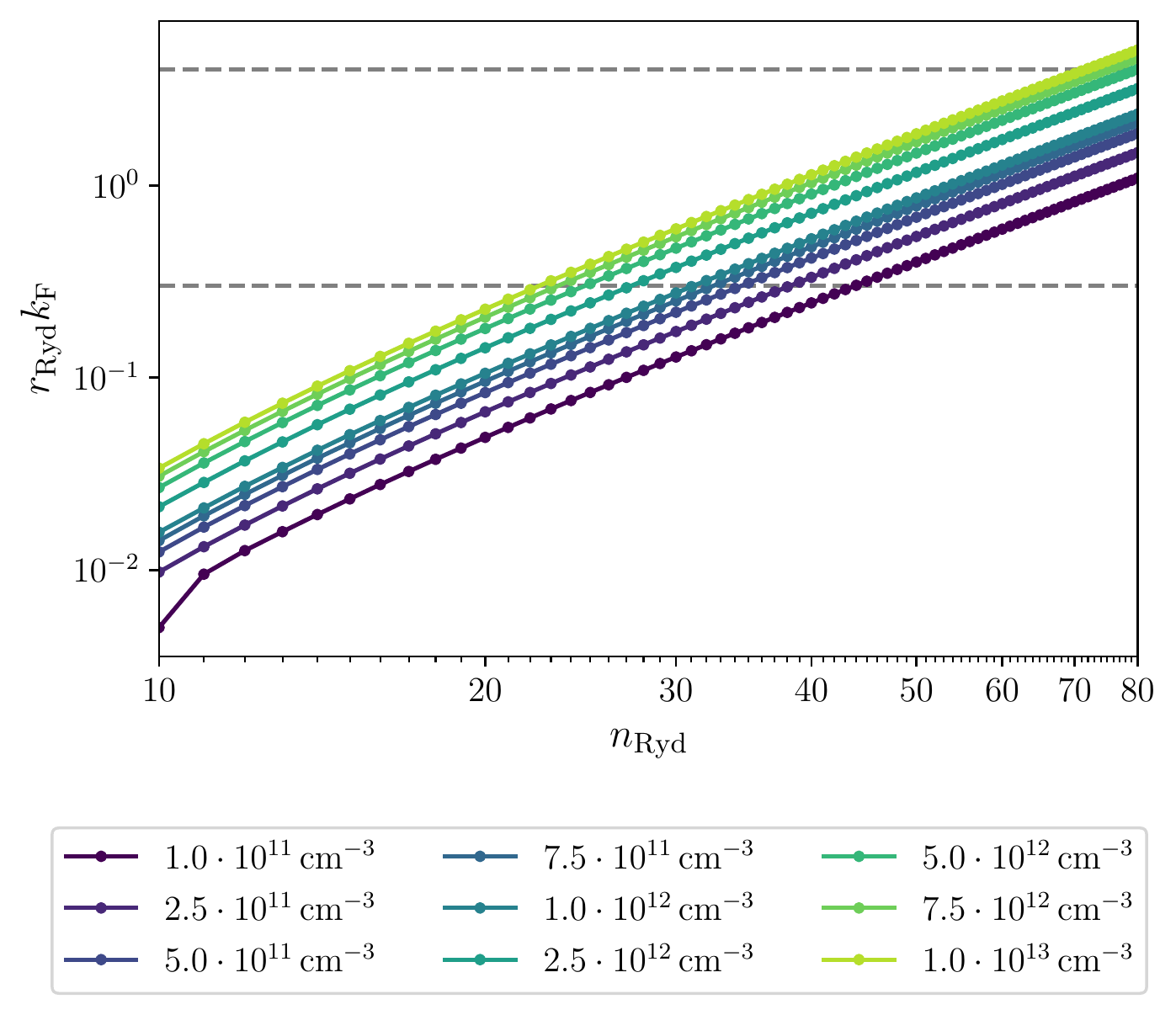}
	\raggedright
	(b)\\
	\vspace{-0.4cm}
	\centering
	\vspace{-15pt}
	\includegraphics[width=1.06\linewidth]{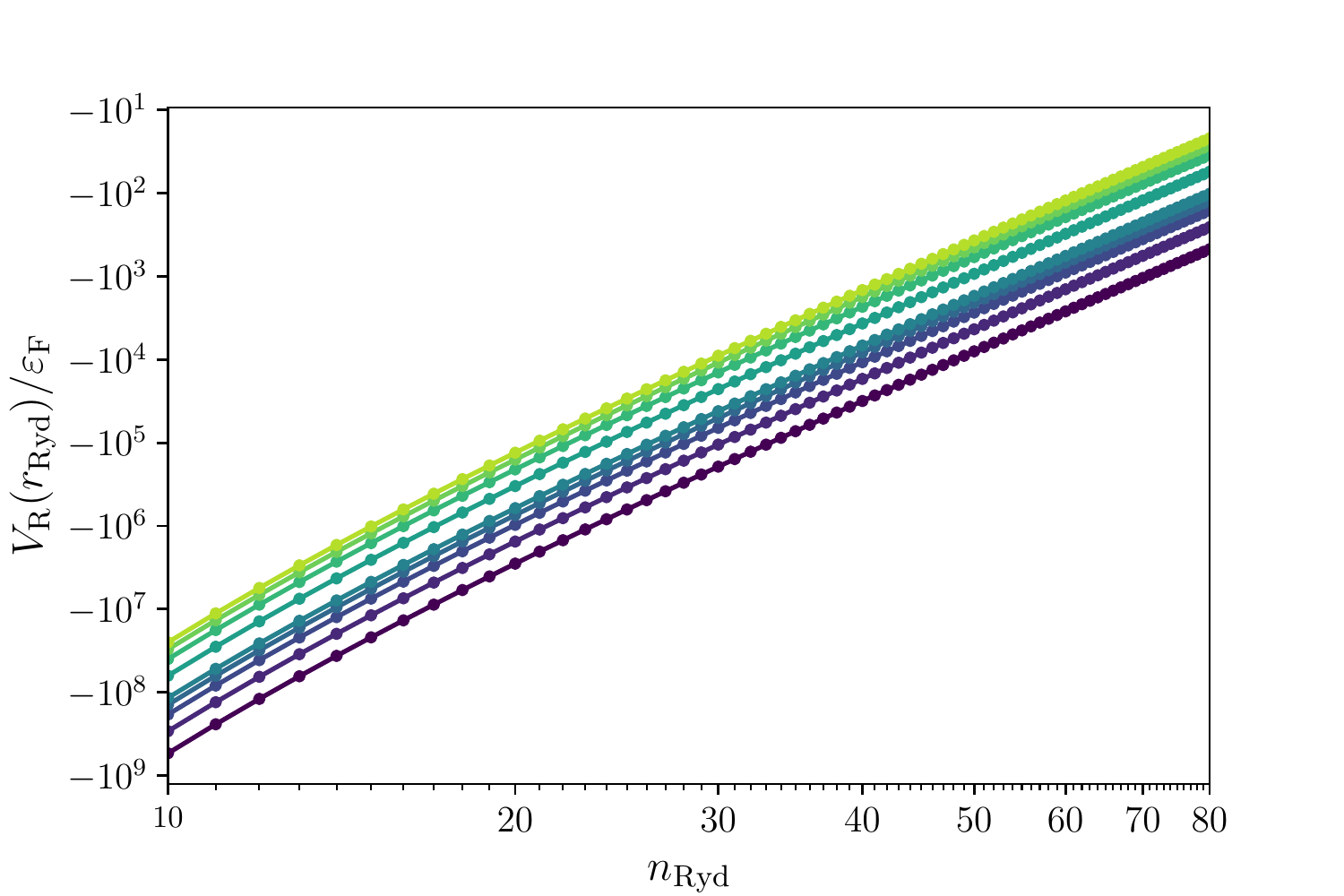}
	\caption{Characteristics of the Rydberg potential $V_\mathrm{R}(r)$ of a $^{87}$Rb impurity in a gas of $^{40}{\mathrm{K}}$ particles measured in units of $\varepsilon_\mathrm{F}$ for different principal quantum numbers $n_\mathrm{Ryd}$ of the outermost electron and different values of densities $\rho_0$ (colors). The upper panel shows the positions $r_\mathrm{Ryd}\sim n_\mathrm{Ryd}^2$ of the dimer minimum and the lower  panel the corresponding potential depth $V_\mathrm{R}(r_\mathrm{Ryd})$. These plots help to choose the right principal numbers $n_\mathrm{Ryd}$ and densities $\rho_0$ for the experimental procedure.}
	\label{fig:Ryderg-potential}
\end{figure}

We consider the spherically symmetric potential $V_{\mathrm{R}}(r)$ for an $s$-wave Rydberg electron, which in total yields:
\begin{align}
	V_\mathrm{R}(r) = \frac{\hbar^2a_e}{2m_e r^2}	\frac{\{W_{\nu,1/2}[2r/(\nu a_0)]\}^2}{a_0\nu^2\Gamma(\nu+1)\Gamma(\nu)}, \quad \nu = n_\mathrm{Ryd}-\delta_0.
	\label{eq:Rydberg-potential-details}
\end{align}
The electron scattering length with $^{40}\mathrm{K}$ atoms is $a_e = -15 a_0$ \cite{eiles2018formation}. The potential $V_\mathrm{R}(r)$ \Eq{eq:Rydberg-potential-details} diverges at the center and is highly oscillating for large principal numbers $n_\mathrm{Ryd}$. It is characterized by a deep minimum located at the Rydberg radius $r_\mathrm{Ryd}$. This minimum induces the localized bound state corresponding to the Rydberg molecule. The position of this bound state $r_\mathrm{Ryd}$ as well as the corresponding potential depth $V_\mathrm{R}(r_\mathrm{Ryd})$ are crucial for our proposed measurement technique. They can be tuned by the principal number $n_\mathrm{Ryd}$ or the density $\rho_0$ measured in units of $k_\mathrm{F}^3$. In \Fig{fig:Ryderg-potential} we show the dependencies of the Rydberg radius $r_\mathrm{Ryd}$ and the potential depth $V_\mathrm{R}(r_\mathrm{Ryd})$ on the principal quantum number $n_\mathrm{Ryd}$ and the overall density $\rho_0$ in units of the Fermi momentum $k_\mathrm{F}$ for gas particles of $^{40}\mathrm{K}$. For observing a polaron's density profile, the region $r_\mathrm{Ryd}k_\mathrm{F}=0.3{-}4.0$ is of particular interest. From \Fig{fig:Ryderg-potential}, we conclude that reasonable principal numbers for our measurement technique are around $n_\mathrm{Ryd}=25-80$ while densities should be at around $\rho_0 = 10^{10}{-}10^{13}\,\mathrm{cm}^{-3}$.

In the presence of the Rydberg potential $V_\mathrm{R}(r)$ \Eq{eq:Rydberg-potential-details}, the radial part of the Schrödinger equation \eqref{eq:Schrödinger-radial} is solved by exact diagonalization of the single-particle Hamiltonian $\hat h_\mathrm{R}$ for a discrete position grid $r_i$ with $i=1,2,...,N$ where $r_1 = 0$ and $r_N = R$. To satisfy the boundary condition $u_{\alpha l}(r_N) = 0$, we only diagonalize the Hamiltonian $[\hat h_\mathrm{R}]_{ij}$ with $i,j = 1,2,...,N-1$. As we allow for a combination of linear grids $r_i$ with higher resolution in the center, we take into account different steps $\delta r_i^+ = r_{i+1}-r_i$ and $\delta r_i^-=r_i-r_{i-1}$ with $\delta r_{1}^+=\delta r_1^-$. Concretely, we use the following discretized Laplacian:
\begin{align}
	\nonumber [\hat h_\mathrm{R}]_{i,i} &= \frac{1}{4m}\!\left[\frac{1}{(\delta r_i^+)^2}+\frac{2}{\delta r_i^+\delta r_i^-}+\frac{1}{(\delta r_i^-)^2}\right]\!+\! V_{\sigma,\mathrm{eff}}(r_i),\\
	\nonumber [\hat h_\mathrm{R}]_{i,i-1} &= - \frac{1}{4m}\left[\frac{1}{(\delta r_i^-)^2}+\frac{1}{\delta r_i^+\delta r_i^-}\right],\\
	[\hat h_\mathrm{R}]_{i,i+1} &= - \frac{1}{4m}\left[\frac{1}{(\delta r_i^+)^2}+\frac{1}{\delta r_i^+\delta r_i^-}\right].
\end{align}
Furthermore, we divide the eigenfunctions $u_{\alpha l}(r_i)$ by the discrete integral $\sqrt{\int_r(u_{\alpha l}(r))^2}$ in order to fulfill their normalization.

\begin{figure}
	\includegraphics[width=\linewidth]{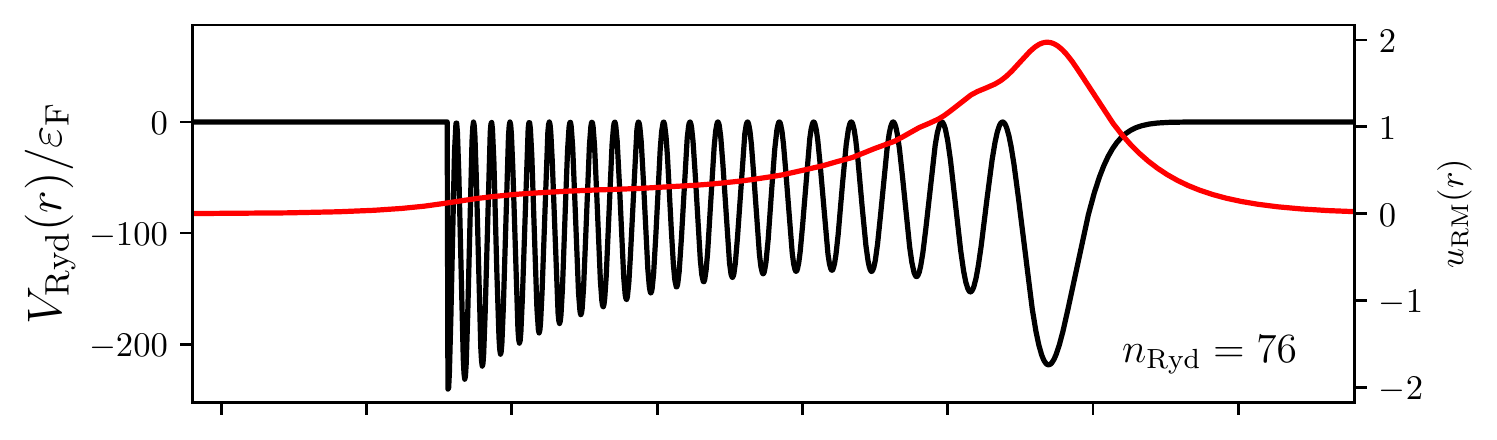}
	\includegraphics[width=\linewidth]{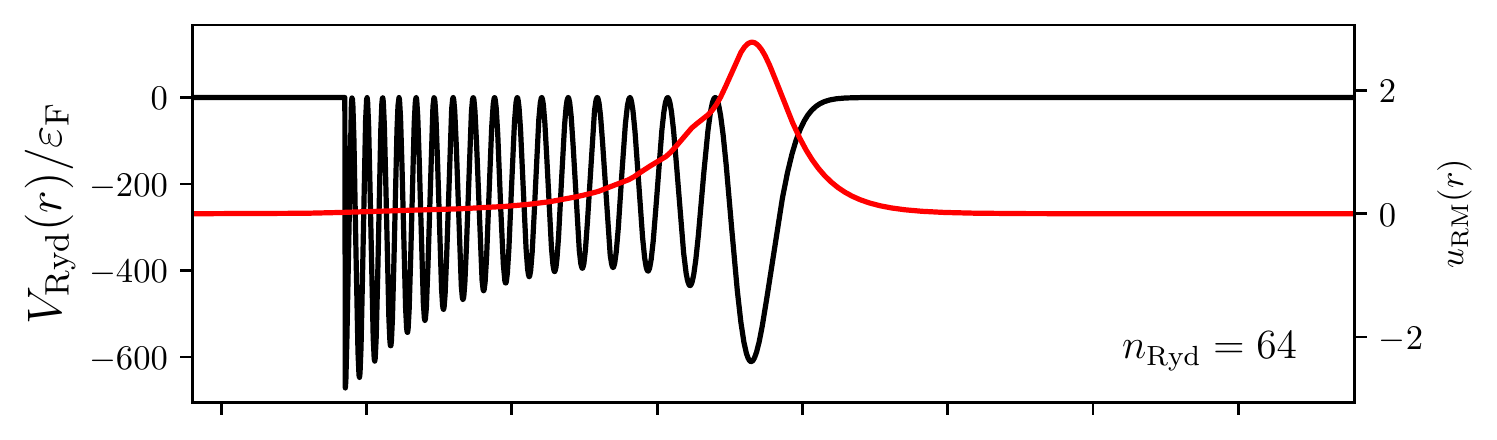}
	\includegraphics[width=\linewidth]{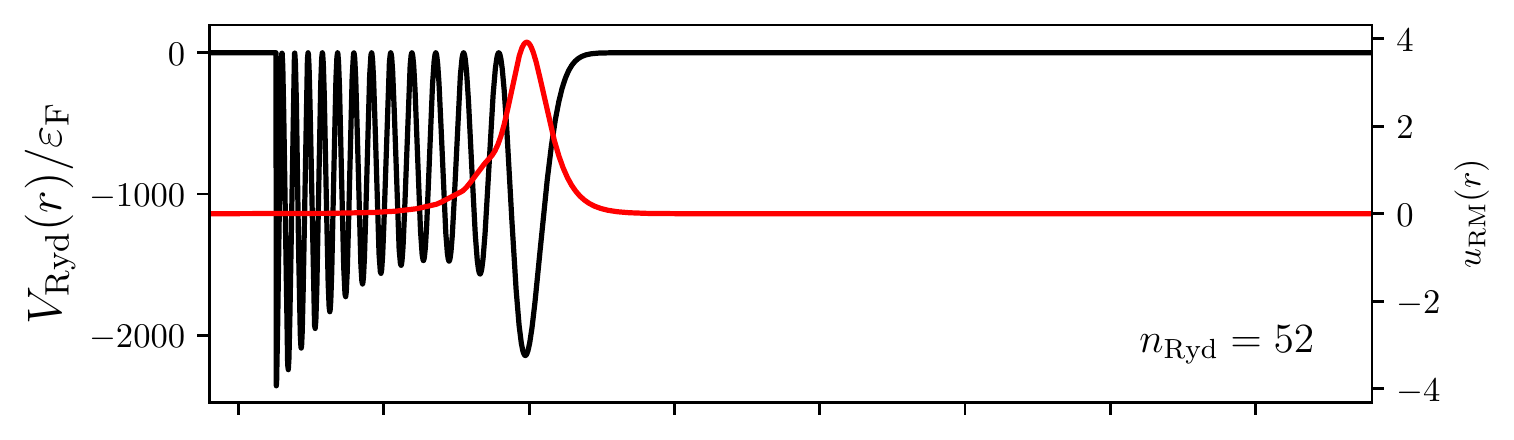}
	\includegraphics[width=\linewidth]{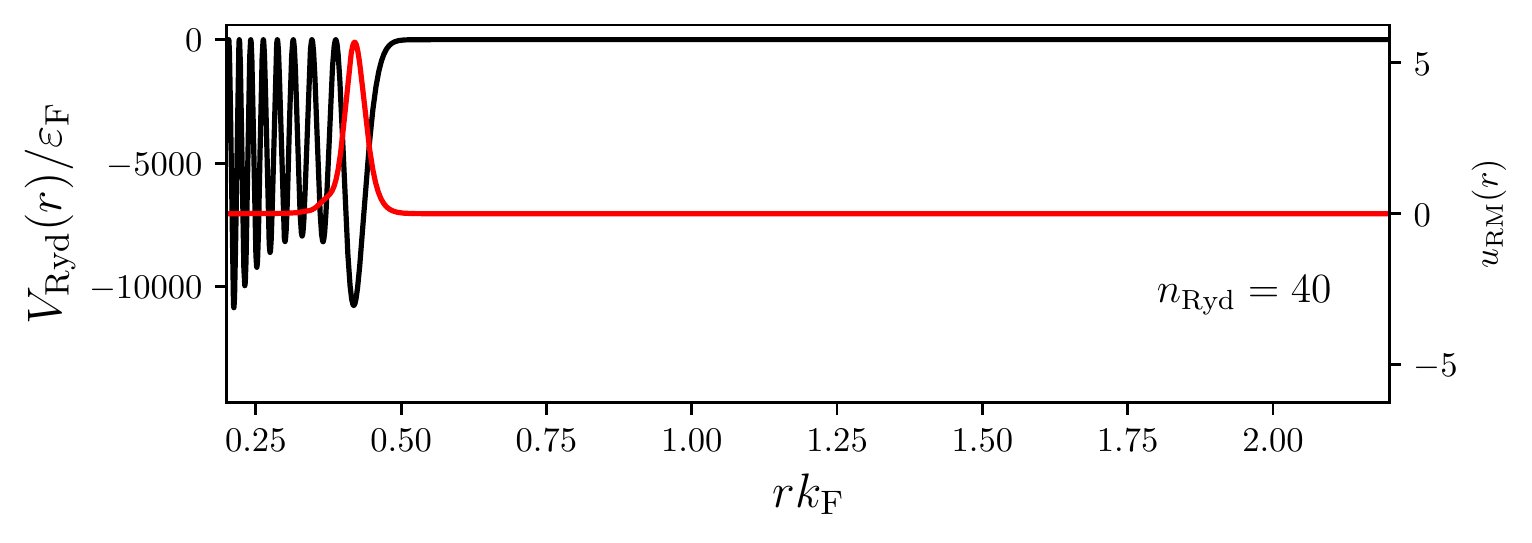}
	\caption{Radial wave functions $u_{\mathrm{RM},l=0}(r)$ of the bound state corresponding to the last minimum in the Rydberg potential $V_\mathrm{R}(r)$ for different principal numbers $n_\mathrm{Ryd}$. The data are calculated for a Rydberg impurity of $^{87}$Rb in a gas of $^{40}$K atoms at particle density $\rho_0 = 5\times 10^{11}\,\mathrm{cm}^{-3}$.
	} 
	\label{fig:Rydberg-potentials}
\end{figure}

\Fig{fig:Rydberg-potentials} shows the Rydberg potential $V_\mathrm{R}(r)$ for different principal numbers $n_\mathrm{Ryd}$ as well as the outermost bound state $u_\mathrm{RM}(r)$, which is located at the Rydberg radius. We see that for higher principal numbers the wave function $u_\mathrm{RM}(r)$ is smeared out. This leads to a less precise value for the position $r$ in the reconstructed density profiles. As we are primarily interested in the dimer peak corresponding to the outermost bound state and not in deeply bound states closer to the center at higher energy scales, our description of the Rydberg potential $V_\mathrm{R}(r)$ does not need to be precise close to the center. In order to lower the numerical effort, we therefore cut the potential in the center as seen in \Fig{fig:Rydberg-potentials}. This does not have an impact on our spectra.

\section{Functional determinant approach}\label{sec:Klich}

Klich's formula \cite{klich2003elementary} allows the calculation of a trace $\mathrm{tr}(\ee^{\hat X})$, where $\hat X$ is a bilinear operator, i.e.,
\begin{align}
	\hat X = \sum_{i,j}\langle i|\hat x|j\rangle\hat c_i^\dagger\hat c_j.
\end{align}
Here, $\hat x$ is the corresponding single-particle operator. For the case of fermionic operators $\hat c_i^\dagger$, $\hat c_j$, Klich's formula takes the form:
\begin{align}
	\mathrm{tr}(\ee^{\hat X}) = \det(\hat\doubleI + \ee^{\hat x}).
\end{align}
This makes it possible to calculate expectation values over products of exponentials, i.e.,
\begin{align}
	\nonumber S_{\sigma_1\sigma_2}(t) &= \langle\ee^{\ii\hat H_{\sigma_1} t}\ee^{-\ii\hat H_{\sigma_2} t} \rangle_{\sigma_1}\\
	\nonumber &= \frac{\mathrm{tr}[\ee^{-\beta(\hat H_{\sigma_1}-\mu\hat N)}\ee^{\ii\hat H_{\sigma_1} t}\ee^{-\ii\hat H_{\sigma_2} t}]}{\mathrm{tr}[\ee^{-\beta(\hat H_{\sigma_1}-\mu\hat N)}]}\\
	&= \mathrm{det}[\hat\doubleI - n_\mathrm{F}(\hat h_{\sigma_1})+n_\mathrm{F}(\hat h_1)\ee^{\ii\hat h_{\sigma_1} t}\ee^{-\ii\hat h_{\sigma_2} t}].
	\label{eq:Ramsey-signal-general}
\end{align}
$S_{\sigma_1\sigma_2}(t)$ is the Ramsey signal when the impurity is prepared in state $|\sigma_1\rangle$ and then switched to the state $|\sigma_2\rangle$.

Furthermore, the Klich formula,
\begin{align}
	\mathrm{tr}(\hat c_i^\dagger \hat c_j\ee^{\hat X}) = \langle j|\frac{\ee^{\hat x}}{\hat{\doubleI}+\ee^{\hat x}}|i\rangle\det(\hat\doubleI+\ee^{\hat x}),
\end{align}
allows the calculation of densities. If the state is prepared at an impurity $|0\rangle$ with density matrix $\hat\rho_0$ and then evolved in presence of the impurity $|\sigma\rangle$, the corresponding density is evaluated as
\begin{align}
	\nonumber\rho_\sigma(\bm{r},t) &= \mathrm{tr}[\hat\rho_{0\sigma}(t)\hat c^\dagger_{\bm{r}}\hat c_{\bm{r}}]\\
	\nonumber &= \frac{\mathrm{tr}[\ee^{-\ii\hat H_\sigma t}\ee^{-\beta(\hat H_0-\mu\hat N)}\ee^{\ii\hat H_\sigma t}\hat c^\dagger_{\bm{r}}\hat c_{\bm{r}}]}{\mathrm{tr}[\ee^{-\beta(\hat H_0-\mu\hat N)}]}\\
	&= \langle \bm{r}|\ee^{-\ii\hat h_\sigma t}n_\mathrm{F}(\hat h_0)\ee^{\ii\hat h_\sigma t}|\bm{r}\rangle.
	\label{eq:density-general}
\end{align}
\Eqs{eq:Ramsey-signal-general} and \eqref{eq:density-general} are the relations we use to calculate the densities and absorption spectra in the main text.

\section{Details on density profiles}\label{sec:Density-details}

In this section, we provide  details for the calculation of the density profiles of the polaron cloud.

The stationary density of the polaron $\rho_\mathrm{pol}(\bm{r})$ given in the main text is calculated by an expansion of the single-particle eigenstates $\langle\bm{r}|\bm{\alpha}\rangle = Y_{lm}(\Omega_{\bm{r}})u_{\alpha l}(r)/r$ with eigenenergies $E_{\bm{\alpha}}$ of the respective impurity problem
\begin{align}
	\nonumber\rho_\mathrm{pol}(\bm{r}) &= \sum_{\bm\alpha}\langle\bm{r}|\bm{\alpha}\rangle n_\mathrm{F}(E_{\bm{\alpha}})\langle\bm{\alpha}|\bm{r}\rangle\\ 
	&= \frac{1}{4\pi r^2} \sum_{\alpha l}(2l+1)n_\mathrm{F}(E_{\alpha l})|u_{\alpha l}(r)|^2.
	\label{eq:stationary-density-final}
\end{align}
Here, the sum over $m$ has been executed as a sum over spherical harmonics so only the radial contributions $u_{\alpha l}(r)$ are left, which are discussed in detail in \Sec{sec:Centro-symmetric-potentials}.

The number of particles within a certain radius $r$ is given by the integral over the density and can be expressed as an integral over the radial wave functions
\begin{align}
	N_\mathrm{pol}(r) &= \sum_{\alpha l}(2l+1)n_\mathrm{F}(E_{\alpha l})\int_0^r\dd r'\,|u_{\alpha l}(r')|^2.
	\label{eq:number_final}
\end{align}
This expression is used to determine the number of particles $N_c$ contributing to the polaron clouds, shown in the main text. The respective expressions for the integrated wave functions are provided by \Eqs{eq:general-integral-for-number} and \eqref{eq:integrated-wavefunctions-delta}.

\begin{figure}
	\includegraphics[width=\linewidth]{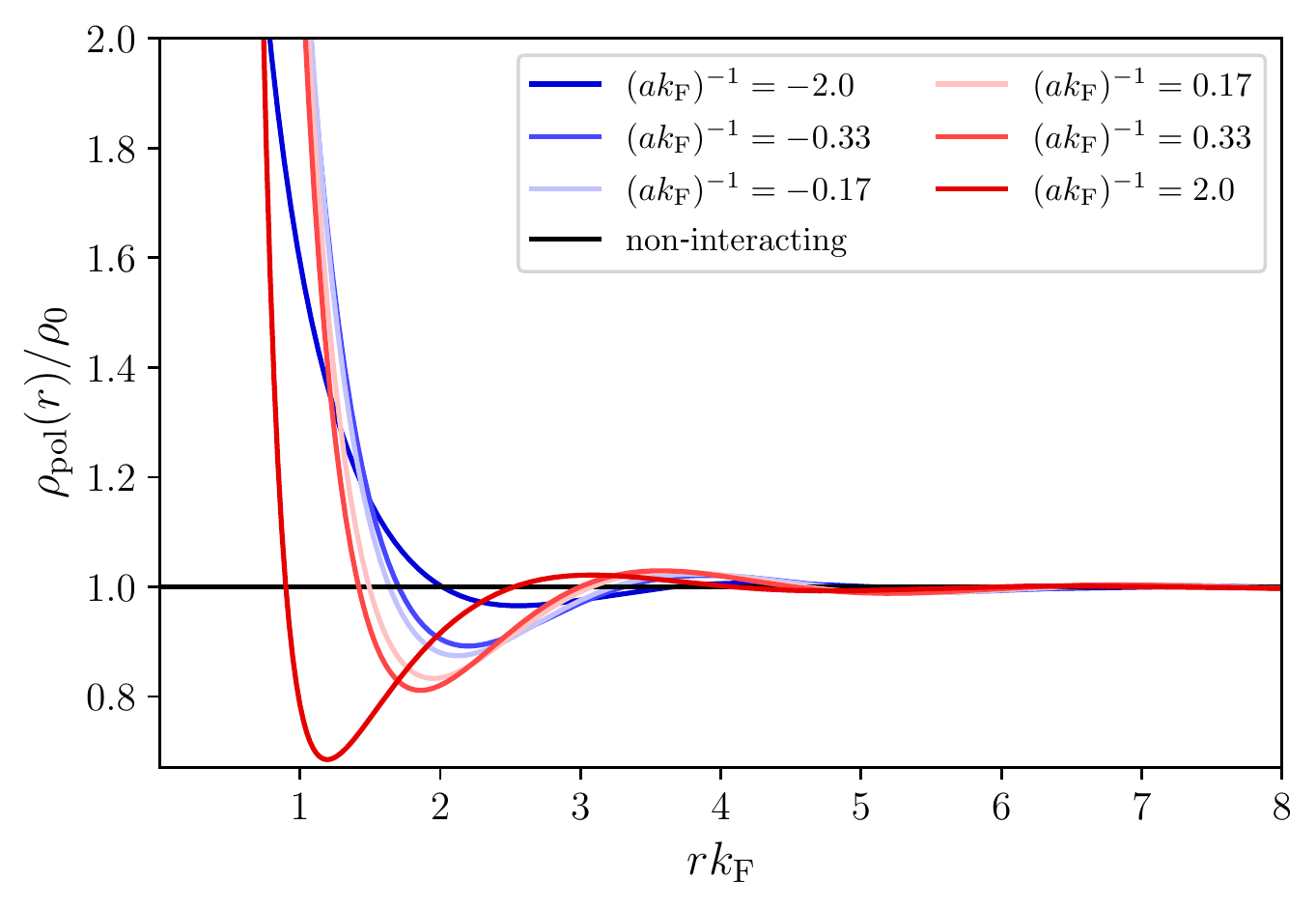}
	\caption{Density profiles $\rho_\mathrm{pol}(r)$ \Eq{eq:stationary-density-final} in the presence of a delta impurity with different inverse $s$-wave scattering lengths $(ak_\mathrm{F})^{-1}$.
	}  
	\label{fig:densities-delta-as}
\end{figure}
\Fig{fig:densities-delta-as} shows density profiles of polaron clouds for different inverse scattering lengths $(ak_\mathrm{F})^{-1}$. These curves supplement the plots provided in the main text. We see the extreme increase of the density around the center. This is the region we identify as the ``polaron cloud''. Further away from the impurity some Friedel-like oscillations appear which fade away quite fast. At distances $rk_\mathrm{F}\gtrsim 5$, the effect of the interacting impurity is not visible anymore such that the polaron densities $\rho_\mathrm{pol}$ coincide with the plateau value $\rho_0$.

\subsection{Particle number from Fumi's theorem}

This section discusses the usefulness of our empirical definition for the polaron cloud (cf.~Fig.~2 in the main text). We define the polaron radius $r_c$ as the first crossing between the density profile $\rho_\mathrm{pol}(r)$ with the background density $\rho_0$, i.e., $\rho_\mathrm{pol}(r_c)=\rho_0$. In the thermodynamic limit, the particle number $N_c$ inside the polaron cloud can be derived from Fumi's theorem (cf. App.~C in \cite{schmidt2018universal}).

For $a<0$, the single-particle energies $E_{\alpha}$ are lowered compared to the corresponding values $\varepsilon_{n}$ of the noninteracting system. The energy of the attractive polaron is then determined by the sum over all occupied energy differences $E_{\alpha=n}-\varepsilon_n$. This can be expressed in terms of the phase shift $\delta_\alpha$ \Eq{eq:phase-shift}
\begin{align}
	E_\mathrm{pol} = \sum_n n_\mathrm{F}(\varepsilon_n)\cdot\left(E_{\alpha=n}-\varepsilon_n\right) \simeq -\frac{2\pi}{R}\sum_n n_\mathrm{F}(\varepsilon_n) \,n\delta_n.
\end{align}
Here, in the last step $\delta_n^2\ll 2n\pi|\delta_n|$ is used as the phase shift is bounded by $\pi$. In the thermodynamic limit, i.e., $R\to\infty$, the energy differences $\Delta\varepsilon_n\equiv\varepsilon_n-\varepsilon_{n-1}={(2n-1)}(\pi/R)^2$ are infinitesimal and we arrive at Fumi's theorem \cite{schmidt2018universal}, which is an integral expression including the polaron energy and the phase shift
\begin{align}
	E_\mathrm{pol} \underset{R\to\infty}{=} -\frac{1}{\pi}\int_0^\infty\dd\varepsilon\,n_\mathrm{F}(\varepsilon)\delta(\varepsilon).
	\label{eq:Fumi's-theorem}
\end{align}

Suppose the chemical potential $\mu$ is fixed and the impurity is switched from $|0\rangle$ to $|1\rangle$. The free energy is reduced by an amount of $E_\mathrm{pol}$, which generates an increase of $N_c$ particles in the system. $N_c$ again can be calculated by a thermodynamic relation of the grand-canonical ensemble
\begin{align}
	N_c = -\frac{\partial E_\mathrm{pol}}{\partial\mu} = \frac{1}{\pi} \int_0^\infty \dd\varepsilon\,\frac{\partial n_\mathrm{F}(\varepsilon)}{\partial\mu}\delta(\varepsilon).
\end{align}
For zero temperature, this yields the phase shift at the Fermi energy $\varepsilon_\mathrm{F}$:
\begin{align}
	N_c = \frac{\delta(\varepsilon_\mathrm{F})}{\pi} = -\frac{1}{\pi}\arctan(k_\mathrm{F}a).
	\label{eq:N_pol-thermodynamics}
\end{align}

For $a > 0$, due to the occupation of the additional boundstate, the energy of the attractive polaron $E_\mathrm{pol}$ \Eq{eq:Fumi's-theorem} is includes the binding energy and $N_c$ \Eq{eq:N_pol-thermodynamics} is increased by $1$.

\begin{figure}
	\includegraphics[width=\linewidth]{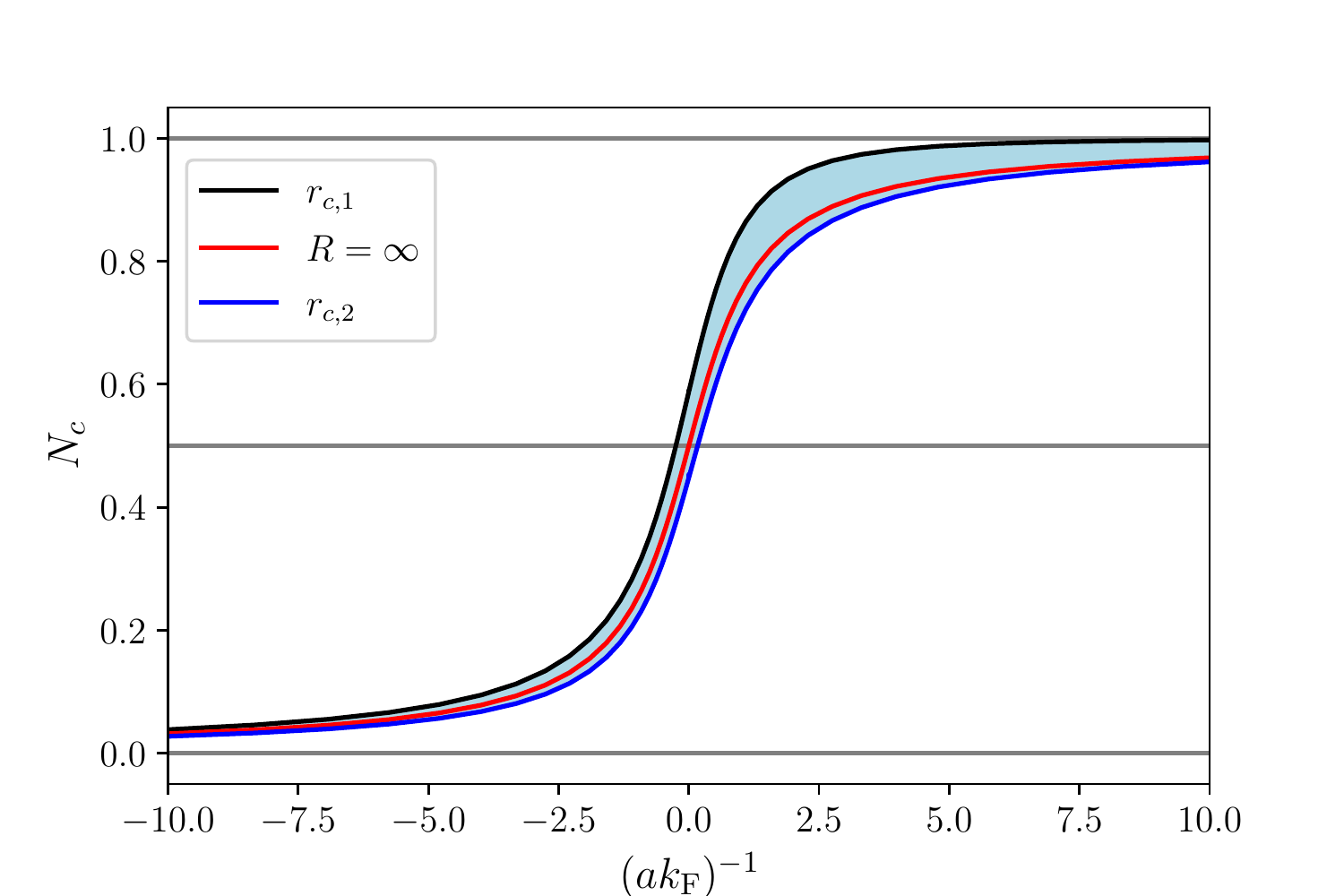}
	\caption{Integrated number of particles inside the polaron cloud $N_c = N_\mathrm{pol}(r_c)-N_0(r_c)$ [cf. \Eq{eq:number_final}] compared to the number at $R\to\infty$ \Eq{eq:N_pol-thermodynamics} extracted from thermodynamic considerations (red line). The black curve corresponds to the integrated number up to the first crossing $\rho_\mathrm{pol}(r_c)=\rho_0$ and the blue curve to the second crossing.}
	\label{fig:polaron_cloud_numbers}
\end{figure}

\Fig{fig:polaron_cloud_numbers} compares the particle number $N_c$ from the main text [cf. \Eq{eq:number_final}], which empirically is determined by an integral over the density up to the polaron cloud radius $r_c$, with the thermodynamic equivalent $N_c$ \Eq{eq:N_pol-thermodynamics}. The qualitative agreement is undeniable. However, an integration only up to the first crossing $\rho_\mathrm{pol}(r_c)=\rho_0$ overestimates the particle number in the polaron whereas an integration up to the second crossing underestimates it.

To conclude, our defined polaron cloud, namely the region of extreme density increase up to the radius $r_c$, may not be exactly identified with the actual polaron defined from thermodynamic properties. Still, the similarity between the integrated number $N_c$ with the corresponding number from Fumi's theorem suggests that the region of increased density gives a useful description of the polaron, which is actually a many-body state of the entire system.

\subsection{Effect of finite angular-momentum states}
\label{sec:lmax}

\begin{figure}
	\includegraphics[width=\linewidth]{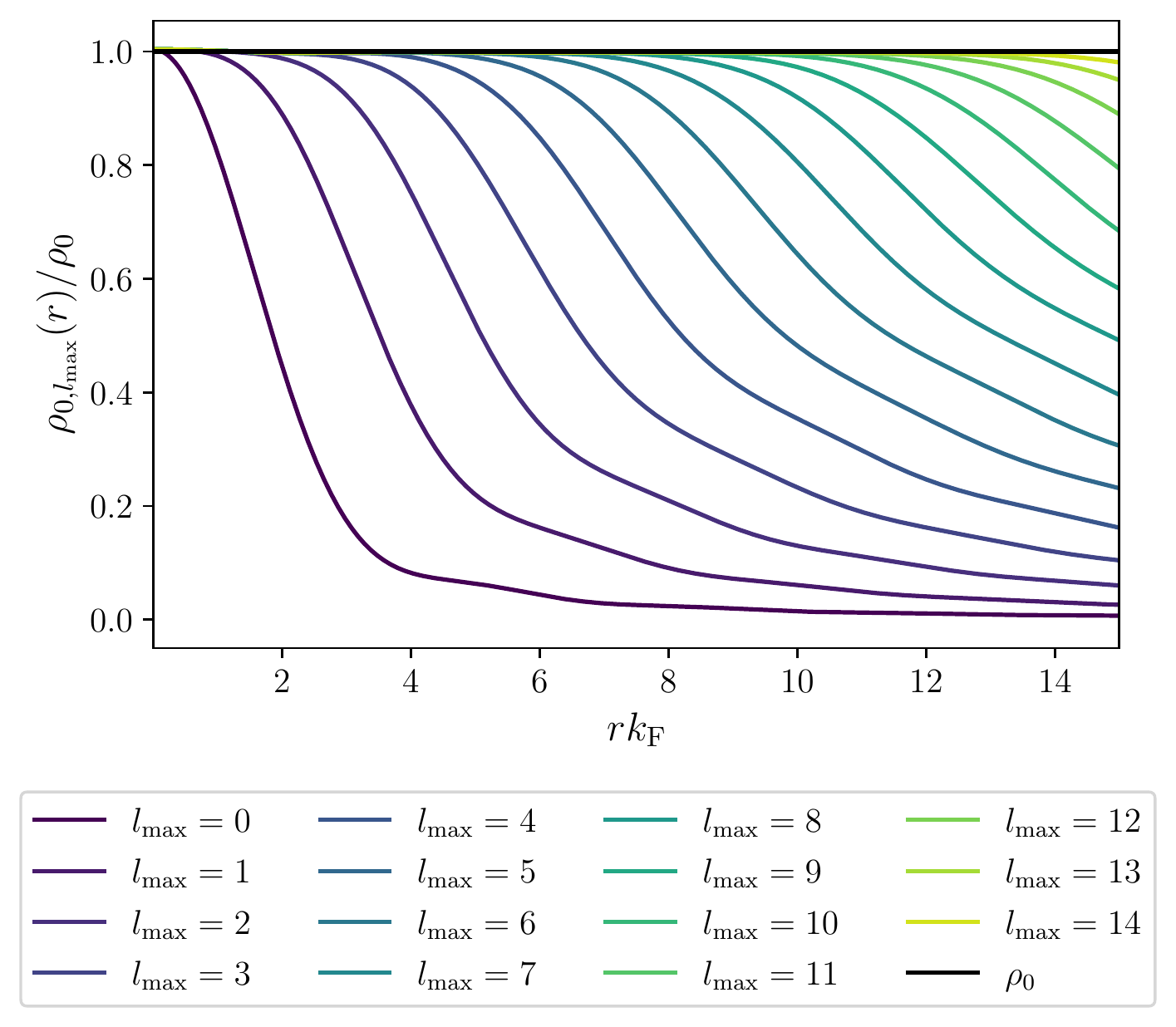}
	\caption{Noninteracting densities $\rho_0(r)$ are shown when the summation in \Eq{eq:stationary-density-final} is limited by different maximal angular momenta $l_{\max}$. The data are compared to the exact value $\rho_0=k_\mathrm{F}^3/(6\pi)^2$.}  
	\label{fig:non-interacting-densities}
\end{figure}
\Fig{fig:non-interacting-densities} shows how many angular momentum states need to be taken into account in the summation in \Eq{eq:stationary-density-final} in order to reach the predicted density plateau $\rho_0$ for the case of a noninteracting impurity $|0\rangle$. We conclude that in order to reach a plateau of radius $r$, one has to take into account about $l/(k_\mathrm{F}r)$ angular momenta. This is important for the calculation of Rydberg spectra. Since the Rydberg radii in our calculations do not exceed $r_\mathrm{Ryd}k_\mathrm{F}=3.0$, it is sufficient to consider only angular momentum states up to $l_\mathrm{max}=8$ in all our calculations.

\subsection{Effect of finite system size}
\label{sec:R}

The finite system size $R$ also modifies the density plateau.  At the center, i.e., $r=0$, only the $s$-wave states contribute to the density. Furthermore, as $\lim_{r\to 0}\sin^2(k_nr)/r^2 = k_n^2 = (n\pi)^2/R^2$ [cf.~\Eq{eq:u_n0}], we find a closed expression for the density summation in \Eq{eq:stationary-density-final} at zero temperature
\begin{align}
	\nonumber \rho_0(r=0) &= \frac{\pi}{2R^3}\sum_n\Theta(\mu-\frac{n^2\pi^2}{2mR^2})n^2\\
	&= \frac{\pi}{12R^3}n_{\max}(n_{\max}+1)(2n_{\max}+1),
\end{align}
where $n_{\max} = \lfloor k_\mathrm{F}R/\pi\rfloor$.
We want to take large system sizes $R$ such that the constant density value $\rho_0 = k_\mathrm{F}^3/(6\pi^2)$ is reached at the center. In order to reach $1\%$ accuracy for the densities around the center, we take a system size of $Rk_\mathrm{F} = 400$.

\subsection{Effect of finite impurity range}

\begin{figure}
	\includegraphics[width=\linewidth]{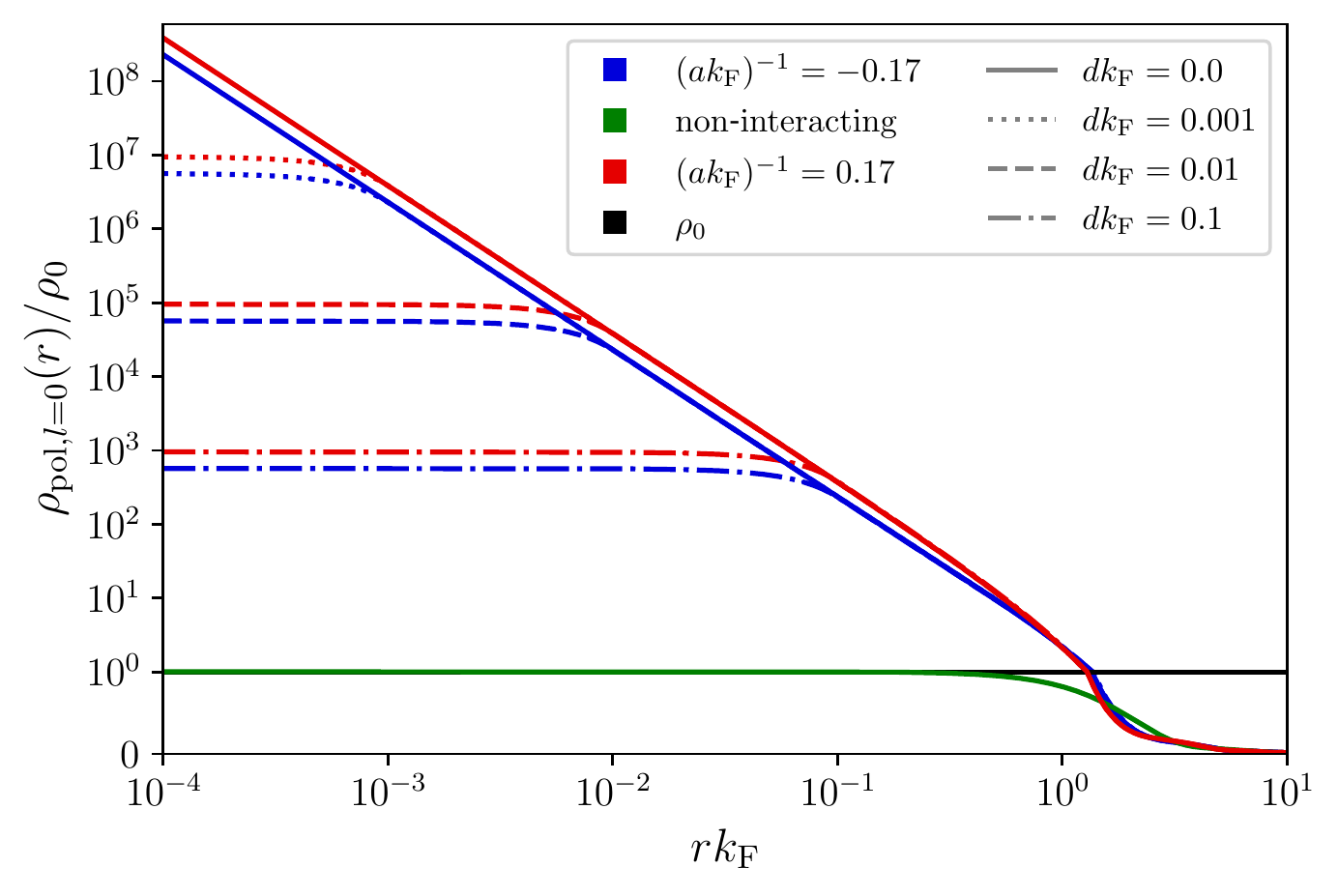}
	\caption{The $s$-wave contributions $\rho_{\mathrm{pol},l=0}(r)$ of the polaron density \Eq{eq:stationary-density-final} are plotted for a spherical-well impurity potential with different interaction radii $d$. Here, $d=0.0$ refers to the solution of the delta impurity. The colors mark different inverse scattering lengths $(ak_\mathrm{F})^{-1}$ and the line styles represent the interaction radii $d$. As a reference, the density plateau $\rho_0$ is shown in black. We conclude that within the range of a finite interaction radius $d$ the density value is just cut off.}  
	\label{fig:sqwl_densities}
\end{figure}
Finally, let us elaborate that the divergent density at the center $r=0$ can be cured by a finite interaction range of the impurity. To do so, in \Fig{fig:sqwl_densities} we show the $s$-wave contributions of the density \Eq{eq:stationary-density-final} when using the radial wave functions $u_{\alpha l}(r)$ for a spherical-well potential of different extents $d>0$ [cf.~\Eqs{eq:u_alpha_SQWL} and \eqref{eq:u_b_SQWL}]. We conclude that in the range of the square-well potential $r<d$, the densities are cut to a finite value so the divergent densities are under control. However, this does not lead to a prominent change of the resulting absorption spectra. Hence, it is sufficient to keep the eigenbasis in presence of the delta impurity.

\subsection{Polaron cloud formation}

\begin{figure}
	\includegraphics[width=\linewidth]{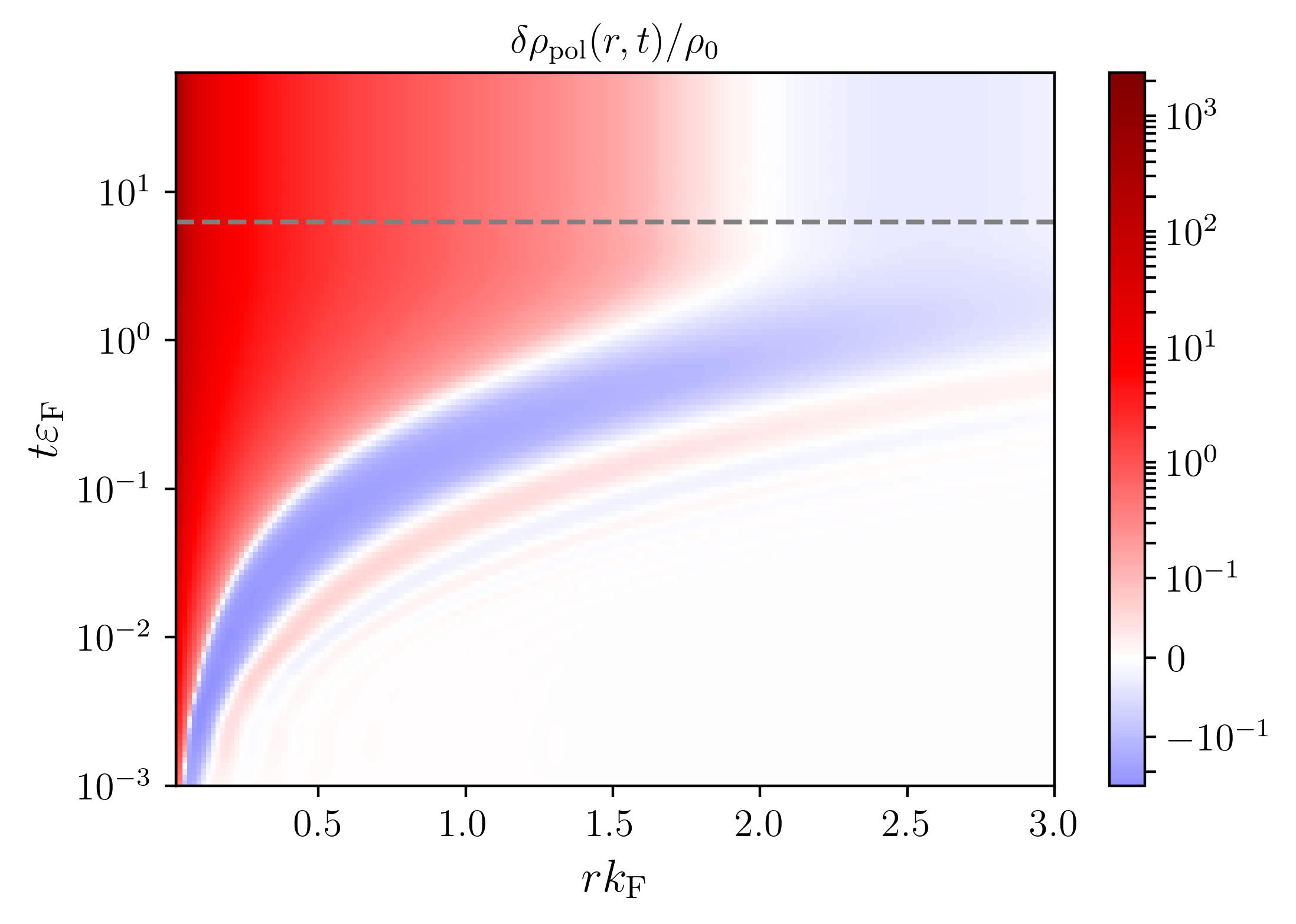}
	\caption{Time-dependent density profile $\delta\rho_\mathrm{pol}(r,t)=\rho_\mathrm{pol}(r,t)-\rho_0$ \Eq{eq:time-dependent-density-final} for an inverse scattering length $(ak_\mathrm{F})^{-1}=-2.0$. The actual polaron cloud formation is complete at $t=2\pi/\varepsilon_\mathrm{F}$, which is marked by the gray dashed line.}  
	\label{fig:cloud_formation}
\end{figure}

We want to elaborate on the statement that the stationary density profile of the polaron cloud \Eq{eq:stationary-density-final} indeed emerges in the long-time limit $t\gg 1/\varepsilon_\mathrm{F}$. For this, we use the more general time-dependent density $\rho(\bm{r},t)$ \Eq{eq:density-general}. An expansion in single-particle wave functions leads to:
\begin{align}
	\nonumber\rho(\bm{r},t) &= \sum_{\bm\alpha,\bm\alpha',\bm n}\langle\bm{r}|\bm\alpha\rangle \ee^{-\ii E_{\bm\alpha} t}\langle\bm\alpha|\bm n\rangle n_\mathrm{F}(\varepsilon_{\bm n})\langle \bm n|\bm\alpha'\rangle\ee^{\ii E_{\bm\alpha'}t}\langle\bm\alpha'|\bm{r}\rangle\\
	&=\sum_{\bm n} n_\mathrm{F}(\varepsilon_{\bm n})\left\vert\sum_{\bm\alpha}\ee^{-\ii E_{\bm\alpha} t}\langle\bm\alpha|\bm n\rangle\langle\bm{r}|\bm \alpha\rangle\right\vert^2.
	\label{eq:time-dependent-density-intermediate}
\end{align}
In analogy to \Eq{eq:stationary-density-final}, the sum over $m$ can be executed by a sum over spherical harmonics and the density only depends on the radial component
\begin{align}
	\nonumber\rho(r,t) &= \frac{1}{4\pi r^2} \sum_{nl}(2l+1)n_\mathrm{F}(\varepsilon_{nl})\\
	&\phantom{= \frac{1}{4\pi r^2} \sum_{nl}}\times\left\vert\sum_{\alpha}\ee^{-\ii E_{\alpha l}t}\langle\alpha l|n l\rangle u_{\alpha l}(r)\right\vert^2.
	\label{eq:time-dependent-density-final}
\end{align}
For a delta impurity, of course, only the $s$-wave contribution is affected by the time dependence. \Fig{fig:cloud_formation} illustrates the formation of a polaron cloud in real time for an exemplary scattering length. The polaron cloud grows with increasing time and is surrounded by fading oscillations until at a time scale inverse to the Fermi energy $t\lesssim 2\pi/\varepsilon_\mathrm{F}$ it reaches its final size. Hence, for a high enough hold time $t_1\gg 1/\varepsilon_\mathrm{F}$, we expect the system to be in a steady state, where \Eq{eq:stationary-density-final} is applicable for the density profile.

\section{Details on absorption spectra}\label{sec:Details-spectra}

\subsection{Relation between $S(t)$ and $A(\omega)$}
To show the relation between the Ramsey signal $S_{\sigma_1\sigma_2}$ \Eq{eq:Ramsey-signal-general} and the absorption spectrum $A_\mathrm{pol}(\omega)$ given in the main text, we follow Ref.~\cite{schmidt2018theory}, only here the impurity is switched from $|1\rangle$ to $|\mathrm{R}\rangle$. The absorption spectrum is calculated by Fermi's golden rule where the initial state is the polaron $|\psi_i\rangle = |1\rangle\otimes|\mathrm{pol}\rangle$ and the final state is in the presence of the Rydberg impurity $|\psi_f\rangle = |\mathrm{R}\rangle\otimes|f\rangle$. The perturbation flips the state of the impurity into the Rydberg state and vice versa, i.e., $\hat\Omega = (|1\rangle\langle\mathrm{R}|+\mathrm{h.c.})\otimes\hat\doubleI$. By rewriting the delta function in Fermi's golden rule as a time integral, one obtains:
\begin{align}
	\nonumber A_\mathrm{pol}(\omega) &= 2\pi \sum_f |\langle\psi_f|\hat\Omega|\psi_i\rangle|^2\delta[\omega-(E_f-E_i)]\\
	&= \sum_f\int_{-\infty}^\infty\dd t\,\ee^{\ii\omega t}\langle \psi_i|\ee^{\ii E_it}\hat\Omega|\psi_f\rangle\langle \psi_f|\ee^{-\ii E_ft}\hat\Omega|\psi_i\rangle.
\end{align}
Alternatively, by inserting $\hat{\Omega}$ and using the fact that the impurity does not contribute to the total energies $E_i, E_f$, the form given in the main text is obtained. $|\psi_i\rangle$ and $|\psi_f\rangle$ are eigenstates with eigenenergies $E_i$ and $E_f$, respectively. Furthermore, the time evolution of the gas in the presence of the impurity $|\sigma\rangle$ is given by the Hamiltonian $\hat{H}_\sigma$ acting only on the subspace of the Fermi gas
\begin{subequations}
	\begin{align}
		\langle \psi_i|\ee^{\ii E_i t} &= \langle 1| \otimes \langle\mathrm{pol}| \ee^{\ii\hat H_1 t},\\
		\langle \psi_f|\ee^{-\ii E_f t} &= \langle R| \otimes \langle f| \ee^{-\ii\hat H_\mathrm{R}t},
	\end{align}
\end{subequations}
such that the overall absorption spectrum reads
\begin{align}
	A_\mathrm{pol}(\omega) &= \sum_f \int_{-\infty}^\infty\dd t\, \ee^{\ii\omega t} \langle\mathrm{pol}|\ee^{\ii\hat{H}_1t}|f\rangle\langle f|\ee^{-\ii\hat H_\mathrm{R}t}|\mathrm{pol}\rangle.
\end{align}
The sum over all possible final configurations for the gas $|f\rangle$ gives an identity so that,  in total, the absorption spectrum is the Fourier transform of the Ramsey signal $S_{1\mathrm{R}}(t)$ \Eq{eq:Ramsey-signal-general}. By separating the integration into negative and positive times, it is straightforward to bring the expression into the form which is given in the main text
\begin{align}
	A_\mathrm{pol}(\omega) &= 2\mathrm{Re}\int_0^\infty\dd t\,e^{\ii\omega t}S_{1\mathrm{R}}(t).
	\label{eq:A_pol-FT}
\end{align}
For numerical stability, we multiply the integrand with an exponential decay $f_\gamma(t) = \ee^{-\gamma t}$ (in our numerical calculations we use $\gamma=0.03\,\varepsilon_\mathrm{F}$). \Eq{eq:A_pol-FT} is calculated by fast Fourier transformation (FFT) up to a maximal time $t_\mathrm{max}$, which needs to be chosen high enough to give a good resolution in frequency space (cf.~\Sec{sec:Numerical_accuracy}).

\subsection{Details on the Ramsey signal}

The determinant for the Ramsey signal [cf.~\Eq{eq:Ramsey-signal-general}],
\begin{align}
	S_{1\mathrm{R}}(t) = \mathrm{det}[\hat\doubleI - n_\mathrm{F}(\hat h_1)+n_\mathrm{F}(\hat h_1)\ee^{\ii\hat h_1 t}\ee^{-\ii\hat h_\mathrm{R} t}],
\end{align}
can be calculated by introducing the single-particle eigenstates $|\bm{\alpha}_\sigma\rangle$ in presence of the local polaronic impurity $\sigma = 1$ and the long-ranged Rydberg impurity $\sigma = \mathrm{R}$ with corresponding eigenenergies $E_{\alpha_\sigma l}$, i.e., $\hat{h}_\sigma|\bm{\alpha}_\sigma\rangle = E_{\alpha_\sigma l}|\bm{\alpha}_\sigma\rangle$. In the eigenbasis of the polaron states $|\bm{\alpha}_1\rangle$, the matrix elements for the calculation of the determinant are:
\begin{align}
	\nonumber&\langle\bm{\alpha}_1|[\hat\doubleI-n_\mathrm{F}(\hat h_1)+ n_\mathrm{F}(\hat h_1)\ee^{\ii\hat h_1 t}\ee^{-\ii\hat h_\mathrm{R} t}]|\bm{\alpha}'_1\rangle \\
	\nonumber &=  \delta_{ll'}\delta_{mm'}\left[[1 - n_\mathrm{F}(E_{\alpha_1 l})]\delta_{\alpha_1\alpha'_1} \phantom{\sum_{\alpha_\mathrm{R}}}\right.\\ &\phantom{\delta_{ll'}\delta_{mm'}}\left. +n_\mathrm{F}(E_{\alpha_1 l})\ee^{\ii E_{\alpha_1 l}t}\sum_{\alpha_\mathrm{R} }\langle \alpha_1 l|\alpha_\mathrm{R} l\rangle  \ee^{-\ii E_{\alpha_\mathrm{R} l}t} \langle \alpha_\mathrm{R} l |\alpha'_1 l\rangle\right].
	\label{eq:Ramsey-polaron-details}
\end{align}
The overlaps $\langle \alpha_1 l|\alpha_\mathrm{R} l\rangle$ and $\langle \alpha_\mathrm{R} l |\alpha'_1 l\rangle$, respectively, are analogous to the expressions in \Eqs{eq:overlap-n-alpha-delta} and \eqref{eq:overlap-n-b-delta}, only for the eigensystem in the presence of the Rydberg impurity. They are obtained by numerical integration.

The whole expression \Eq{eq:Ramsey-polaron-details} is diagonal in the angular-momentum quantum numbers $l$ and $m$ due to the conversation of angular momentum. 
Let us call the term within the brackets in \Eq{eq:Ramsey-polaron-details} $M_{\alpha_1\alpha_1'l}$. As it is the same for each of the $2l+1$ blocks corresponding to the magnetic quantum number $m$, the total determinant in \Eq{eq:Ramsey-polaron-details} is calculated analogously to that in Ref.~\cite{sous2020rydberg}
\begin{align}
	S(t) &= \det(\delta_{ll'}\delta_{mm'}M_{\alpha_1\alpha_1'l}) = \prod_{l=0}^\infty \left[\det(M_{\alpha\alpha'l})\right]^{2l+1}.
\end{align}

\begin{figure}
	\includegraphics[width=\linewidth]{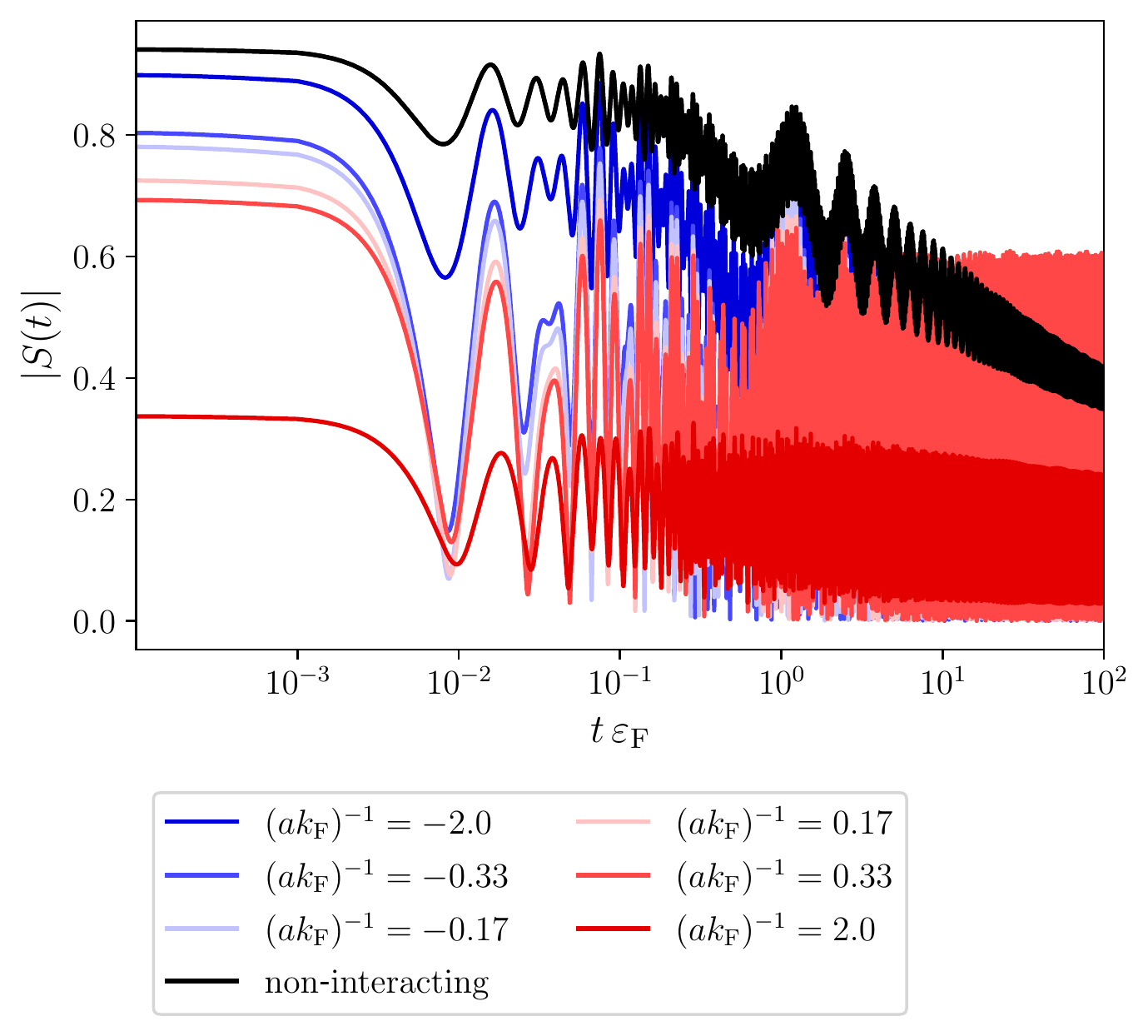}
	\caption{Ramsey signals $S(t)$ of a Rydberg atom with $n_\mathrm{Ryd}=60$ in a polaron formed at different inverse scattering lengths $(ak_\mathrm{F})^{-1}$. The fine oscillations with periods $\tau = 2\pi/|\omega_\mathrm{peak}|$ correspond to the dimer peak of the Rydberg molecules. The loss of spectral weight at $t=0$ is due to the omission of highly energetic bound states, but does not affect the quality of the dimer peaks.} 
	\label{fig:Ramsey-polaron}
\end{figure}

We provide the Ramsey signals of a Rydberg atom in different polaron clouds in \Fig{fig:Ramsey-polaron}. The fine oscillations have a period of $\tau = 2\pi/|\omega_\mathrm{peak}|$ and  correspond to the dimer peaks at $\omega_\mathrm{peak}$ in the absorption spectra. In order to resolve absorption peaks at high frequency values, it is essential to resolve the time dependence of the Ramsey signal $S(t)$ optimally (cf.~\Sec{sec:Numerical_accuracy}).

We see that our calculated Ramsey signals do not take into account enough eigenstates in order to reach the exact value $S(t=0)=1$. This is due to the fact that the Rydberg impurity generates a lot of more deeply bound states with more overlap to the polaron's bound states or the polaron's scattering states, which we do not take into account (cf.~\Sec{sec:Rydberg-system}). As we are primarily concerned about the overlap with the outermost Rydberg molecular state, the missing weight in the global spectrum does not affect our final results.

\subsection{Reconstructed density by tuning $\rho_0$}
\begin{figure}
	\includegraphics[width=\linewidth]{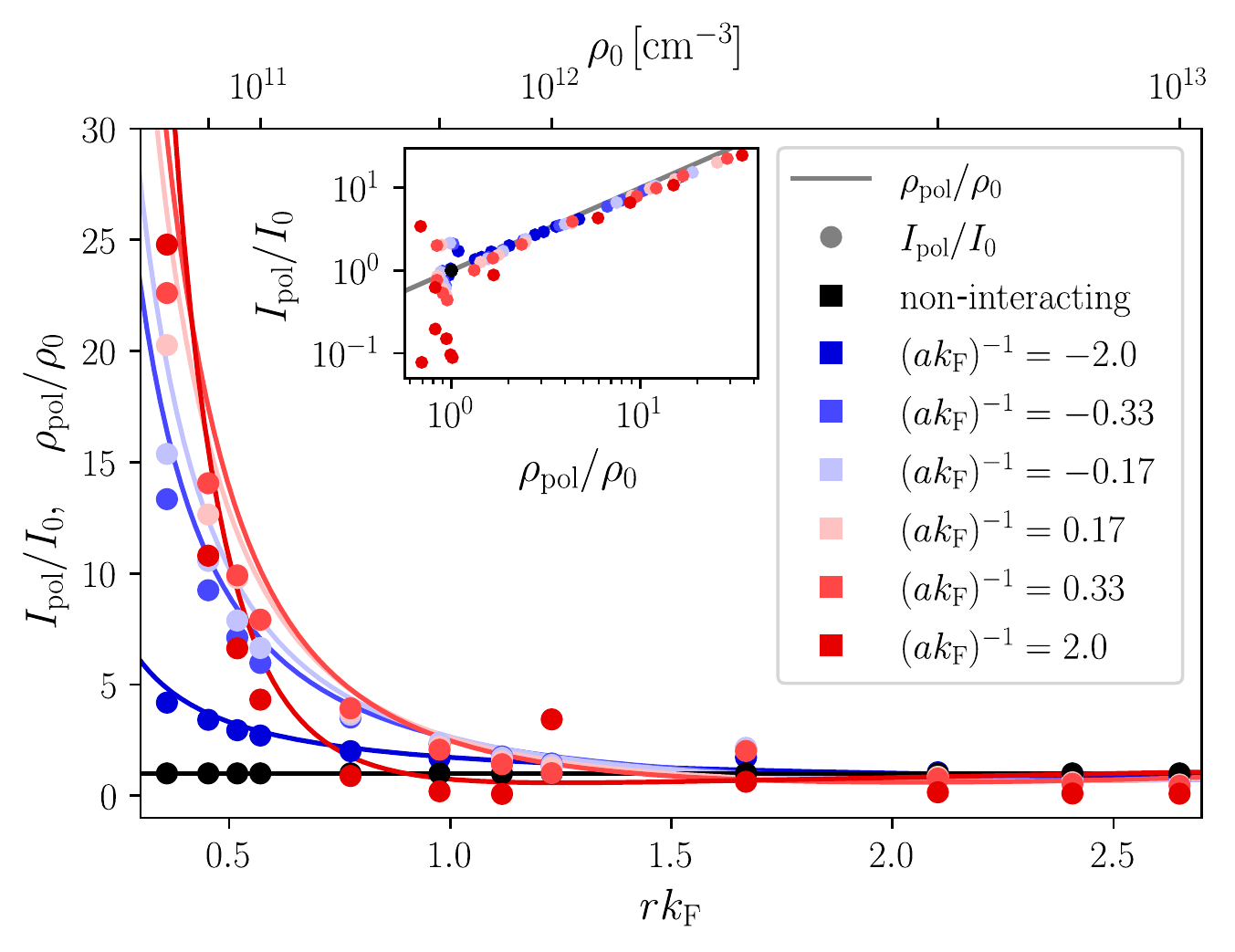}
	\caption{Normalized density profiles $\rho_\mathrm{pol}(r)/\rho_0$ for different polaron clouds with inverse scattering lengths $(a k_\mathrm{F})^{-1}$ (solid lines) are compared to the integrated dimer peaks $I_\mathrm{pol}(r_\mathrm{Ryd})/I_0(r_\mathrm{Ryd})$ (dots), which correspond to the Rydberg radius through the respective overall density, i.e., $r_\mathrm{Ryd}(\rho_0)$. The inset shows the analysis of the dependence between the integrated absorption peak weights $I_\mathrm{pol}/I_0$ and the corresponding densities $\rho_\mathrm{pol}/\rho_0$.}
	\label{fig:Rydberg-densities-delta-as_rhos}
\end{figure}

In the main text, we have mentioned that the polaron's density can also be reconstructed from Rydberg spectroscopy for fixed principal number $n_\mathrm{Ryd}$ and different background densities $\rho_0$. Such a plot is provided in \Fig{fig:Rydberg-densities-delta-as_rhos}.

In the experimental setup, the Rydberg radius $r_\mathrm{Ryd}$ is of course fixed for a specific atom type and principal number $n_\mathrm{Ryd}$, regardless of the density $\rho_0$ of the surrounding medium. However, the polaron's density profile is determined by the Fermi momentum of the gas. As we give physical quantities  in units of the Fermi momentum, the relative size between the Rydberg impurity at fixed $n_\mathrm{Ryd}$ and the polaron with radius $r_c\sim k_\mathrm{F}^{-1}$ changes when tuning the background density $\rho_0 = \rho_0(k_\mathrm{F})$. This has already been clarified in \Fig{fig:Ryderg-potential}, where different density values $\rho_0$ lead to different Rydberg radii $r_\mathrm{Ryd}$ in units of the Fermi momentum.

\Fig{fig:Rydberg-densities-delta-as_rhos} is analogous to the plot given in the main text, only here we have fixed the principal number to $n_\mathrm{Ryd}=59$ and associated the position dependence $rk_\mathrm{F}$ by changing the overall density $\rho_0$ as discussed before. When changing the overall density, the shape of the dimer peaks differ much more from one to the other than when changing the principal numbers \cite{sous2020rydberg}. That is why overall the data in \Fig{fig:Rydberg-densities-delta-as_rhos} are more noisy than those provided in the main text. However, the agreement between integrated weights and density values is still obvious. The two complementary methods  give more flexibility in the experimental realization and underline the validity of our technique.

\subsection{Effect of finite temperature}

In this section, we elaborate that a finite temperature does not have a significant impact on our data. Thus, our method is robust against temperatures used in typical ultracold atomic gases. First of all, FDA does not have any conceptual limitation to zero temperature. Temperature only enters in the Fermi distribution $n_\mathrm{F}(\varepsilon)$, which is included in the density $\rho_\mathrm{pol}(\bm{r})$ \Eq{eq:stationary-density-final} and the Ramsey signal $S_{\sigma_1,\sigma_2}(t)$ \Eq{eq:Ramsey-signal-general}.

\begin{figure}
	\includegraphics[width=\linewidth]{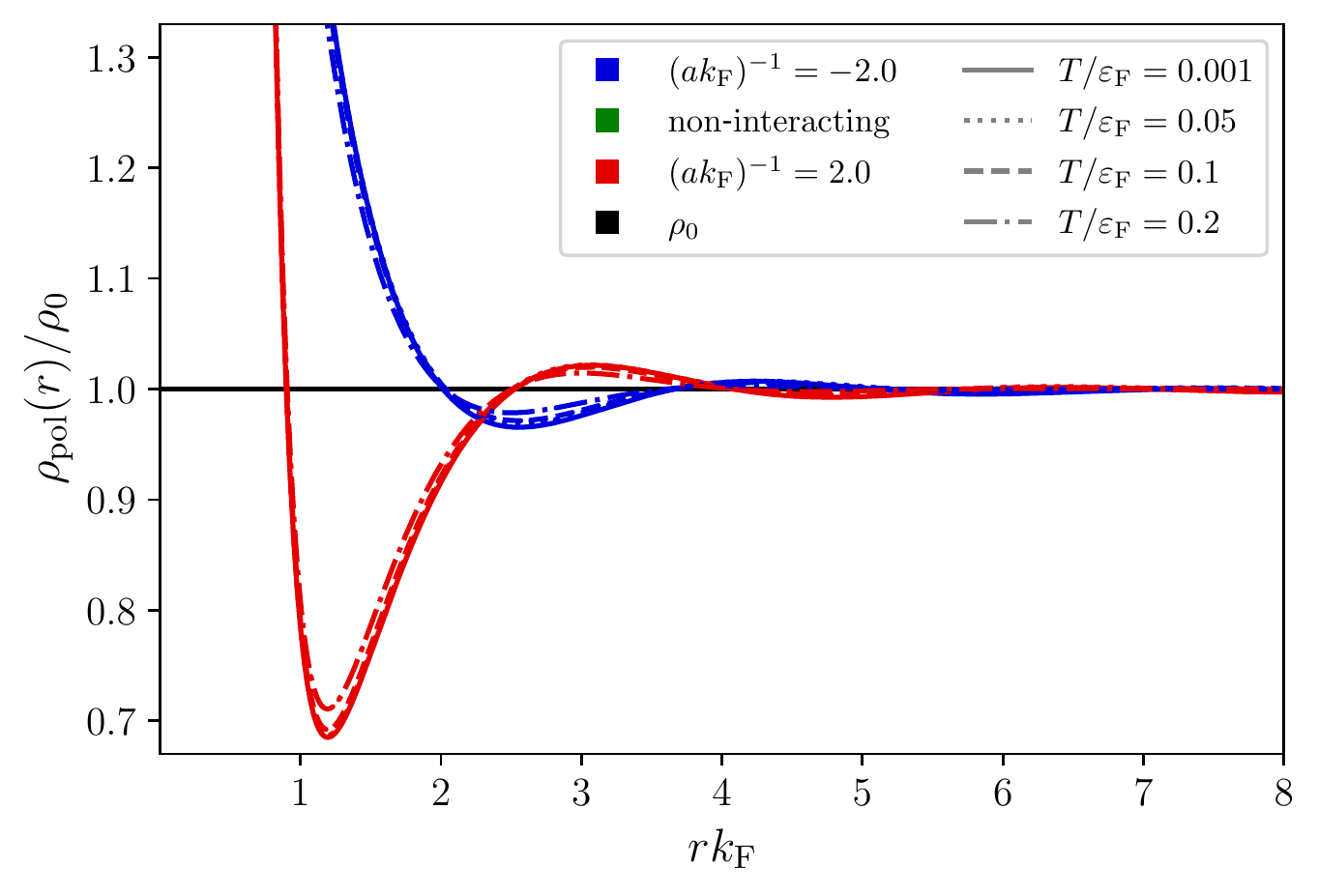}
	\caption{Density profiles $\rho_\mathrm{pol}(r)$ \Eq{eq:stationary-density-final} in the presence of a delta impurity with different inverse $s$-wave scattering lengths $(ak_\mathrm{F})^{-1}$ and temperatures $T$.}
	\label{fig:densities-delta-as_temp}
\end{figure}

\Fig{fig:densities-delta-as_temp} shows the density profiles of polaron clouds $\rho_\mathrm{pol}(r)$ for different temperatures $T$. For higher temperatures, the chemical potential $\mu$ is lowered such that the background density $\rho_0 = k_\mathrm{F}^3/(6\pi^2)$ is kept constant. At $T=0.05\,\varepsilon_\mathrm{F}$, which is basically the state of the art for fermionic quantum mixtures \cite{lous2017thermometry}, the difference to $T=0.001\,\varepsilon_\mathrm{F}$, which is used in the rest of our work, is barely visible. We compare these density values to those at $T=0.1\,\varepsilon_\mathrm{F}$ and $T=0.2\,\varepsilon_\mathrm{F}$, which are typical temperatures of Fermi polaron experiments. With increasing temperature, the density enhancement in the center as well as the accompanied oscillations are softened. This can be explained by the averaging over various statistical realizations. However, already here we see that this is only a slight effect so our probe barely changes with temperature.

\begin{figure}
	\raggedright
	(a)\\
	\vspace{-0.4cm}
	\centering 
	\includegraphics[width=\linewidth]{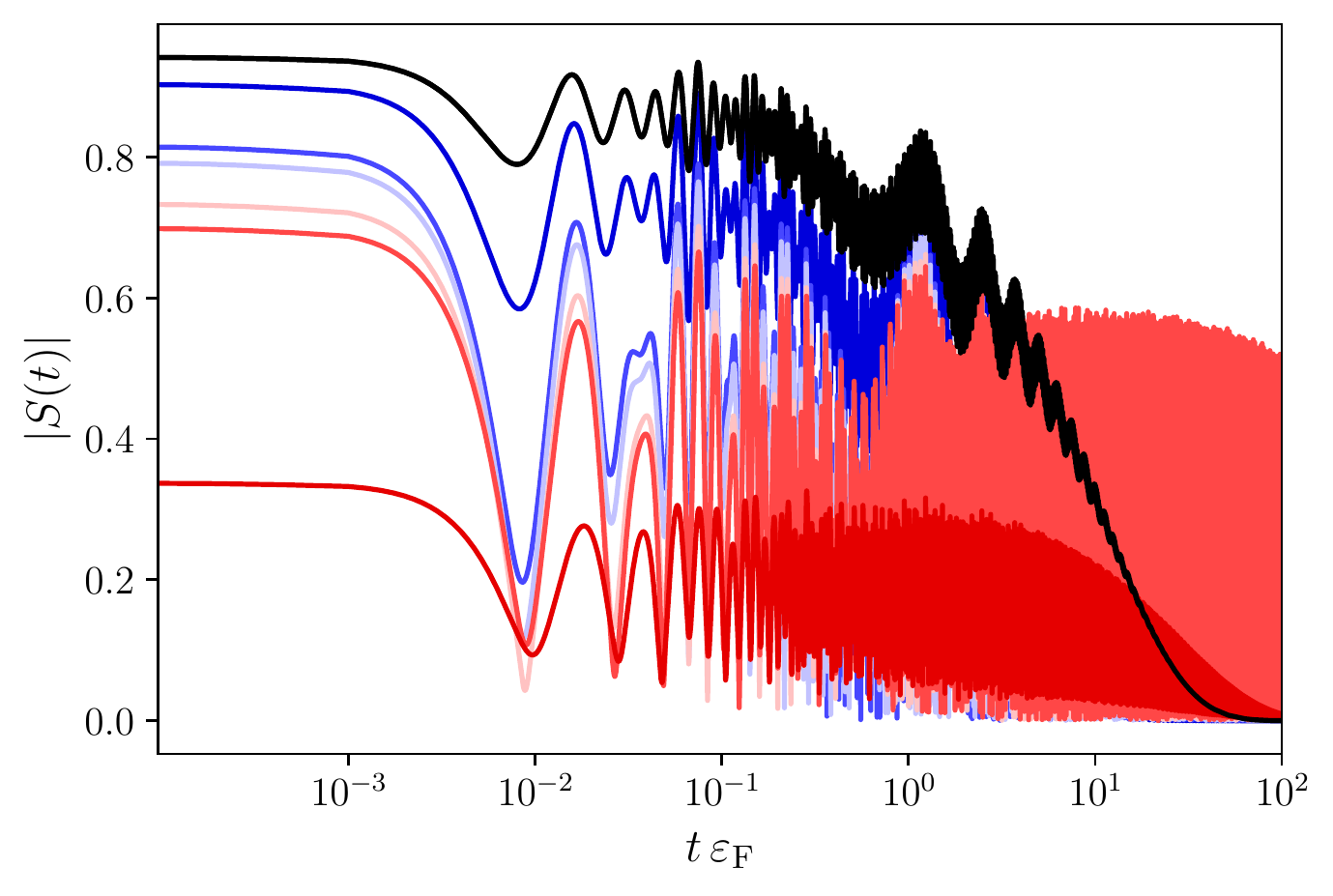}
	\raggedright
	(b)\\
	\vspace{-0.4cm}
	\centering
	\includegraphics[width=\linewidth]{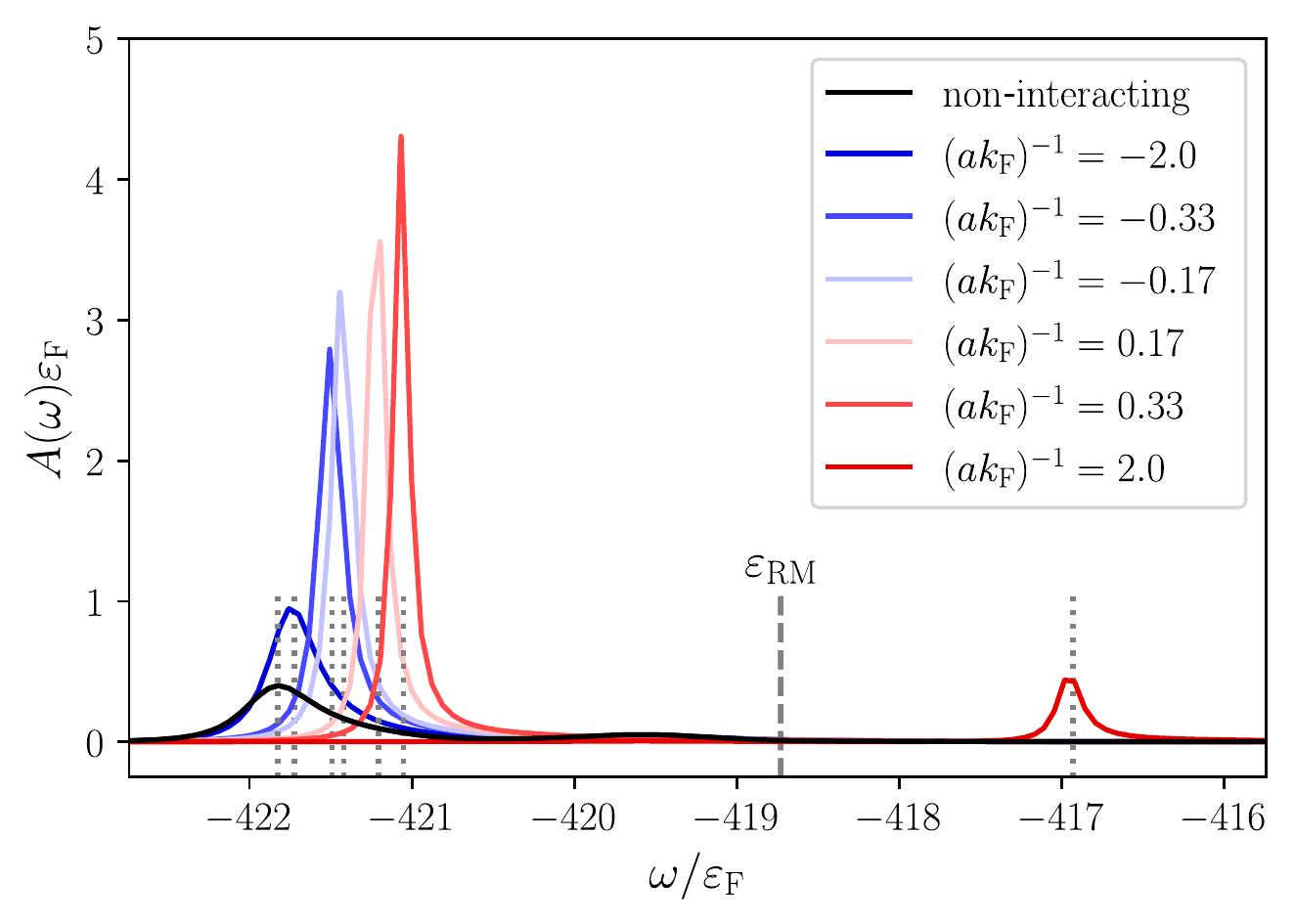}
	\caption{(a) Ramsey signals $S(t)$ of a Rydberg atom with $n_\mathrm{Ryd}=60$ in a polaron formed at different inverse scattering lengths $(ak_\mathrm{F})^{-1}$ for a temperature $T=0.2\,\varepsilon_\mathrm{F}$.
		(b) RM peaks of the corresponding absorption spectrum. The peak positions $\omega_\mathrm{peak}$ are marked by gray doted lines and the binding energy $\varepsilon_\mathrm{RM}$ by a dashed gray line.} 
	\label{fig:Spectra-for-finite-T}
\end{figure}

\begin{figure}
	\includegraphics[width=\linewidth]{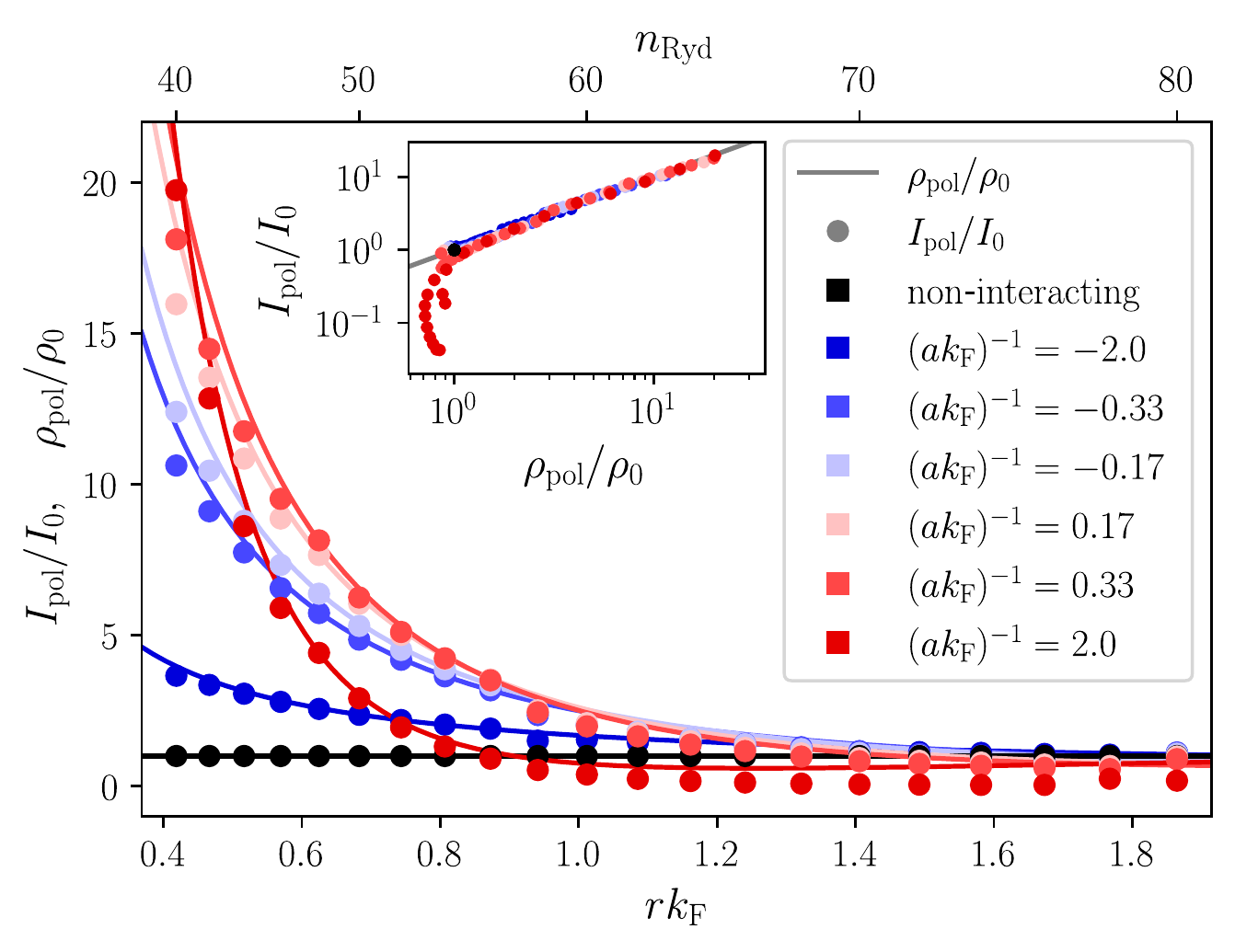}
	\caption{Normalized density profiles $\rho_\mathrm{pol}(r)/\rho_0$ for different polaron clouds with inverse scattering lengths $(ak_\mathrm{F})^{-1}$ (solid lines) are compared to the integrated dimer peaks $I_\mathrm{pol}(r_\mathrm{Ryd})/I_0(r_\mathrm{Ryd})$ (dots). The latter corresponds to the Rydberg radius through the respective principal numbers, i.e., $r_\mathrm{Ryd}(n_\mathrm{Ryd})$. We use a fixed $\rho_0={5\times 10^{11}\,\mathrm{cm}^{-3}}$ and a finite temperature $T=0.2\,\varepsilon_\mathrm{F}$. The inset shows the dependence between $I_\mathrm{pol}/I_0$ and the densities $\rho_\mathrm{pol}/\rho_0$. }
	\label{fig:Rydberg-densities-delta-as_temp}
\end{figure}

To validate that our method of determining density profiles from Rydberg atom spectroscopy stays stable, we provide Ramsey signals, absorption spectra and the reconstructed density profiles (cf.~\Figs{fig:Spectra-for-finite-T} and \ref{fig:Rydberg-densities-delta-as_temp}) in analogy to those given in the main text, only here for a temperature $T=0.2\,\varepsilon_\mathrm{F}$.

For a finite temperature, the Ramsey signals $S_\mathrm{pol}(t)$ decay faster [cf.~\Fig{fig:Spectra-for-finite-T}(a)] as the orthogonality catastrophe is not fulfilled anymore \cite{knap2012time}. This leads to a broadening of the dimer response peaks in the absorption spectra [cf.~\Fig{fig:Spectra-for-finite-T}(b)]. Note that, due to the finite temperature, the peaks are no longer asymmetrically cut on the left, but their shape is more Gaussian-like. The positions of the peak's maxima, however, do not change and reveal the polaron energy [cf.~Eq.~(7) in the main text].

The tight relation between the integrated dimer response $I_\mathrm{pol}$ and the actual density $\rho_\mathrm{pol}$ is still recovered for $T=0.2\,\varepsilon_\mathrm{F}$. This is illustrated in \Fig{fig:Rydberg-densities-delta-as_temp}, whose similarity to the plot at $T=0.001\,\varepsilon_\mathrm{F}$ given in the main text is undeniable. This is due to the fact that our method only relies on the integrated spectral weight of the dimer response $I_\mathrm{pol}$, which is basically unaffected by a finite temperature in contrast to the actual shape of the peak.

In a more realistic setting, the absorption spectra might be broadened also through other effects like the finite lifetime of the Rydberg excitation or the mobility of the impurity. Similar to the effect of a finite temperature, we expect our method to be robust as long as these effects do not significantly change the spectral weight of the dimer response.

\section{Numerical accuracy}\label{sec:Numerical_accuracy}

In this section, we briefly mention how we choose our numerical parameters in order to achieve high accuracy in the calculated quantities. The choice of $l_\mathrm{max}$ and $R$ is already motivated in Secs.~\ref{sec:lmax} and \ref{sec:R}, respectively.

The maximal number $\alpha_\mathrm{max}$ of single-particle scattering states $u_{\alpha l}(r)$ is chosen  such that enough states above the Fermi energy are taken into account. For a brief estimate, we consider the noninteracting impurity $|0\rangle$ at zero temperature and angular momentum. The highest radial quantum number for $n$, which needs to be considered in the calculations, is determined by the Fermi momentum
\begin{align}
	\frac{k_{n_\mathrm{max}}^2}{2m}=\frac{n_\mathrm{max}^2\pi^2}{2mR^2}\geq z\,\varepsilon_\mathrm{F} \Rightarrow n_\mathrm{max} \geq \sqrt{z}\frac{k_\mathrm{F}R}{\pi}.
\end{align}
To increase precision, we want to take into account the doubled amount of maximal energy, i.e., $z=2$, such that $n_\mathrm{max}=\lfloor\sqrt{2}k_\mathrm{F}R/\pi\rfloor$ states are considered. In our calculations, we use $Rk_\mathrm{F}=400$ and $n_\mathrm{max}=250=\alpha_\mathrm{max}$.

Let us now discuss the maximal time, we can take into account in the Ramsey signals $S(t)$ used in the Fourier transform for the absorption spectrum \Eq{eq:A_pol-FT}. For this, we again consider the noninteracting system. Because of the finite system size, there is a difference of discrete momenta, i.e., $\delta k = \pi/R$, which leads to a differences of energies, i.e., $\delta E = k/m \cdot \delta k = \pi k/(mR)$. The largest difference in energy $\delta\varepsilon_\mathrm{max}$ provides an upper time limit for the Ramsey signal $t_\mathrm{max} \leq 2\pi/\delta \varepsilon_\mathrm{max}$ before finite-size effects occur. With $k_\mathrm{max} = k_\mathrm{F}$, we thus have:
\begin{align}
	t_\mathrm{max}\leq\frac{2mR}{k_\mathrm{F}}.
\end{align}
In natural units, i.e., $2m = 1 = k_\mathrm{F}$, the maximal time is just given by the system size $t_{\max}\leq R$. For our data we use $t_\mathrm{max}=100$.

The maximal time, however, corresponds to a finite frequency resolution $\delta\omega$ when taking the numerical Fourier transform. Thus, the absorption spectra $A(\omega)$ are not resolved more accurately than the maximal energy accuracy. In natural units, i.e., $2m = 1 = \varepsilon_\mathrm{F}$, this takes the values $\delta\omega \sim 2\pi/ R$.

On the other hand, the resolution of the Ramsey signals (cf.~\Fig{fig:Ramsey-polaron}) corresponds to a maximal frequency $\omega_\mathrm{max}$ in the absorption spectrum reached by the FFT. This maximal frequency has to be larger than the binding energy of the Rydberg molecule, i.e., $\omega_\mathrm{max} > |\varepsilon_\mathrm{RM}|$ such that the dimer peak is included in the spectrum. This corresponds to a minimal time resolution of the Ramsey signal
\begin{align}
	\delta t < \frac{2\pi}{\omega_\mathrm{max}} = \frac{2\pi}{|\varepsilon_\mathrm{RM}|}.
\end{align}

In fact, the resolution of our data is affected by the system size $R$. In addition, we need extremely well resolved Ramsey signals in order to reach high accuracy in the absorption spectra for the highly energetic values of the dimer peaks.

\end{document}